\documentclass[twocolumn,preprintnumbers,prl,amsmath,amssymb,aps]{revtex4-2}

\usepackage{amsmath}
\usepackage{amssymb}
\usepackage{graphicx}
\usepackage{dcolumn}
\usepackage{bm}
\usepackage{color}
\usepackage{lineno}

\newcommand{\beq}{\begin{equation}}

\newcommand{\eeq}{\end{equation}}

\newcommand{\bea}{\begin{eqnarray}}
\newcommand{\eea}{\end{eqnarray}}

\newcommand{\aap}{Astron.\ Astrophys.}
\newcommand{\mnras}{Mon.\ Not.\ R.\ Astron.\ Soc.}
\newcommand{\mnrasl}{Mon.\ Not.\ R.\ Astron.\ Soc.: Lett.}
\newcommand{\physrep}{Phys.\ Rep.}
\newcommand{\aj}{Astron.\ J.}
\newcommand{\apjl}{Astrophys.\ J.\ Lett.}
\newcommand{\apjs}{Astrophys.\ J.\ Suppl.\ Ser.}

\newcommand{\astronlett}{Astron.\ Lett.}
\newcommand{\jhea}{J.\ High Energy Astrophys.}

\newcommand{\apss}{Astrophys.\ Space Sci.}
\newcommand{\sovast}{Sov.\ Astron.}

\newcommand{\pasj}{Pub.\ Astron.\ Soc.\ Jpn.}
\newcommand{\pasa}{Publ.\ Astron.\ Soc.\ Aust.}
\newcommand{\llr}{Living Rev.\ Relativ.}

\newcommand{\jpg}{J.\ Phys.\ G}
\newcommand{\nrm}{Nat.\ Rev.\ Mater.}
\newcommand{\natmater}{Nat.\ Mater.}

\newcommand{\natastron}{Nat.\ Astron.}
\newcommand{\natcomm}{Nat. Commun.}
\newcommand{\currsci}{Curr.\ Sci.}
\newcommand{\ini}{\mathrm{ini}}

\begin{document}

\title{Crust glass formation reveals the neutron star birth properties in IGR J17480-2446}

\author{D. A.\ Baiko}
\email{baiko.astro@mail.ioffe.ru}
\author{A. I.\ Chugunov}
\email{andr.astro@mail.ioffe.ru}
\affiliation{Ioffe Institute, Politekhnicheskaya 26, 194021 Saint Petersburg, Russia}

\date{\today}


\begin{abstract} 
IGR J17480-2446 is a low-mass X-ray binary,
harboring an exceptional accreting pulsar (a neutron star)
with an unusual spin frequency of 11 Hz
and a very slow post-outburst crust cooling.
The former may imply that it is observed at 
an early stage of recycling, 
while the latter was shown to 
indicate the presence in the outer crust 
of a low thermal conductivity layer, 
possibly made of glass.  
Here we argue that the glass layer formation 
is a natural result of accretion induced failure
of pristine cold crystal crust. This allows us to determine 
the mass of the accreted material as $\Delta M \approx 2.4\times 10^{-6}~M_\odot$, 
confirming very early accretion stage for this neutron star.
An analysis of spin and thermal state reveals 
a peculiar set of neutron star birth properties which 
is commonly associated with `recycled' neutron stars, 
i.e.\ those that have been experiencing
prolonged periods of accretion from a companion.
We speculate that such birth properties
may represent the outcome of neutron star formation in 
an electron-capture supernova. 
\end{abstract} 

\maketitle

{\it Introduction.} Neutron stars (NSs) are fascinating objects with a typical mass of 
$\sim 1.4 \, M_\odot$ and a radius of just 
$\sim 12 \, {\rm km} \approx 1.7 \times 10^{-5} \, R_\odot$ . 
With central densities well in excess of the standard nuclear saturation 
density 
and main properties of matter
(composition, thermodynamics, kinetics) governed by strong interaction,
NSs are excellent laboratories for nuclear physics 
\cite{hpy07,LP07_NSobservations,Baym+18,Compstar19,Lattimer21_NS_NSEOS,oks26_NS_EOS_review}.
 
The superdense core of an NS is enshrouded by a relatively thin 
($\sim 1$ km thick) crust whose density decreases outwards, from nuclear to 
typical terrestrial densities, covering a wide range of 
physical conditions not accessible in a laboratory. 
These include
matter composed of
exotic (neutron-rich, possibly extremely 
aspherical) atomic nuclei, unbound neutrons, and electrons; 
strongly coupled
fully ionized 
Coulomb plasma; 
various partially ionized states with strong magnetic field effects.    
Understanding crust properties is 
crucial for correct interpretation of the information related to the NS core 
and for verification of theoretical plasma models 
\cite{ch08,ch17,mdkse18,BlaschkeChame18_PhasesOfDenseMatter}.

NSs are also of fundamental importance for astrophysics
as they are thought to be responsible for numerous striking 
observational phenomena
such as radiopulsars, millisecond pulsars (MSPs), magnetars, central compact objects etc
\cite{Kaspi10_GandUnif_NS,DeLuca17_CCO,DemorestGoss24_MSP_discov_Hist,BCZ26_NSzoo_encicl,
Wahl26_NS_and_Pulsars_Encicl,RDG26_Magnetars_Encicl}.
Several scenarios of NS formation have been proposed. The most
standard is a collapse of the iron core of a massive star (hereafter FeCC) promoted
by iron photo-disintegration and giving rise to spectacular supernovae
events 
\cite{CD_ER18_NSformation_CCSN,CCSN25_encicl}. 
An alternative route to an NS is a collapse of a degenerate ONeMg core 
\cite{M+80,Wang_ea26_ECSN_review}. The latter is promoted by a loss of
degenerate electron pressure due to electron captures on $^{24}$Mg and $^{20}$Ne, 
if the core exceeds a critical mass of $\sim 1.37 \, M_\odot$. 
Presently, it is poorly understood whether particular NS formation scenarios
produce certain observational classes of NSs or
NSs with specific birth properties. 
See End Matter (EM) for a brief review of the available information. 

NSs can be found in binary systems 
with or without mass transfer onto an NS 
\cite{Lorimer08_LRR}. 
Among mass-transferring systems, 
of special interest
are low-mass X-ray binaries (LMXBs).
These systems are thought to be progenitors of MSPs
via the recycling scenario
\cite{BK74,BvdH91}.
There is however an LMXB/MSP birthrate problem 
\cite{kn88,Kulkarni+90,FW07_MSP_birth_prop}, 
and an alternative scenario
of direct MSP birth via accretion-induced collapse (AIC) has been proposed
\cite{BG90_NS_MSP_from_AIC,H+10}.

Transiently accreting LMXBs 
constitute an important subclass 
\cite{bbr98,ylpgc04,Brown_eal18_RapidNeutrinoCooling_MXB,pcc19,Heinke_ea25_OutburstCatalogue}. 
For $\sim 10$ of these sources, real time cooling was observed 
\cite{WDP17_CoolingAccreted},
which provided an opportunity for a 
detailed analysis. 
This cooling was interpreted as thermal relaxation of the crust, which was heated out 
of thermal equilibrium with the core by accretion; modeling this process allows one to 
deduce dense matter properties such as
heat capacity and thermal conductivity 
\cite{UR01_TimeVariableEmission_Transients,Ruthledge_etal02_KS,bc09,WDP17_CoolingAccreted}.

IGR J17480-2446 (hereafter J1748; $\mathrm{NS_{J1748}}$ its NS), the main
object
of the present paper, 
belongs to this last group
\cite{Papitto_ea11_IGR17480,Testa_ea12_IGR17480_IR,Patruno_ea12_IGR17480,Wijnands_ea13_partaccr, MC14_CompConv_Transients}.
An interpretation of unique kinetic properties of the crust of
$\mathrm{NS_{J1748}}$
\cite{Ootes_ea19,Potekhin_ea25_transients}, 
allows us to address NS birth properties
in a degenerate collapse as well as to contribute to understanding
of magneto-rotational evolution of NSs in binaries.

{\it J1748.}
Only one episode of active accretion can be certainly attributed to
J1748 \cite{Ootes_ea19}. 
It took place in 2010 \cite{DW11_AccretionHeatedCrust_J17480_2446}. 
Subsequent X-ray observations detected
thermal emission with decreasing temperature interpreted as crust cooling
\cite{DBW11_CrustCooling_J17480_2446,Ootes_ea19}.
The observed cooling was unusually slow. This was shown to 
be inconsistent with theoretical models of thermal evolution, unless 
a region with low thermal conductivity was phenomenologically introduced 
in the outer crust \cite{Ootes_ea19,Potekhin_ea25_transients}.

We remind that matter in 
the outer crust consists of bare atomic
nuclei and an almost uniform charge-neutralizing 
degenerate electron gas 
\cite{hpy07,ch08}.
In the standard model, the nuclei are arranged in a crystal lattice,
and electron conductivities are very high.
An attempt was made \cite{Ootes_ea19} to link the poor conduction 
with multicomponent nuclear composition of the crust.
The latter was described by the impurity parameter
$Q \equiv \langle Z^2\rangle-\langle Z\rangle^2$,
where $Z$ is the charge number and $\langle \ldots \rangle$ here denotes averaging
over nuclear species.
However, a study
within the Markov
chain Monte Carlo framework had shown that the observational data 
required $Q\sim 10^3$
\cite{Ootes_ea19}, which
implied a very peculiar composition and exceeded $Q$ predicted in detailed 
models based on nuclear reaction networks \cite{Lau_ea18,Shchechilin_ea21} by 
more than an order of magnitude.

Accordingly, it was proposed \cite{Ootes_ea19,Potekhin_ea25_transients}
that the thermal conductivity was low not due 
to a strong impurity contamination, but because of a lack of a long-range 
(i.e.\ crystalline) order in ion positions so that the matter in the 
layer was in an amorphous (glass) state.
To fit the cooling data, the authors
of Ref.\ \cite{Potekhin_ea25_transients} placed the poorly conducting layer 
at densities 
$\rho_1 \lesssim \rho \lesssim \rho_2$, where
$\rho_1=3\times 10^{10}$ g/cc, $\rho_2=3\times 10^{11}$ g/cc were set as fiducial values.
Let us stress, that Ref.\ \cite{Potekhin_ea25_transients} proposed the glass layer 
on pure phenomenological grounds, providing neither model of glass 
formation nor of its density range.

In this work, we present 
a natural model of glass formation 
and analyze its astrophysical consequences.

{\it Model.} The idea of a glassy state of matter in NS crust has been put forward 
previously \cite{Jones99_AmorphousCrust,Jones01}, but the ultimate 
source of disorder
in these works was a strong nuclear charge heterogeneity.
Since we here consider consequences of the lack of a positional 
long-range order, we shall assume that at any given density 
there is only one sort of nuclei.   
A realistic model of matter is then a one-component Coulomb plasma
or Yukawa plasma
in the weak-screening regime
(point-like 
nuclei on almost uniform 
neutralizing background of electrons).

It is well known that a glassy state of terrestrial materials can be 
obtained by a rapid quench of a liquid \cite{Sun_ea16_NatRM_MetallicGlass}.
The same was demonstrated  
for Coulomb
or Yukawa 
plasmas \cite{Ogata92,Hammerberg_ea94}.
In these simulations, a cooled disordered plasma was not able to `find' a 
crystal state which clearly had lower free energy, due to
the fact that too low kinetic energy of ions did not allow them to fully
explore the configuration space.

We are not aware of any physical process which could produce a rapid 
quench of a liquid NS crust. Simulations of crust heating and 
relaxation \cite{Potekhin_ea25_transients} indicate that the 
layer in question is below crystallization 
temperature $T_{\rm m}$ both during and after the outburst.
Thus, a
plausible path to a glassy state
must 
start from a cold crystalline state. 
We suggest that this path is associated with crystal breaking under an excessive stress.
Indeed, if the crystal failure results in a disordered state 
and if the temperature is sufficiently 
low, it would take exponentially long for matter to nucleate a new crystal.
The stress in question can be due to the weight of the accreted material.
A similar process of stress-induced amorphization of crystals
is known in terrestrial 
conditions \cite{RH97_Pressure_induced_amorphization,Greaves_ea03_NatMat_amorphization,
Bu_ea24_NatCom_Amorphization}
and was briefly discussed for Yukawa plasma \cite{HK09_breaking}.
 
\begin{figure}
\begin{center}
\leavevmode
\includegraphics[width=87mm]{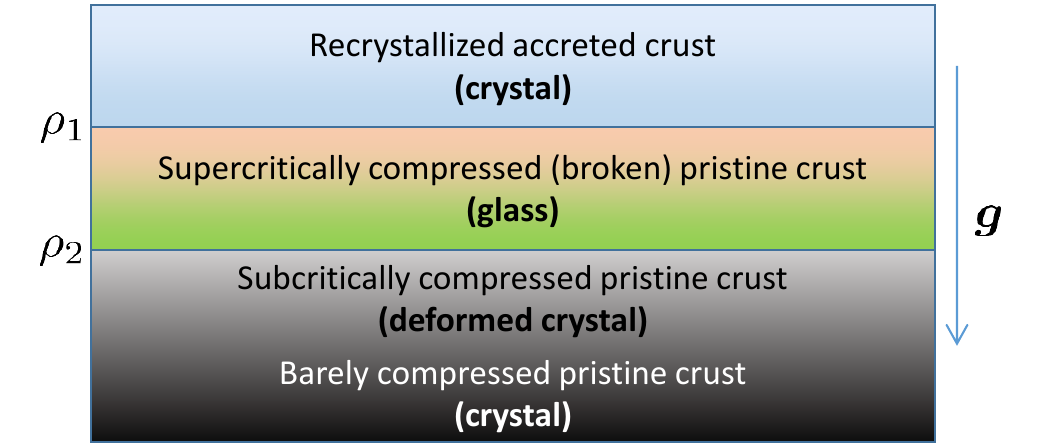}
\end{center}
\vspace{-0.3cm}
\caption[ ]{A cartoon of glass layer formation in NS crust.}
\label{Fig:Scheme}
\end{figure}

The initial accretion stage can be viewed as loading of the crystallized 
pristine crust by the accreted material (Fig.\ \ref{Fig:Scheme}).
The crystal strength decreases with depth decrease. At a given time, 
the loading produces critical stress at a density, which 
can be identified with $\rho_2$. At 
$\rho<\rho_2$,
the critical stress
has been exceeded and the crystal is broken;
at $\rho>\rho_2$, the original 
(albeit deformed) crystal survives
(grey region in Fig.\ \ref{Fig:Scheme}; darker color 
illustrates 
higher densities and weaker deformation).

To be consistent with observations, a broken crystal at 
$\rho<\rho_1$ should have been able to form a new crystal state. 
To understand the origin of 
a specific $\rho_1$ value, let us introduce
the dimensionless 
coupling parameter
\beq
  \Gamma = \left(\frac{4\pi e^6}{3m_{\rm u}}\right)^{1/3} 
   \frac{(Z^5 Y \rho)^{1/3}}{T}~,
\label{Gamma}
\eeq
where $e$ is the elementary charge, $m_{\rm u}$ is the atomic mass unit,
$Y\equiv Z/A$, $A$ is the mass number, $T$ is the temperature.
Crystallization 
occurs at 
$\Gamma = \Gamma_{\rm m} \approx 175.7$
\cite{BC22}.

Glass transition in a one-component plasma is characterized by a value $\Gamma_{\rm g}$ above which
a crystal does not arise from a disordered state. 
Theoretical predictions for $\Gamma_{\rm g}$ have considerable 
scatter
and can be roughly divided into higher and lower values,
$\Gamma_{\rm g} \gtrsim 400-600$ \cite{Ogata92,Hammerberg_ea94,CT21_GlassOCP}
and $\Gamma_{\rm g} \lesssim 300-400$ \cite{Yazdi_ea14_Glass_Yuk,CT21_GlassOCP}.
However, the explicit dependence of $\Gamma$ on density (Eq.\ \ref{Gamma}),
is weak, and the temperature is a smooth function
of $\rho$ due to high thermal conductivity. A natural physical mechanism 
which would set a particular value of 
$\rho_1$ could be 
a jump of $Z$, and indeed, such a jump is predicted for the boundary 
between the pristine and accreted nuclear compositions of the crust.

According to Kepler or Superburst 
outer crust models of Ref.\ \cite{Shchechilin_ea21},
accreted matter at $\rho_1$ is expected to have average
charge number $\langle Z \rangle\approx 21-22$ 
(average mass number $\langle A \rangle\approx 52-55$, $Y\approx 0.4$).
The pristine crust in the immediate vicinity 
of $\rho_1$ has
large $Z$ \cite{Chamel20}:  
starting from $Z=36$ ($A=86$, $Y\approx 0.42$), 
it gradually drops to $Z=28$, then jumps to $Z=44$, and eventually, 
at $\rho_2$, settles at $Z=38$ ($A=120$, $Y\approx 0.32$). 
Thus, 
at $T\sim (2-3)\times 10^8$ K and $\rho=\rho_1$, 
accreted matter has $\Gamma \sim 290-430$ i.e.\ $\lesssim \Gamma_{\rm g}$, whereas 
for pristine
matter, $\Gamma \sim 690-1030 > \Gamma_{\rm g}$,
locking supercritically compressed (i.e.\ broken) pristine crust matter 
in a glassy state. 

Summarizing, we predict the glass layer to be bounded by conditions:
a) the layer above it is accreted and thus has low $Z$; b) the region
below it is made of subcritically compressed pristine matter (crystal is not broken); 
the glass region is shown by green color in Fig.\ \ref{Fig:Scheme}.

{\it Other transiently accreting NS.}
It is
appropriate to
confront our glass formation model with 
available constraints on the crustal transport properties in other 
transiently accreting LMXBs with observed NS cooling.

It has been shown \cite{Cackett_ea06_CoolingKS_MXB}
that crustal thermal conductivity
of another transient source, KS 1731-260, is high, in particular, 
that an `amorphous' NS crust is not consistent with observations of its 
post-outburst relaxation \cite{Cackett_ea06_CoolingKS_MXB,syhp07,Jain_ea25_KS1731}.
Similarly, fitting  thermal relaxation of the transient source MXB 1659-29 
\cite{Potekhin_ea23_transients,Potekhin_ea25_transients}
yields impurity parameter $Q$ in full agreement with
theoretical models of multicomponent accreted crust composition 
\cite{Shchechilin_ea21,Shchechilin_ea23_InnerCrustComposition}
and does not require low-conductivity regions in the outer crust.

This may seem to contradict our model but, actually, it is not so, as the model
implies
that the glassy layer is formed from 
pristine crust, and it sinks to
higher densities as more matter is accreted.
In particular, the
glassy layer should be swept out from the outer crust, as soon as pristine material 
is replaced 
by accreted one. The latter takes place for accreted mass 
$\sim 5\times 10^{-5}~M_\odot$, which is much less than a typical mass 
of accreted matter during LMXB stage. Thus, it is not surprising, that other 
transiently accreting NSs do not have glassy layers in the outer crust. 
In fact,
KS 1731-260 and MXB 1659-29 have spin frequencies of 524~Hz and 567~Hz, 
respectively, plausibly implying that they have by now accreted much more
matter and angular momentum. 

In principle, for a fully accreted crust, a glassy layer can 
form in its innermost layer.
In the course of pristine crust replacement by continued accretion, the
glass inner boundary, $\rho_2$, increases until it reaches sub-nuclear 
densities where
nuclei disintegrate \cite{GC20_DiffEq,GC24_nHD_Shells}.
The outer boundary, 
$\rho_1$, also increases, until  inter-ion spacing becomes 
sufficiently small to compensate (cf. Eq. \ref{Gamma}) for the smallness 
of $Z^5 Y$ of the accreted matter in the inner crust; numerically, it is expected to occur at 
$\sim 10^{13}$ g/cc.
Note, that continuing relaxation 
of KS 1731-260 seems to favor
a poorly conducting layer at 
high densities presently explained by a hypothetical
disordered nuclear pasta phase \cite{Diebel_ea17_LateCool_Pasta,Jain_ea25_KS1731}.
Alternatively, this low-conductivity layer could be yet
another manifestation of the 
glass state and
amorphization.

{\it Accreted mass.}
The above logic
allows one to determine the accreted mass $\Delta M$ in 
J1748
by applying 
the hydrostatic equilibrium condition at $\rho_1$:
\begin{equation}
     4\pi R^2 P(\rho_1) = g \Delta M~. 
\label{Hydrostat}
\end{equation}
Henceforth, we neglect general relativity corrections, 
$g=GM/R^2$ is the gravitational acceleration,
$P$ is the pressure, 
$M$ and $R$ are the NS mass and radius, and $G$ is the gravitational constant.
This yields
\beq
  \Delta M \approx 2.4 \times 10^{-6} \, M_\odot \, \left(\frac{1.4\, M_\odot}{M}\right) 
  \left(\frac{R}{12 \, {\rm km}} \right)^4~,
\label{dM}
\eeq
where we took
$P (\rho_1)\approx 3.4\times 10^{28}$ 
dyn/cm$^2$, corresponding 
to 
$Y=0.4$ at the bottom of the accreted crust.

Let us note, that a	possibility of a partially accreted 
(also called hybrid)
crust in $\mathrm{NS_{J1748}}$
has been discussed previously 
\cite{Wijnands_ea13_partaccr,Chaikin_ea18}
on the basis of different
arguments and with a much higher anticipated boundary density
(e.g.\ $\rho\sim 10^{12}$ g/cc was suggested \cite{Chaikin_ea18} as 
a boundary between accreted and ground-state layers).
Our estimate of $\Delta M$ based on the glass formation condition 
turns out to be $2-4$ orders of magnitude lower than that 
predicted in these works. 
As we demonstrate below, 
it carves very specific constraints on the 
NS formation scenario.

Another hydrostatic equilibrium condition can be written at $\rho_2$, where 
the same accreted mass 
causes an additional stress which is
equilibrated by a critical compression of the crystal. This allows one to
infer a crustal breaking parameter at $\rho_2$ which turns out to be
in a good agreement with theory (see EM).

{\it Thermal and spin state of $\mathrm{NS_{J1748}}$.}
According to 
binary evolution models, 
accretion starts long time after NS formation,
meaning that the NS had enough time to cool down to very low temperatures.
After the beginning of accretion, long-term 
thermal state of the NS is determined by a balance 
between heating processes in the crust (e.g., deep crustal heating) 
and cooling due to photon emission from the surface and neutrino 
emission from the interior \cite{bbr98,Colpi_ea01_T_of_SXRT,Yakovlev_etal03_transients}.
Our model predicts an extremely low amount of accreted matter, prompting
a question whether it was possible
to heat up the NS by 
such $\Delta M$ to the 
observed pre-outburst state?

To estimate the thermal content 
of an NS, one chiefly needs to specify its mass and equation of state, in particular, 
adopt a model of nucleon superfluidity
which suppresses nucleon heat capacity (see EM).
Under the assumptions off a relatively low
NS mass ($M \lesssim 1.4 \, M_\odot$),
strongly suppressed nucleon 
heat capacity, and 
partial core heating by the shallow heating mechanism, there is no contradiction 
between the observed thermal state of
$\mathrm{NS_{J1748}}$
and the $\Delta M$ value Eq.\ (\ref{dM}).

It is widely accepted that 
newly born NSs spin down until
accretion 
from a binary companion 
accelerates their rotation \cite{PR72,L+73}.
Based on the time-average accretion rate estimate of 
$\langle \dot{M} \rangle \sim 2\times 10^{-11} \, M_\odot$/yr 
deduced from 
30-year observations  \cite{Ootes_ea19,pcc19}, 
accretion onto the NS has started very recently,
$\Delta M/ \langle \dot{M} \rangle \sim 120$ kyr ago.
During the 2010 outburst, 
spin-up rate 
of $\dot \nu \approx 1.4\times 10^{-12}$ Hz/s 
was measured \cite{Papitto_ea11_IGR17480,Cavecchi_ea11_J17480-2446_burst_oscill}.
Assuming that it was proportional 
to the accretion rate during the outburst 
$\dot{M}_{\rm b}\sim 3 \times 10^{-9} \, M_\odot$/yr and integrating over 
the accreted mass, we obtain 
spin frequency increase during the entire
accretion phase as 
$\Delta \nu=\dot \nu (\Delta M/\dot{M}_{\rm b})\sim 0.035$ Hz. This 
is much lower than the observed 
NS spin frequency 
$\nu_{\rm obs}\approx 11$ Hz (see also EM).

We therefore have an NS that had spun down 
essentially to its present frequency
during the previous evolutionary phase
and is now at the very beginning of a spin-up.
In view of the low $\Delta \nu/\nu_{\rm obs}$ 
ratio, this conclusion cannot be altered by uncertainties
of $\Delta M$ and $\Delta \nu$,
stemming from the lack of specific
information on the NS mass, radius, and moment of inertia.
Due to weak 
$B$-dependence of the torque, $N\propto B^{2/7}$,
the conclusion also cannot be altered 
by any reasonable supposition of a much greater $B$-field before the accretion onset.
It crucially constrains the spin-down phase
and thus the birth parameters of this NS.

Indeed, adopting
a model of spin-down (while allowing for a possible magnetic field evolution) 
and taking into account the source age
and its present day spin frequency, one can constrain (see EM)
the magneto-rotational parameters of the NS at birth, $B_\ini$ and $\nu_\ini$.
In the standard scenario, thanks to rapid 
evolution of massive stars, 
$\mathrm{NS_{J1748}}$ must have been born not later than
$2-4$ Gyr ago, the time of the last star formation 
episode 
in Terzan 5 \cite{Crociati_ea24_StarFormHist_Ter5}. 
A lower NS age would imply a non-standard delayed collapse scenario
(see EM),
whereas the maximum allowed NS age is comparable to the Hubble time.

\begin{figure}
\begin{center}
	\includegraphics[width=87mm]{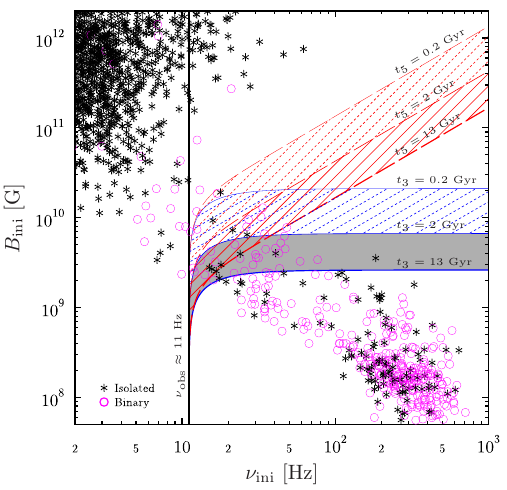}
\end{center}
\vspace{-0.3cm}
\caption[]{
Initial magnetic field $B_\ini$ vs initial spin frequency $\nu_\ini$
for three fiducial ages $t=2\times 10^8$, $2\times 10^9$, and $1.3\times 10^{10}$ yrs 
(from thinner to thicker lines)
assuming braking index $n=3$ (solid) and $n=5$ (dashes).
The allowed NS birth age ($2\times 10^9<t<1.3\times 10^{10}$ yrs) is 
filled by gray color (hatched by solid lines) for $n=3$ ($n=5$) models.
The region $2\times 10^8<t<2\times 10^9$ yrs is hatched by dashes, 
indicating 
a significant NS birth delay.
Asteriscs (circles) are present day parameters of 
isolated (binary) pulsars from the ATNF catalog \cite{atnf_paper,atnf_web263}.
}
\label{Fig:InitParams}
\end{figure}

In Fig.\ \ref{Fig:InitParams}, we plot possible NS birth parameters  
($B_\ini$ vs $\nu_\ini$) by thin, intermediate, and thick curves,
for three fiducial NS ages $t=2\times 10^8$, $2\times 10^9$, and 
$1.3\times 10^{10}$ yrs, respectively,
and two 
spin-down models
characterized by braking 
indices $n=3$ (solid) and $n=5$ (dashes).
Also shown are present day 
parameters\footnote{For consistency, the magnetic fields of pulsars 
in Fig.\ \ref{Fig:InitParams}
are calculated as $B=2.6\times 10^{19} (P\dot{P})^{1/2}$ G,
where $P$ and $\dot{P}$ are the period in seconds and 
its derivative; these data were taken from the ATNF catalog \cite{atnf_paper}
(version 2.6.3 \cite{atnf_web263}; note, the numerical factor for $B$ is 1.7 times 
lower than that in the 
vacuum formula, used by default in the ATNF catalog for historical reasons).
For pulsars in GCs, the data were updated according to P.\ Freire's
catalog \cite{Freire_GC_Pulsar_Cat_y25m07d25}
(downloaded on July 25, 2025); it allowed
us to extract $\dot{P}$ values for certain pulsars.
The observed (not intrinsic) values of $\dot{P}$ were used in all cases.}
of isolated (asteriscs) and binary (circles) pulsars from the ATNF 
catalog \cite{atnf_paper}.
The cluster of sources in the top left corner shows  
normal rotation-powered pulsars whereas 
MSPs accumulate
on the lower right.

{\it Formation scenario.}
The range of $\mathrm{NS_{J1748}}$ birth properties shown in Fig.\ \ref{Fig:InitParams}
is at variance with theoretical expectations
for a newly born NS, originating via
the standard path of NS
formation, i.e.\ via a FeCC. 
These NSs, as exemplified
by a population of normal 
radiopulsars
\cite{FK06,KK16,Igoshev_ea22_InitP_B}, are not expected to have at birth 
weak $B$-fields $\ll 2\times 10^{11}$ G and/or millisecond periods.
The pulsars with 
similar
parameters are observed (cf.\ Fig.\ \ref{Fig:InitParams}), 
but they are believed to be recycled
(i.e., spun up by accretion).

This can be taken as a sign of
a non-standard formation path
for $\mathrm{NS_{J1748}}$ which would be in line with the long held view
\cite{BG90_NS_MSP_from_AIC,BvdH91,P+02,KP06,Ivanova_etal08} that the majority of NSs retained in 
globular clusters (GCs)
originate in a collapse of a degenerate ONeMg core (see 
EM),
predominantly via AIC of a white dwarf (WD) or electron-capture supernova (ECSN).
In these cases, the expected NS mass is $\sim 1.25 \, M_\odot$ \cite{P+05},
thus our tentative conclusion regarding 
low $\mathrm{NS_{J1748}}$ mass is compatible with this idea.
 
The typically discussed mechanism of NS magnetic field origin in AIC
relies on a conservation of a WD 
magnetic flux \cite{H+10,B+11,Tauris_ea13_AIC}.
In this way, NS magnetic fields of $\sim 10^{11} - 10^{13}$ G typical for
regular pulsars can be obtained from the upper range of observed WD 
fields of $\sim 10^6 - 10^8$ G (moreover, simulations of AIC of ultramagnetized 
WDs predict ultramagnetized NSs \cite{D+07,C+25}). But lower resulting NS 
fields of $10^8-10^{10}$ G should plausibly be more common \cite{BG90_NS_MSP_from_AIC}.
We are not aware of studies of magnetic field generation in ECSN.
They also cannot be constrained from observations of Galactic field 
pulsars \cite{Igoshev_ea22_InitP_B}. 
It seems reasonable to assume that, in the degenerate pre-collapse
core, the field is not greater than that of an ordinary ONe WD.
Consequently, after the collapse, the field is likely lower than those expected
for AIC NSs. 

$\mathrm{NS_{J1748}}$ birth via AIC has been suggested 
previously \cite{Patruno_ea12_IGR17480}, 
however no restrictions
on the NS birth properties were deduced.
The AIC scenario 
was further elaborated \cite{Tauris_ea13_AIC}  
by an argument
that the presently observed tight binary was not the one in which 
AIC occurred. The latter had to be disrupted and the NS
had to acquire a new companion via a dynamical interaction.

In our model, delayed collapse is neither required nor excluded.
However, AIC 
is typically followed by a brief detachment phase, 
after which an intense 
accretion (already onto 
NS) resumes \cite{Tauris_ea13_AIC}. 
This has to produce an NS
with a substantially replaced crust (with low $Z$)
which is incompatible with the glass layer in the outer crust
and thus with our interpretation of the cooling data. 
To avoid
that, the post-AIC binary must have been disrupted by a dynamical
interaction shortly after the collapse or 
by a natal NS kick (strong enough to unbind the binary but weak enough to keep the NS
in the GC) with NS picking up the present companion later on. Both
possibilities seem to require some fine-tuning,
making
AIC 
less likely. 
	
If $\mathrm{NS_{J1748}}$
were formed via 
a merger-induced collapse (MIC), it also had to capture the present 
companion. While probability of such 
an event in Terzan 5 may be acceptable, 
NS formation via MIC is a less common 
process in GCs \cite{Ivanova_etal08}.
It plausibly results in 
a higher NS mass and $B$-field
\cite{K+01,L+06,Ruiter_ea19_AIC}, 
rendering this path also unlikely.

The remaining NS formation channel is ECSN.
It has been proposed decades ago 
\cite{M+80}
and was actively studied ever 
since 
\cite{N84, N87, P+04, K+06, P+08, J+08, 
J+13, M+14, D+17, GessnerJanka18_HD_ECSN, Z+21}.
Quite recently, SN2018zd has been argued to be of ECSN 
type \cite{Z+20,H+21,S+24}.
Thanks to low kick velocity it is predicted to be the main source  
of NSs retained in GCs 
\cite{KP06,Ivanova_etal08}. 
As a result, we consider ECSN the most appropriate
$\mathrm{NS_{J1748}}$ formation channel.

{\it Conclusions.} 
The observed slow NS cooling after an outburst in 
J1748 is explained by super-critical deformation and 
amorphization of pristine crystal NS crust 
caused by 
weight of the accreted material. 
This idea relies on the fact that the glass transition temperature 
of the ground-state matter is noticeably higher than that of the accreted matter, 
and is further supported by 
a good agreement of
the crustal breaking parameter $\delta_{\rm crit}$
derived from observations of
J1748 
with its theoretical value. 
Our 
reasoning 
is fully compatible with 
earlier studies
of other transiently accreting LMXBs,
which have found
high crustal thermal conductivity for their NSs. It also hints at a
natural mechanism of late-time cooling, observed in some of these sources.  

Moreover,
the proposed picture allows
us to determine total accreted mass
($\Delta M \approx 2.4\times 10^{-6} \, M_\odot$), which 
strongly constrains
the evolutionary scenario of
$\mathrm{NS_{J1748}}$.
In particular, we 
demonstrate that the NS birth properties were such as those typically attributed
to recycled pulsars. 
It is plausible that these birth properties represent those of ECSN NSs, indicating that the initial spin and magnetic field distributions may be different for degenerate-collapse NSs and FeCC NSs.

We expect the proposed evolutionary scenario for $\mathrm{NS_{J1748}}$ to stimulate renewed interest in NS cooling in LMXBs and in statistics of LMXB and MSP populations. Beyond astrophysics, it motivates further research in condensed-matter and plasma physics on glass transition and structure of matter stressed supercritically.

{\it Acknowledgements---}%
Dedicated to the memory of Yurii Anatolievich Shibanov.
A.I.C. acknowledges support of this work by Russian Science Foundation  (Grant No.~22-12-00048-P).

{\it Data availability---}%
All data generated or analysed during this study are included in this published article.


\begin{thebibliography}{144}%
	\makeatletter
	\providecommand \@ifxundefined [1]{%
		\@ifx{#1\undefined}
	}%
	\providecommand \@ifnum [1]{%
		\ifnum #1\expandafter \@firstoftwo
		\else \expandafter \@secondoftwo
		\fi
	}%
	\providecommand \@ifx [1]{%
		\ifx #1\expandafter \@firstoftwo
		\else \expandafter \@secondoftwo
		\fi
	}%
	\providecommand \natexlab [1]{#1}%
	\providecommand \enquote  [1]{``#1''}%
	\providecommand \bibnamefont  [1]{#1}%
	\providecommand \bibfnamefont [1]{#1}%
	\providecommand \citenamefont [1]{#1}%
	\providecommand \href@noop [0]{\@secondoftwo}%
	\providecommand \href [0]{\begingroup \@sanitize@url \@href}%
	\providecommand \@href[1]{\@@startlink{#1}\@@href}%
	\providecommand \@@href[1]{\endgroup#1\@@endlink}%
	\providecommand \@sanitize@url [0]{\catcode `\\12\catcode `\$12\catcode
		`\&12\catcode `\#12\catcode `\^12\catcode `\_12\catcode `\%12\relax}%
	\providecommand \@@startlink[1]{}%
	\providecommand \@@endlink[0]{}%
	\providecommand \url  [0]{\begingroup\@sanitize@url \@url }%
	\providecommand \@url [1]{\endgroup\@href {#1}{\urlprefix }}%
	\providecommand \urlprefix  [0]{URL }%
	\providecommand \Eprint [0]{\href }%
	\providecommand \doibase [0]{https://doi.org/}%
	\providecommand \selectlanguage [0]{\@gobble}%
	\providecommand \bibinfo  [0]{\@secondoftwo}%
	\providecommand \bibfield  [0]{\@secondoftwo}%
	\providecommand \translation [1]{[#1]}%
	\providecommand \BibitemOpen [0]{}%
	\providecommand \bibitemStop [0]{}%
	\providecommand \bibitemNoStop [0]{.\EOS\space}%
	\providecommand \EOS [0]{\spacefactor3000\relax}%
	\providecommand \BibitemShut  [1]{\csname bibitem#1\endcsname}%
	\let\auto@bib@innerbib\@empty
	\bibitem [{\citenamefont {Haensel}\ \emph {et~al.}(2007)\citenamefont
		{Haensel}, \citenamefont {Potekhin},\ and\ \citenamefont {Yakovlev}}]{hpy07}%
	\BibitemOpen
	\bibfield  {author} {\bibinfo {author} {\bibfnamefont {P.}~\bibnamefont
			{Haensel}}, \bibinfo {author} {\bibfnamefont {A.}~\bibnamefont {Potekhin}},\
		and\ \bibinfo {author} {\bibfnamefont {D.}~\bibnamefont {Yakovlev}},\
	}\href@noop {} {\emph {\bibinfo {title} {Neutron Stars 1: Equation of State
				and Structure}}},\ Astrophysics and Space Science Library\ (\bibinfo
	{publisher} {Springer-Verlag},\ \bibinfo {address} {Berlin},\ \bibinfo {year}
	{2007})\BibitemShut {NoStop}%
	\bibitem [{\citenamefont {{Lattimer}}\ and\ \citenamefont
		{{Prakash}}(2007)}]{LP07_NSobservations}%
	\BibitemOpen
	\bibfield  {author} {\bibinfo {author} {\bibfnamefont {J.~M.}\ \bibnamefont
			{{Lattimer}}}\ and\ \bibinfo {author} {\bibfnamefont {M.}~\bibnamefont
			{{Prakash}}},\ }\bibfield  {title} {\bibinfo {title} {{Neutron star
				observations: Prognosis for equation of state constraints}},\ }\href
	{https://doi.org/10.1016/j.physrep.2007.02.003} {\bibfield  {journal}
		{\bibinfo  {journal} {\physrep}\ }\textbf {\bibinfo {volume} {442}},\
		\bibinfo {pages} {109} (\bibinfo {year} {2007})},\ \Eprint
	{https://arxiv.org/abs/astro-ph/0612440} {arXiv:astro-ph/0612440 [astro-ph]}
	\BibitemShut {NoStop}%
	\bibitem [{\citenamefont {{Baym}}\ \emph {et~al.}(2018)\citenamefont {{Baym}},
		\citenamefont {{Hatsuda}}, \citenamefont {{Kojo}}, \citenamefont {{Powell}},
		\citenamefont {{Song}},\ and\ \citenamefont {{Takatsuka}}}]{Baym+18}%
	\BibitemOpen
	\bibfield  {author} {\bibinfo {author} {\bibfnamefont {G.}~\bibnamefont
			{{Baym}}}, \bibinfo {author} {\bibfnamefont {T.}~\bibnamefont {{Hatsuda}}},
		\bibinfo {author} {\bibfnamefont {T.}~\bibnamefont {{Kojo}}}, \bibinfo
		{author} {\bibfnamefont {P.~D.}\ \bibnamefont {{Powell}}}, \bibinfo {author}
		{\bibfnamefont {Y.}~\bibnamefont {{Song}}},\ and\ \bibinfo {author}
		{\bibfnamefont {T.}~\bibnamefont {{Takatsuka}}},\ }\bibfield  {title}
	{\bibinfo {title} {{From hadrons to quarks in neutron stars: a review}},\
	}\href {https://doi.org/10.1088/1361-6633/aaae14} {\bibfield  {journal}
		{\bibinfo  {journal} {Reports on Progress in Physics}\ }\textbf {\bibinfo
			{volume} {81}},\ \bibinfo {eid} {056902} (\bibinfo {year} {2018})},\ \Eprint
	{https://arxiv.org/abs/1707.04966} {arXiv:1707.04966 [astro-ph.HE]}
	\BibitemShut {NoStop}%
	\bibitem [{\citenamefont {Rezzolla}\ \emph {et~al.}(2019)\citenamefont
		{Rezzolla}, \citenamefont {Pizzochero}, \citenamefont {Jones}, \citenamefont
		{Rea},\ and\ \citenamefont {Vida{\~n}a}}]{Compstar19}%
	\BibitemOpen
	\bibfield  {author} {\bibinfo {author} {\bibfnamefont {L.}~\bibnamefont
			{Rezzolla}}, \bibinfo {author} {\bibfnamefont {P.}~\bibnamefont
			{Pizzochero}}, \bibinfo {author} {\bibfnamefont {D.}~\bibnamefont {Jones}},
		\bibinfo {author} {\bibfnamefont {N.}~\bibnamefont {Rea}},\ and\ \bibinfo
		{author} {\bibfnamefont {I.}~\bibnamefont {Vida{\~n}a}},\ }\href
	{https://books.google.ru/books?id=SRyDDwAAQBAJ} {\emph {\bibinfo {title} {The
				Physics and Astrophysics of Neutron Stars}}},\ Astrophysics and Space Science
	Library\ (\bibinfo  {publisher} {Springer International Publishing},\
	\bibinfo {year} {2019})\BibitemShut {NoStop}%
	\bibitem [{\citenamefont {{Lattimer}}(2021)}]{Lattimer21_NS_NSEOS}%
	\BibitemOpen
	\bibfield  {author} {\bibinfo {author} {\bibfnamefont {J.~M.}\ \bibnamefont
			{{Lattimer}}},\ }\bibfield  {title} {\bibinfo {title} {{Neutron Stars and the
				Nuclear Matter Equation of State}},\ }\href
	{https://doi.org/10.1146/annurev-nucl-102419-124827} {\bibfield  {journal}
		{\bibinfo  {journal} {Annual Review of Nuclear and Particle Science}\
		}\textbf {\bibinfo {volume} {71}},\ \bibinfo {pages} {433} (\bibinfo {year}
		{2021})}\BibitemShut {NoStop}%
	\bibitem [{\citenamefont {Ofengeim}\ \emph {et~al.}(2026)\citenamefont
		{Ofengeim}, \citenamefont {Kolomeitsev},\ and\ \citenamefont
		{Shternin}}]{oks26_NS_EOS_review}%
	\BibitemOpen
	\bibfield  {author} {\bibinfo {author} {\bibfnamefont {D.~D.}\ \bibnamefont
			{Ofengeim}}, \bibinfo {author} {\bibfnamefont {E.~E.}\ \bibnamefont
			{Kolomeitsev}},\ and\ \bibinfo {author} {\bibfnamefont {P.~S.}\ \bibnamefont
			{Shternin}},\ }\bibfield  {title} {\bibinfo {title} {Neutron star equation of
			state},\ }in\ \href
	{https://doi.org/https://doi.org/10.1016/B978-0-443-21439-4.00137-1} {\emph
		{\bibinfo {booktitle} {Encyclopedia of Astrophysics (First Edition)}}},\
	\bibinfo {editor} {edited by\ \bibinfo {editor} {\bibfnamefont
			{I.}~\bibnamefont {Mandel}}}\ (\bibinfo  {publisher} {Elsevier},\ \bibinfo
	{address} {Oxford},\ \bibinfo {year} {2026})\ \bibinfo {edition} {first
		edition}\ ed.,\ pp.\ \bibinfo {pages} {160--204}\BibitemShut {NoStop}%
	\bibitem [{\citenamefont {{Chamel}}\ and\ \citenamefont
		{{Haensel}}(2008)}]{ch08}%
	\BibitemOpen
	\bibfield  {author} {\bibinfo {author} {\bibfnamefont {N.}~\bibnamefont
			{{Chamel}}}\ and\ \bibinfo {author} {\bibfnamefont {P.}~\bibnamefont
			{{Haensel}}},\ }\bibfield  {title} {\bibinfo {title} {Physics of neutron star
			crusts},\ }\href@noop {} {\bibfield  {journal} {\bibinfo  {journal} {Liv.
				Rev. Relativ.}\ }\textbf {\bibinfo {volume} {11}},\ \bibinfo {pages} {10}
		(\bibinfo {year} {2008})}\BibitemShut {NoStop}%
	\bibitem [{\citenamefont {Caplan}\ and\ \citenamefont {Horowitz}(2017)}]{ch17}%
	\BibitemOpen
	\bibfield  {author} {\bibinfo {author} {\bibfnamefont {M.~E.}\ \bibnamefont
			{Caplan}}\ and\ \bibinfo {author} {\bibfnamefont {C.~J.}\ \bibnamefont
			{Horowitz}},\ }\bibfield  {title} {\bibinfo {title} {Colloquium:
			Astromaterial science and nuclear pasta},\ }\href
	{https://doi.org/10.1103/RevModPhys.89.041002} {\bibfield  {journal}
		{\bibinfo  {journal} {Rev. Mod. Phys.}\ }\textbf {\bibinfo {volume} {89}},\
		\bibinfo {pages} {041002} (\bibinfo {year} {2017})}\BibitemShut {NoStop}%
	\bibitem [{\citenamefont {{Meisel}}\ \emph {et~al.}(2018)\citenamefont
		{{Meisel}}, \citenamefont {{Deibel}}, \citenamefont {{Keek}}, \citenamefont
		{{Shternin}},\ and\ \citenamefont {{Elfritz}}}]{mdkse18}%
	\BibitemOpen
	\bibfield  {author} {\bibinfo {author} {\bibfnamefont {Z.}~\bibnamefont
			{{Meisel}}}, \bibinfo {author} {\bibfnamefont {A.}~\bibnamefont {{Deibel}}},
		\bibinfo {author} {\bibfnamefont {L.}~\bibnamefont {{Keek}}}, \bibinfo
		{author} {\bibfnamefont {P.}~\bibnamefont {{Shternin}}},\ and\ \bibinfo
		{author} {\bibfnamefont {J.}~\bibnamefont {{Elfritz}}},\ }\bibfield  {title}
	{\bibinfo {title} {Nuclear physics of the outer layers of accreting neutron
			stars},\ }\href {https://doi.org/10.1088/1361-6471/aad171} {\bibfield
		{journal} {\bibinfo  {journal} {\jpg}\ }\textbf {\bibinfo {volume} {45}},\
		\bibinfo {pages} {093001} (\bibinfo {year} {2018})},\ \Eprint
	{https://arxiv.org/abs/1807.01150} {arXiv:1807.01150 [astro-ph.HE]}
	\BibitemShut {NoStop}%
	\bibitem [{\citenamefont {{Blaschke}}\ and\ \citenamefont
		{{Chamel}}(2018)}]{BlaschkeChame18_PhasesOfDenseMatter}%
	\BibitemOpen
	\bibfield  {author} {\bibinfo {author} {\bibfnamefont {D.}~\bibnamefont
			{{Blaschke}}}\ and\ \bibinfo {author} {\bibfnamefont {N.}~\bibnamefont
			{{Chamel}}},\ }\bibfield  {title} {\bibinfo {title} {{Phases of Dense Matter
				in Compact Stars}},\ }in\ \href {https://doi.org/10.1007/978-3-319-97616-7_7}
	{\emph {\bibinfo {booktitle} {Astrophysics and Space Science Library}}},\
	\bibinfo {series} {Astrophysics and Space Science Library}, Vol.\ \bibinfo
	{volume} {457},\ \bibinfo {editor} {edited by\ \bibinfo {editor}
		{\bibfnamefont {L.}~\bibnamefont {{Rezzolla}}}, \bibinfo {editor}
		{\bibfnamefont {P.}~\bibnamefont {{Pizzochero}}}, \bibinfo {editor}
		{\bibfnamefont {D.~I.}\ \bibnamefont {{Jones}}}, \bibinfo {editor}
		{\bibfnamefont {N.}~\bibnamefont {{Rea}}},\ and\ \bibinfo {editor}
		{\bibfnamefont {I.}~\bibnamefont {{Vida{\~n}a}}}}\ (\bibinfo {year} {2018})\
	p.\ \bibinfo {pages} {337},\ \Eprint {https://arxiv.org/abs/1803.01836}
	{arXiv:1803.01836 [nucl-th]} \BibitemShut {NoStop}%
	\bibitem [{\citenamefont {{Kaspi}}(2010)}]{Kaspi10_GandUnif_NS}%
	\BibitemOpen
	\bibfield  {author} {\bibinfo {author} {\bibfnamefont {V.~M.}\ \bibnamefont
			{{Kaspi}}},\ }\bibfield  {title} {\bibinfo {title} {{Grand unification of
				neutron stars}},\ }\href {https://doi.org/10.1073/pnas.1000812107} {\bibfield
		{journal} {\bibinfo  {journal} {Proceedings of the National Academy of
				Science}\ }\textbf {\bibinfo {volume} {107}},\ \bibinfo {pages} {7147}
		(\bibinfo {year} {2010})},\ \Eprint {https://arxiv.org/abs/1005.0876}
	{arXiv:1005.0876 [astro-ph.HE]} \BibitemShut {NoStop}%
	\bibitem [{\citenamefont {De~Luca}(2017)}]{DeLuca17_CCO}%
	\BibitemOpen
	\bibfield  {author} {\bibinfo {author} {\bibfnamefont {A.}~\bibnamefont
			{De~Luca}},\ }\bibfield  {title} {\bibinfo {title} {Central compact objects
			in supernova remnants},\ }\href
	{https://doi.org/10.1088/1742-6596/932/1/012006} {\bibfield  {journal}
		{\bibinfo  {journal} {Journal of Physics: Conference Series}\ }\textbf
		{\bibinfo {volume} {932}},\ \bibinfo {pages} {012006} (\bibinfo {year}
		{2017})}\BibitemShut {NoStop}%
	\bibitem [{\citenamefont {{Demorest}}\ and\ \citenamefont
		{{Goss}}(2024)}]{DemorestGoss24_MSP_discov_Hist}%
	\BibitemOpen
	\bibfield  {author} {\bibinfo {author} {\bibfnamefont {P.~B.}\ \bibnamefont
			{{Demorest}}}\ and\ \bibinfo {author} {\bibfnamefont {W.~M.}\ \bibnamefont
			{{Goss}}},\ }\bibfield  {title} {\bibinfo {title} {{The discovery of
				millisecond pulsars: Don Backer and the response to the unexpected}},\ }\href
	{https://doi.org/10.48550/arXiv.2407.18194} {\bibfield  {journal} {\bibinfo
			{journal} {Journal of Astronomical History and Heritage}\ }\textbf {\bibinfo
			{volume} {27}},\ \bibinfo {pages} {465} (\bibinfo {year} {2024})},\ \Eprint
	{https://arxiv.org/abs/2407.18194} {arXiv:2407.18194 [astro-ph.HE]}
	\BibitemShut {NoStop}%
	\bibitem [{\citenamefont {Borghese}\ and\ \citenamefont {{Coti
				Zelati}}(2026)}]{BCZ26_NSzoo_encicl}%
	\BibitemOpen
	\bibfield  {author} {\bibinfo {author} {\bibfnamefont {A.}~\bibnamefont
			{Borghese}}\ and\ \bibinfo {author} {\bibfnamefont {F.}~\bibnamefont {{Coti
					Zelati}}},\ }\bibfield  {title} {\bibinfo {title} {The zoo of isolated
			neutron stars},\ }in\ \href
	{https://doi.org/https://doi.org/10.1016/B978-0-443-21439-4.00095-X} {\emph
		{\bibinfo {booktitle} {Encyclopedia of Astrophysics (First Edition)}}},\
	\bibinfo {editor} {edited by\ \bibinfo {editor} {\bibfnamefont
			{I.}~\bibnamefont {Mandel}}}\ (\bibinfo  {publisher} {Elsevier},\ \bibinfo
	{address} {Oxford},\ \bibinfo {year} {2026})\ \bibinfo {edition} {first
		edition}\ ed.,\ pp.\ \bibinfo {pages} {145--159}\BibitemShut {NoStop}%
	\bibitem [{\citenamefont {Wahl}(2026)}]{Wahl26_NS_and_Pulsars_Encicl}%
	\BibitemOpen
	\bibfield  {author} {\bibinfo {author} {\bibfnamefont {H.~M.}\ \bibnamefont
			{Wahl}},\ }\bibfield  {title} {\bibinfo {title} {Neutron star and pulsar
			fundamentals},\ }in\ \href
	{https://doi.org/https://doi.org/10.1016/B978-0-443-21439-4.00149-8} {\emph
		{\bibinfo {booktitle} {Encyclopedia of Astrophysics (First Edition)}}},\
	\bibinfo {editor} {edited by\ \bibinfo {editor} {\bibfnamefont
			{I.}~\bibnamefont {Mandel}}}\ (\bibinfo  {publisher} {Elsevier},\ \bibinfo
	{address} {Oxford},\ \bibinfo {year} {2026})\ \bibinfo {edition} {first
		edition}\ ed.,\ pp.\ \bibinfo {pages} {132--144}\BibitemShut {NoStop}%
	\bibitem [{\citenamefont {Rea}\ and\ \citenamefont {{De
				Grandis}}(2026)}]{RDG26_Magnetars_Encicl}%
	\BibitemOpen
	\bibfield  {author} {\bibinfo {author} {\bibfnamefont {N.}~\bibnamefont
			{Rea}}\ and\ \bibinfo {author} {\bibfnamefont {D.}~\bibnamefont {{De
					Grandis}}},\ }\bibfield  {title} {\bibinfo {title} {Magnetars},\ }in\ \href
	{https://doi.org/https://doi.org/10.1016/B978-0-443-21439-4.00096-1} {\emph
		{\bibinfo {booktitle} {Encyclopedia of Astrophysics (First Edition)}}},\
	\bibinfo {editor} {edited by\ \bibinfo {editor} {\bibfnamefont
			{I.}~\bibnamefont {Mandel}}}\ (\bibinfo  {publisher} {Elsevier},\ \bibinfo
	{address} {Oxford},\ \bibinfo {year} {2026})\ \bibinfo {edition} {first
		edition}\ ed.,\ pp.\ \bibinfo {pages} {205--222}\BibitemShut {NoStop}%
	\bibitem [{\citenamefont {{Cerda-Duran}}\ and\ \citenamefont
		{{Elias-Rosa}}(2018)}]{CD_ER18_NSformation_CCSN}%
	\BibitemOpen
	\bibfield  {author} {\bibinfo {author} {\bibfnamefont {P.}~\bibnamefont
			{{Cerda-Duran}}}\ and\ \bibinfo {author} {\bibfnamefont {N.}~\bibnamefont
			{{Elias-Rosa}}},\ }\bibfield  {title} {\bibinfo {title} {{Neutron Stars
				Formation and Core Collapse Supernovae}},\ }in\ \href
	{https://doi.org/10.1007/978-3-319-97616-7_1} {\emph {\bibinfo {booktitle}
			{Astrophysics and Space Science Library}}},\ \bibinfo {series} {Astrophysics
		and Space Science Library}, Vol.\ \bibinfo {volume} {457},\ \bibinfo {editor}
	{edited by\ \bibinfo {editor} {\bibfnamefont {L.}~\bibnamefont {{Rezzolla}}},
		\bibinfo {editor} {\bibfnamefont {P.}~\bibnamefont {{Pizzochero}}}, \bibinfo
		{editor} {\bibfnamefont {D.~I.}\ \bibnamefont {{Jones}}}, \bibinfo {editor}
		{\bibfnamefont {N.}~\bibnamefont {{Rea}}},\ and\ \bibinfo {editor}
		{\bibfnamefont {I.}~\bibnamefont {{Vida{\~n}a}}}}\ (\bibinfo {year} {2018})\
	p.~\bibinfo {pages} {1},\ \Eprint {https://arxiv.org/abs/1806.07267}
	{arXiv:1806.07267 [astro-ph.HE]} \BibitemShut {NoStop}%
	\bibitem [{\citenamefont {Jerkstrand}\ \emph {et~al.}(2026)\citenamefont
		{Jerkstrand}, \citenamefont {Milisavljevic},\ and\ \citenamefont
		{Muller}}]{CCSN25_encicl}%
	\BibitemOpen
	\bibfield  {author} {\bibinfo {author} {\bibfnamefont {A.}~\bibnamefont
			{Jerkstrand}}, \bibinfo {author} {\bibfnamefont {D.}~\bibnamefont
			{Milisavljevic}},\ and\ \bibinfo {author} {\bibfnamefont {B.}~\bibnamefont
			{Muller}},\ }\bibfield  {title} {\bibinfo {title} {Core-collapse
			supernovae},\ }in\ \href
	{https://doi.org/https://doi.org/10.1016/B978-0-443-21439-4.00090-0} {\emph
		{\bibinfo {booktitle} {Encyclopedia of Astrophysics (First Edition)}}},\
	\bibinfo {editor} {edited by\ \bibinfo {editor} {\bibfnamefont
			{I.}~\bibnamefont {Mandel}}}\ (\bibinfo  {publisher} {Elsevier},\ \bibinfo
	{address} {Oxford},\ \bibinfo {year} {2026})\ \bibinfo {edition} {first
		edition}\ ed.,\ pp.\ \bibinfo {pages} {639--668}\BibitemShut {NoStop}%
	\bibitem [{\citenamefont {{Miyaji}}\ \emph {et~al.}(1980)\citenamefont
		{{Miyaji}}, \citenamefont {{Nomoto}}, \citenamefont {{Yokoi}},\ and\
		\citenamefont {{Sugimoto}}}]{M+80}%
	\BibitemOpen
	\bibfield  {author} {\bibinfo {author} {\bibfnamefont {S.}~\bibnamefont
			{{Miyaji}}}, \bibinfo {author} {\bibfnamefont {K.}~\bibnamefont {{Nomoto}}},
		\bibinfo {author} {\bibfnamefont {K.}~\bibnamefont {{Yokoi}}},\ and\ \bibinfo
		{author} {\bibfnamefont {D.}~\bibnamefont {{Sugimoto}}},\ }\bibfield  {title}
	{\bibinfo {title} {Supernova triggered by electron captures},\ }\href@noop {}
	{\bibfield  {journal} {\bibinfo  {journal} {\pasj}\ }\textbf {\bibinfo
			{volume} {32}},\ \bibinfo {pages} {303} (\bibinfo {year} {1980})}\BibitemShut
	{NoStop}%
	\bibitem [{\citenamefont {Wang}\ \emph {et~al.}(2026)\citenamefont {Wang},
		\citenamefont {Liu}, \citenamefont {Guo},\ and\ \citenamefont
		{Han}}]{Wang_ea26_ECSN_review}%
	\BibitemOpen
	\bibfield  {author} {\bibinfo {author} {\bibfnamefont {B.}~\bibnamefont
			{Wang}}, \bibinfo {author} {\bibfnamefont {D.}~\bibnamefont {Liu}}, \bibinfo
		{author} {\bibfnamefont {Y.}~\bibnamefont {Guo}},\ and\ \bibinfo {author}
		{\bibfnamefont {Z.}~\bibnamefont {Han}},\ }\bibfield  {title} {\bibinfo
		{title} {The formation of electron-capture supernovae: A review},\ }\href
	{https://doi.org/10.1088/1674-4527/ae2d0e} {\bibfield  {journal} {\bibinfo
			{journal} {Research in Astronomy and Astrophysics}\ }\textbf {\bibinfo
			{volume} {26}},\ \bibinfo {pages} {032001} (\bibinfo {year}
		{2026})}\BibitemShut {NoStop}%
	\bibitem [{\citenamefont {{Lorimer}}(2008)}]{Lorimer08_LRR}%
	\BibitemOpen
	\bibfield  {author} {\bibinfo {author} {\bibfnamefont {D.~R.}\ \bibnamefont
			{{Lorimer}}},\ }\bibfield  {title} {\bibinfo {title} {Binary and millisecond
			pulsars},\ }\href {https://doi.org/10.12942/lrr-2008-8} {\bibfield  {journal}
		{\bibinfo  {journal} {\llr}\ }\textbf {\bibinfo {volume} {11}},\ \bibinfo
		{eid} {8} (\bibinfo {year} {2008})},\ \Eprint
	{https://arxiv.org/abs/0811.0762} {arXiv:0811.0762 [astro-ph]} \BibitemShut
	{NoStop}%
	\bibitem [{\citenamefont {{Bisnovatyi-Kogan}}\ and\ \citenamefont
		{{Komberg}}(1974)}]{BK74}%
	\BibitemOpen
	\bibfield  {author} {\bibinfo {author} {\bibfnamefont {G.~S.}\ \bibnamefont
			{{Bisnovatyi-Kogan}}}\ and\ \bibinfo {author} {\bibfnamefont {B.~V.}\
			\bibnamefont {{Komberg}}},\ }\bibfield  {title} {\bibinfo {title} {{Pulsars
				and close binary systems}},\ }\href@noop {} {\bibfield  {journal} {\bibinfo
			{journal} {\sovast}\ }\textbf {\bibinfo {volume} {18}},\ \bibinfo {pages}
		{217} (\bibinfo {year} {1974})}\BibitemShut {NoStop}%
	\bibitem [{\citenamefont {Bhattacharya}\ and\ \citenamefont {{van den
				Heuvel}}(1991)}]{BvdH91}%
	\BibitemOpen
	\bibfield  {author} {\bibinfo {author} {\bibfnamefont {D.}~\bibnamefont
			{Bhattacharya}}\ and\ \bibinfo {author} {\bibfnamefont {E.}~\bibnamefont
			{{van den Heuvel}}},\ }\bibfield  {title} {\bibinfo {title} {Formation and
			evolution of binary and millisecond radio pulsars},\ }\href
	{https://doi.org/https://doi.org/10.1016/0370-1573(91)90064-S} {\bibfield
		{journal} {\bibinfo  {journal} {Physics Reports}\ }\textbf {\bibinfo {volume}
			{203}},\ \bibinfo {pages} {1} (\bibinfo {year} {1991})}\BibitemShut {NoStop}%
	\bibitem [{\citenamefont {{Kulkarni}}\ and\ \citenamefont
		{{Narayan}}(1988)}]{kn88}%
	\BibitemOpen
	\bibfield  {author} {\bibinfo {author} {\bibfnamefont {S.~R.}\ \bibnamefont
			{{Kulkarni}}}\ and\ \bibinfo {author} {\bibfnamefont {R.}~\bibnamefont
			{{Narayan}}},\ }\bibfield  {title} {\bibinfo {title} {Birthrates of low-mass
			binary pulsars and low-mass {X}-ray binaries},\ }\href
	{https://doi.org/10.1086/166964} {\bibfield  {journal} {\bibinfo  {journal}
			{\apj}\ }\textbf {\bibinfo {volume} {335}},\ \bibinfo {pages} {755} (\bibinfo
		{year} {1988})}\BibitemShut {NoStop}%
	\bibitem [{\citenamefont {{Kulkarni}}\ \emph {et~al.}(1990)\citenamefont
		{{Kulkarni}}, \citenamefont {{Narayan}},\ and\ \citenamefont
		{{Romani}}}]{Kulkarni+90}%
	\BibitemOpen
	\bibfield  {author} {\bibinfo {author} {\bibfnamefont {S.~R.}\ \bibnamefont
			{{Kulkarni}}}, \bibinfo {author} {\bibfnamefont {R.}~\bibnamefont
			{{Narayan}}},\ and\ \bibinfo {author} {\bibfnamefont {R.~W.}\ \bibnamefont
			{{Romani}}},\ }\bibfield  {title} {\bibinfo {title} {{The Pulsar Content of
				Globular Clusters}},\ }\href {https://doi.org/10.1086/168828} {\bibfield
		{journal} {\bibinfo  {journal} {\apj}\ }\textbf {\bibinfo {volume} {356}},\
		\bibinfo {pages} {174} (\bibinfo {year} {1990})}\BibitemShut {NoStop}%
	\bibitem [{\citenamefont {{Ferrario}}\ and\ \citenamefont
		{{Wickramasinghe}}(2007)}]{FW07_MSP_birth_prop}%
	\BibitemOpen
	\bibfield  {author} {\bibinfo {author} {\bibfnamefont {L.}~\bibnamefont
			{{Ferrario}}}\ and\ \bibinfo {author} {\bibfnamefont {D.}~\bibnamefont
			{{Wickramasinghe}}},\ }\bibfield  {title} {\bibinfo {title} {{The birth
				properties of Galactic millisecond radio pulsars}},\ }\href
	{https://doi.org/10.1111/j.1365-2966.2006.11365.x} {\bibfield  {journal}
		{\bibinfo  {journal} {\mnras}\ }\textbf {\bibinfo {volume} {375}},\ \bibinfo
		{pages} {1009} (\bibinfo {year} {2007})},\ \Eprint
	{https://arxiv.org/abs/astro-ph/0701444} {arXiv:astro-ph/0701444 [astro-ph]}
	\BibitemShut {NoStop}%
	\bibitem [{\citenamefont {{Bailyn}}\ and\ \citenamefont
		{{Grindlay}}(1990)}]{BG90_NS_MSP_from_AIC}%
	\BibitemOpen
	\bibfield  {author} {\bibinfo {author} {\bibfnamefont {C.~D.}\ \bibnamefont
			{{Bailyn}}}\ and\ \bibinfo {author} {\bibfnamefont {J.~E.}\ \bibnamefont
			{{Grindlay}}},\ }\bibfield  {title} {\bibinfo {title} {{Neutron Stars and
				Millisecond Pulsars from Accretion-induced Collapse in Globular Clusters}},\
	}\href {https://doi.org/10.1086/168602} {\bibfield  {journal} {\bibinfo
			{journal} {\apj}\ }\textbf {\bibinfo {volume} {353}},\ \bibinfo {pages} {159}
		(\bibinfo {year} {1990})}\BibitemShut {NoStop}%
	\bibitem [{\citenamefont {Hurley}\ \emph {et~al.}(2010)\citenamefont {Hurley},
		\citenamefont {Tout}, \citenamefont {Wickramasinghe}, \citenamefont
		{Ferrario},\ and\ \citenamefont {Kiel}}]{H+10}%
	\BibitemOpen
	\bibfield  {author} {\bibinfo {author} {\bibfnamefont {J.~R.}\ \bibnamefont
			{Hurley}}, \bibinfo {author} {\bibfnamefont {C.~A.}\ \bibnamefont {Tout}},
		\bibinfo {author} {\bibfnamefont {D.~T.}\ \bibnamefont {Wickramasinghe}},
		\bibinfo {author} {\bibfnamefont {L.}~\bibnamefont {Ferrario}},\ and\
		\bibinfo {author} {\bibfnamefont {P.~D.}\ \bibnamefont {Kiel}},\ }\bibfield
	{title} {\bibinfo {title} {Formation of binary millisecond pulsars by
			accretion-induced collapse of white dwarfs},\ }\href
	{https://doi.org/10.1111/j.1365-2966.2009.15988.x} {\bibfield  {journal}
		{\bibinfo  {journal} {\mnras}\ }\textbf {\bibinfo {volume} {402}},\ \bibinfo
		{pages} {1437} (\bibinfo {year} {2010})},\ \Eprint
	{https://arxiv.org/abs/https://academic.oup.com/mnras/article-pdf/402/3/1437/3111626/mnras0402-1437.pdf}
	{https://academic.oup.com/mnras/article-pdf/402/3/1437/3111626/mnras0402-1437.pdf}
	\BibitemShut {NoStop}%
	\bibitem [{\citenamefont {{Brown}}\ \emph {et~al.}(1998)\citenamefont
		{{Brown}}, \citenamefont {{Bildsten}},\ and\ \citenamefont
		{{Rutledge}}}]{bbr98}%
	\BibitemOpen
	\bibfield  {author} {\bibinfo {author} {\bibfnamefont {E.~F.}\ \bibnamefont
			{{Brown}}}, \bibinfo {author} {\bibfnamefont {L.}~\bibnamefont
			{{Bildsten}}},\ and\ \bibinfo {author} {\bibfnamefont {R.~E.}\ \bibnamefont
			{{Rutledge}}},\ }\bibfield  {title} {\bibinfo {title} {Crustal heating and
			quiescent emission from transiently accreting neutron stars},\ }\href
	{https://doi.org/10.1086/311578} {\bibfield  {journal} {\bibinfo  {journal}
			{\apjl}\ }\textbf {\bibinfo {volume} {504}},\ \bibinfo {pages} {L95}
		(\bibinfo {year} {1998})},\ \Eprint
	{https://arxiv.org/abs/arXiv:astro-ph/9807179} {arXiv:astro-ph/9807179}
	\BibitemShut {NoStop}%
	\bibitem [{\citenamefont {{Yakovlev}}\ \emph {et~al.}(2004)\citenamefont
		{{Yakovlev}}, \citenamefont {{Levenfish}}, \citenamefont {{Potekhin}},
		\citenamefont {{Gnedin}},\ and\ \citenamefont {{Chabrier}}}]{ylpgc04}%
	\BibitemOpen
	\bibfield  {author} {\bibinfo {author} {\bibfnamefont {D.~G.}\ \bibnamefont
			{{Yakovlev}}}, \bibinfo {author} {\bibfnamefont {K.~P.}\ \bibnamefont
			{{Levenfish}}}, \bibinfo {author} {\bibfnamefont {A.~Y.}\ \bibnamefont
			{{Potekhin}}}, \bibinfo {author} {\bibfnamefont {O.~Y.}\ \bibnamefont
			{{Gnedin}}},\ and\ \bibinfo {author} {\bibfnamefont {G.}~\bibnamefont
			{{Chabrier}}},\ }\bibfield  {title} {\bibinfo {title} {Thermal states of
			coldest and hottest neutron stars in soft {X}-ray transients},\ }\href
	{https://doi.org/10.1051/0004-6361:20034191} {\bibfield  {journal} {\bibinfo
			{journal} {Astron. Astrophys.}\ }\textbf {\bibinfo {volume} {417}},\ \bibinfo
		{pages} {169} (\bibinfo {year} {2004})}\BibitemShut {NoStop}%
	\bibitem [{\citenamefont {{Brown}}\ \emph {et~al.}(2018)\citenamefont
		{{Brown}}, \citenamefont {{Cumming}}, \citenamefont {{Fattoyev}},
		\citenamefont {{Horowitz}}, \citenamefont {{Page}},\ and\ \citenamefont
		{{Reddy}}}]{Brown_eal18_RapidNeutrinoCooling_MXB}%
	\BibitemOpen
	\bibfield  {author} {\bibinfo {author} {\bibfnamefont {E.~F.}\ \bibnamefont
			{{Brown}}}, \bibinfo {author} {\bibfnamefont {A.}~\bibnamefont {{Cumming}}},
		\bibinfo {author} {\bibfnamefont {F.~J.}\ \bibnamefont {{Fattoyev}}},
		\bibinfo {author} {\bibfnamefont {C.~J.}\ \bibnamefont {{Horowitz}}},
		\bibinfo {author} {\bibfnamefont {D.}~\bibnamefont {{Page}}},\ and\ \bibinfo
		{author} {\bibfnamefont {S.}~\bibnamefont {{Reddy}}},\ }\bibfield  {title}
	{\bibinfo {title} {{Rapid Neutrino Cooling in the Neutron Star MXB
				1659-29}},\ }\href {https://doi.org/10.1103/PhysRevLett.120.182701}
	{\bibfield  {journal} {\bibinfo  {journal} {\prl}\ }\textbf {\bibinfo
			{volume} {120}},\ \bibinfo {eid} {182701} (\bibinfo {year} {2018})},\ \Eprint
	{https://arxiv.org/abs/1801.00041} {arXiv:1801.00041 [astro-ph.HE]}
	\BibitemShut {NoStop}%
	\bibitem [{\citenamefont {{Potekhin}}\ \emph {et~al.}(2019)\citenamefont
		{{Potekhin}}, \citenamefont {{Chugunov}},\ and\ \citenamefont
		{{Chabrier}}}]{pcc19}%
	\BibitemOpen
	\bibfield  {author} {\bibinfo {author} {\bibfnamefont {A.~Y.}\ \bibnamefont
			{{Potekhin}}}, \bibinfo {author} {\bibfnamefont {A.~I.}\ \bibnamefont
			{{Chugunov}}},\ and\ \bibinfo {author} {\bibfnamefont {G.}~\bibnamefont
			{{Chabrier}}},\ }\bibfield  {title} {\bibinfo {title} {Thermal evolution and
			quiescent emission of transiently accreting neutron stars},\ }\href
	{https://doi.org/10.1051/0004-6361/201936003} {\bibfield  {journal} {\bibinfo
			{journal} {\aap}\ }\textbf {\bibinfo {volume} {629}},\ \bibinfo {eid} {A88}
		(\bibinfo {year} {2019})},\ \Eprint {https://arxiv.org/abs/1907.08299}
	{arXiv:1907.08299 [astro-ph.HE]} \BibitemShut {NoStop}%
	\bibitem [{\citenamefont {{Heinke}}\ \emph {et~al.}(2025)\citenamefont
		{{Heinke}}, \citenamefont {{Zheng}}, \citenamefont {{Maccarone}},
		\citenamefont {{Degenaar}}, \citenamefont {{Bahramian}}, \citenamefont
		{{Sivakoff}},\ and\ \citenamefont {{Toor}}}]{Heinke_ea25_OutburstCatalogue}%
	\BibitemOpen
	\bibfield  {author} {\bibinfo {author} {\bibfnamefont {C.~O.}\ \bibnamefont
			{{Heinke}}}, \bibinfo {author} {\bibfnamefont {J.}~\bibnamefont {{Zheng}}},
		\bibinfo {author} {\bibfnamefont {T.~J.}\ \bibnamefont {{Maccarone}}},
		\bibinfo {author} {\bibfnamefont {N.}~\bibnamefont {{Degenaar}}}, \bibinfo
		{author} {\bibfnamefont {A.}~\bibnamefont {{Bahramian}}}, \bibinfo {author}
		{\bibfnamefont {G.~R.}\ \bibnamefont {{Sivakoff}}},\ and\ \bibinfo {author}
		{\bibfnamefont {S.}~\bibnamefont {{Toor}}},\ }\bibfield  {title} {\bibinfo
		{title} {{Catalog of Outbursts of Neutron Star Low-mass X-Ray Binaries}},\
	}\href {https://doi.org/10.3847/1538-4365/ade99a} {\bibfield  {journal}
		{\bibinfo  {journal} {\apjs}\ }\textbf {\bibinfo {volume} {279}},\ \bibinfo
		{eid} {57} (\bibinfo {year} {2025})},\ \Eprint
	{https://arxiv.org/abs/2407.18867} {arXiv:2407.18867 [astro-ph.HE]}
	\BibitemShut {NoStop}%
	\bibitem [{\citenamefont {{Wijnands}}\ \emph {et~al.}(2017)\citenamefont
		{{Wijnands}}, \citenamefont {{Degenaar}},\ and\ \citenamefont
		{{Page}}}]{WDP17_CoolingAccreted}%
	\BibitemOpen
	\bibfield  {author} {\bibinfo {author} {\bibfnamefont {R.}~\bibnamefont
			{{Wijnands}}}, \bibinfo {author} {\bibfnamefont {N.}~\bibnamefont
			{{Degenaar}}},\ and\ \bibinfo {author} {\bibfnamefont {D.}~\bibnamefont
			{{Page}}},\ }\bibfield  {title} {\bibinfo {title} {{Cooling of
				Accretion-Heated Neutron Stars}},\ }\href
	{https://doi.org/10.1007/s12036-017-9466-5} {\bibfield  {journal} {\bibinfo
			{journal} {Journal of Astrophysics and Astronomy}\ }\textbf {\bibinfo
			{volume} {38}},\ \bibinfo {eid} {49} (\bibinfo {year} {2017})},\ \Eprint
	{https://arxiv.org/abs/1709.07034} {arXiv:1709.07034 [astro-ph.HE]}
	\BibitemShut {NoStop}%
	\bibitem [{\citenamefont {{Ushomirsky}}\ and\ \citenamefont
		{{Rutledge}}(2001)}]{UR01_TimeVariableEmission_Transients}%
	\BibitemOpen
	\bibfield  {author} {\bibinfo {author} {\bibfnamefont {G.}~\bibnamefont
			{{Ushomirsky}}}\ and\ \bibinfo {author} {\bibfnamefont {R.~E.}\ \bibnamefont
			{{Rutledge}}},\ }\bibfield  {title} {\bibinfo {title} {{Time-variable
				emission from transiently accreting neutron stars in quiescence due to deep
				crustal heating}},\ }\href {https://doi.org/10.1046/j.1365-8711.2001.04515.x}
	{\bibfield  {journal} {\bibinfo  {journal} {\mnras}\ }\textbf {\bibinfo
			{volume} {325}},\ \bibinfo {pages} {1157} (\bibinfo {year} {2001})},\ \Eprint
	{https://arxiv.org/abs/astro-ph/0101141} {arXiv:astro-ph/0101141 [astro-ph]}
	\BibitemShut {NoStop}%
	\bibitem [{\citenamefont {Rutledge}\ \emph {et~al.}(2002)\citenamefont
		{Rutledge}, \citenamefont {Bildsten}, \citenamefont {Brown}, \citenamefont
		{Pavlov}, \citenamefont {Zavlin},\ and\ \citenamefont
		{Ushomirsky}}]{Ruthledge_etal02_KS}%
	\BibitemOpen
	\bibfield  {author} {\bibinfo {author} {\bibfnamefont {R.~E.}\ \bibnamefont
			{Rutledge}}, \bibinfo {author} {\bibfnamefont {L.}~\bibnamefont {Bildsten}},
		\bibinfo {author} {\bibfnamefont {E.~F.}\ \bibnamefont {Brown}}, \bibinfo
		{author} {\bibfnamefont {G.~G.}\ \bibnamefont {Pavlov}}, \bibinfo {author}
		{\bibfnamefont {V.~E.}\ \bibnamefont {Zavlin}},\ and\ \bibinfo {author}
		{\bibfnamefont {G.}~\bibnamefont {Ushomirsky}},\ }\bibfield  {title}
	{\bibinfo {title} {Crustal emission and the quiescent spectrum of the neutron
			star in {KS} 1731--260},\ }\href
	{http://stacks.iop.org/0004-637X/580/i=1/a=413} {\bibfield  {journal}
		{\bibinfo  {journal} {\apj}\ }\textbf {\bibinfo {volume} {580}},\ \bibinfo
		{pages} {413} (\bibinfo {year} {2002})}\BibitemShut {NoStop}%
	\bibitem [{\citenamefont {{Brown}}\ and\ \citenamefont
		{{Cumming}}(2009)}]{bc09}%
	\BibitemOpen
	\bibfield  {author} {\bibinfo {author} {\bibfnamefont {E.~F.}\ \bibnamefont
			{{Brown}}}\ and\ \bibinfo {author} {\bibfnamefont {A.}~\bibnamefont
			{{Cumming}}},\ }\bibfield  {title} {\bibinfo {title} {Mapping crustal heating
			with the cooling light curves of quasi-persistent transients},\ }\href
	{https://doi.org/10.1088/0004-637X/698/2/1020} {\bibfield  {journal}
		{\bibinfo  {journal} {Astrophys. J.}\ }\textbf {\bibinfo {volume} {698}},\
		\bibinfo {pages} {1020} (\bibinfo {year} {2009})}\BibitemShut {NoStop}%
	\bibitem [{\citenamefont {{Papitto}}\ \emph {et~al.}(2011)\citenamefont
		{{Papitto}}, \citenamefont {{D'A{\`\i}}}, \citenamefont {{Motta}},
		\citenamefont {{Riggio}}, \citenamefont {{Burderi}}, \citenamefont {{di
				Salvo}}, \citenamefont {{Belloni}},\ and\ \citenamefont
		{{Iaria}}}]{Papitto_ea11_IGR17480}%
	\BibitemOpen
	\bibfield  {author} {\bibinfo {author} {\bibfnamefont {A.}~\bibnamefont
			{{Papitto}}}, \bibinfo {author} {\bibfnamefont {A.}~\bibnamefont
			{{D'A{\`\i}}}}, \bibinfo {author} {\bibfnamefont {S.}~\bibnamefont
			{{Motta}}}, \bibinfo {author} {\bibfnamefont {A.}~\bibnamefont {{Riggio}}},
		\bibinfo {author} {\bibfnamefont {L.}~\bibnamefont {{Burderi}}}, \bibinfo
		{author} {\bibfnamefont {T.}~\bibnamefont {{di Salvo}}}, \bibinfo {author}
		{\bibfnamefont {T.}~\bibnamefont {{Belloni}}},\ and\ \bibinfo {author}
		{\bibfnamefont {R.}~\bibnamefont {{Iaria}}},\ }\bibfield  {title} {\bibinfo
		{title} {The spin and orbit of the newly discovered pulsar {IGR}
			{{J}17480}-2446},\ }\href {https://doi.org/10.1051/0004-6361/201015974}
	{\bibfield  {journal} {\bibinfo  {journal} {\aap}\ }\textbf {\bibinfo
			{volume} {526}},\ \bibinfo {eid} {L3} (\bibinfo {year} {2011})},\ \Eprint
	{https://arxiv.org/abs/1010.4793} {arXiv:1010.4793 [astro-ph.HE]}
	\BibitemShut {NoStop}%
	\bibitem [{\citenamefont {{Testa}}\ \emph {et~al.}(2012)\citenamefont
		{{Testa}}, \citenamefont {{di Salvo}}, \citenamefont {{D'Antona}},
		\citenamefont {{Menna}}, \citenamefont {{Ventura}}, \citenamefont
		{{Burderi}}, \citenamefont {{Riggio}}, \citenamefont {{Iaria}}, \citenamefont
		{{D'A{\`\i}}}, \citenamefont {{Papitto}},\ and\ \citenamefont
		{{Robba}}}]{Testa_ea12_IGR17480_IR}%
	\BibitemOpen
	\bibfield  {author} {\bibinfo {author} {\bibfnamefont {V.}~\bibnamefont
			{{Testa}}}, \bibinfo {author} {\bibfnamefont {T.}~\bibnamefont {{di Salvo}}},
		\bibinfo {author} {\bibfnamefont {F.}~\bibnamefont {{D'Antona}}}, \bibinfo
		{author} {\bibfnamefont {M.~T.}\ \bibnamefont {{Menna}}}, \bibinfo {author}
		{\bibfnamefont {P.}~\bibnamefont {{Ventura}}}, \bibinfo {author}
		{\bibfnamefont {L.}~\bibnamefont {{Burderi}}}, \bibinfo {author}
		{\bibfnamefont {A.}~\bibnamefont {{Riggio}}}, \bibinfo {author}
		{\bibfnamefont {R.}~\bibnamefont {{Iaria}}}, \bibinfo {author} {\bibfnamefont
			{A.}~\bibnamefont {{D'A{\`\i}}}}, \bibinfo {author} {\bibfnamefont
			{A.}~\bibnamefont {{Papitto}}},\ and\ \bibinfo {author} {\bibfnamefont
			{N.}~\bibnamefont {{Robba}}},\ }\bibfield  {title} {\bibinfo {title} {The
			near-{IR} counterpart of {IGR} {{J}17480-2446} in {Terzan} 5},\ }\href
	{https://doi.org/10.1051/0004-6361/201219904} {\bibfield  {journal} {\bibinfo
			{journal} {\aap}\ }\textbf {\bibinfo {volume} {547}},\ \bibinfo {eid} {A28}
		(\bibinfo {year} {2012})},\ \Eprint {https://arxiv.org/abs/1210.8261}
	{arXiv:1210.8261 [astro-ph.SR]} \BibitemShut {NoStop}%
	\bibitem [{\citenamefont {{Patruno}}\ \emph {et~al.}(2012)\citenamefont
		{{Patruno}}, \citenamefont {{Alpar}}, \citenamefont {{van der Klis}},\ and\
		\citenamefont {{van den Heuvel}}}]{Patruno_ea12_IGR17480}%
	\BibitemOpen
	\bibfield  {author} {\bibinfo {author} {\bibfnamefont {A.}~\bibnamefont
			{{Patruno}}}, \bibinfo {author} {\bibfnamefont {M.~A.}\ \bibnamefont
			{{Alpar}}}, \bibinfo {author} {\bibfnamefont {M.}~\bibnamefont {{van der
					Klis}}},\ and\ \bibinfo {author} {\bibfnamefont {E.~P.~J.}\ \bibnamefont
			{{van den Heuvel}}},\ }\bibfield  {title} {\bibinfo {title} {The peculiar
			evolutionary history of {IGR} {{J}17480-2446} in {Terzan} 5},\ }\href
	{https://doi.org/10.1088/0004-637X/752/1/33} {\bibfield  {journal} {\bibinfo
			{journal} {\apj}\ }\textbf {\bibinfo {volume} {752}},\ \bibinfo {eid} {33}
		(\bibinfo {year} {2012})},\ \Eprint {https://arxiv.org/abs/1112.5315}
	{arXiv:1112.5315 [astro-ph.HE]} \BibitemShut {NoStop}%
	\bibitem [{\citenamefont {{Wijnands}}\ \emph {et~al.}(2013)\citenamefont
		{{Wijnands}}, \citenamefont {{Degenaar}},\ and\ \citenamefont
		{{Page}}}]{Wijnands_ea13_partaccr}%
	\BibitemOpen
	\bibfield  {author} {\bibinfo {author} {\bibfnamefont {R.}~\bibnamefont
			{{Wijnands}}}, \bibinfo {author} {\bibfnamefont {N.}~\bibnamefont
			{{Degenaar}}},\ and\ \bibinfo {author} {\bibfnamefont {D.}~\bibnamefont
			{{Page}}},\ }\bibfield  {title} {\bibinfo {title} {Testing the deep-crustal
			heating model using quiescent neutron-star very-faint {X}-ray transients and
			the possibility of partially accreted crusts in accreting neutron stars},\
	}\href {https://doi.org/10.1093/mnras/stt599} {\bibfield  {journal} {\bibinfo
			{journal} {\mnras}\ }\textbf {\bibinfo {volume} {432}},\ \bibinfo {pages}
		{2366} (\bibinfo {year} {2013})},\ \Eprint {https://arxiv.org/abs/1208.4273}
	{arXiv:1208.4273 [astro-ph.HE]} \BibitemShut {NoStop}%
	\bibitem [{\citenamefont {{Medin}}\ and\ \citenamefont
		{{Cumming}}(2014)}]{MC14_CompConv_Transients}%
	\BibitemOpen
	\bibfield  {author} {\bibinfo {author} {\bibfnamefont {Z.}~\bibnamefont
			{{Medin}}}\ and\ \bibinfo {author} {\bibfnamefont {A.}~\bibnamefont
			{{Cumming}}},\ }\bibfield  {title} {\bibinfo {title} {{A Signature of
				Chemical Separation in the Cooling Light Curves of Transiently Accreting
				Neutron Stars}},\ }\href {https://doi.org/10.1088/2041-8205/783/1/L3}
	{\bibfield  {journal} {\bibinfo  {journal} {\apjl}\ }\textbf {\bibinfo
			{volume} {783}},\ \bibinfo {eid} {L3} (\bibinfo {year} {2014})},\ \Eprint
	{https://arxiv.org/abs/1311.5145} {arXiv:1311.5145 [astro-ph.SR]}
	\BibitemShut {NoStop}%
	\bibitem [{\citenamefont {{Ootes}}\ \emph
		{et~al.}(2019{\natexlab{a}})\citenamefont {{Ootes}}, \citenamefont {{Vats}},
		\citenamefont {{Page}}, \citenamefont {{Wijnands}}, \citenamefont {{Parikh}},
		\citenamefont {{Degenaar}}, \citenamefont {{Wijngaarden}}, \citenamefont
		{{Altamirano}}, \citenamefont {{Bahramian}}, \citenamefont {{Cackett}},
		\citenamefont {{Heinke}}, \citenamefont {{Homan}},\ and\ \citenamefont
		{{Miller}}}]{Ootes_ea19}%
	\BibitemOpen
	\bibfield  {author} {\bibinfo {author} {\bibfnamefont {L.~S.}\ \bibnamefont
			{{Ootes}}}, \bibinfo {author} {\bibfnamefont {S.}~\bibnamefont {{Vats}}},
		\bibinfo {author} {\bibfnamefont {D.}~\bibnamefont {{Page}}}, \bibinfo
		{author} {\bibfnamefont {R.}~\bibnamefont {{Wijnands}}}, \bibinfo {author}
		{\bibfnamefont {A.~S.}\ \bibnamefont {{Parikh}}}, \bibinfo {author}
		{\bibfnamefont {N.}~\bibnamefont {{Degenaar}}}, \bibinfo {author}
		{\bibfnamefont {M.~J.~P.}\ \bibnamefont {{Wijngaarden}}}, \bibinfo {author}
		{\bibfnamefont {D.}~\bibnamefont {{Altamirano}}}, \bibinfo {author}
		{\bibfnamefont {A.}~\bibnamefont {{Bahramian}}}, \bibinfo {author}
		{\bibfnamefont {E.~M.}\ \bibnamefont {{Cackett}}}, \bibinfo {author}
		{\bibfnamefont {C.~O.}\ \bibnamefont {{Heinke}}}, \bibinfo {author}
		{\bibfnamefont {J.}~\bibnamefont {{Homan}}},\ and\ \bibinfo {author}
		{\bibfnamefont {J.~M.}\ \bibnamefont {{Miller}}},\ }\bibfield  {title}
	{\bibinfo {title} {Continued cooling of the accretion-heated neutron star
			crust in the {X}-ray transient {IGR} {J}17480-2446 located in the globular
			cluster {Terzan} 5},\ }\href {https://doi.org/10.1093/mnras/stz1406}
	{\bibfield  {journal} {\bibinfo  {journal} {\mnras}\ }\textbf {\bibinfo
			{volume} {487}},\ \bibinfo {pages} {1447} (\bibinfo {year}
		{2019}{\natexlab{a}})},\ \Eprint {https://arxiv.org/abs/1805.00610}
	{arXiv:1805.00610 [astro-ph.HE]} \BibitemShut {NoStop}%
	\bibitem [{\citenamefont {{Potekhin}}\ \emph {et~al.}(2025)\citenamefont
		{{Potekhin}}, \citenamefont {{Chugunov}}, \citenamefont {{Shchechilin}},\
		and\ \citenamefont {{Gusakov}}}]{Potekhin_ea25_transients}%
	\BibitemOpen
	\bibfield  {author} {\bibinfo {author} {\bibfnamefont {A.~Y.}\ \bibnamefont
			{{Potekhin}}}, \bibinfo {author} {\bibfnamefont {A.~I.}\ \bibnamefont
			{{Chugunov}}}, \bibinfo {author} {\bibfnamefont {N.~N.}\ \bibnamefont
			{{Shchechilin}}},\ and\ \bibinfo {author} {\bibfnamefont {M.~E.}\
			\bibnamefont {{Gusakov}}},\ }\bibfield  {title} {\bibinfo {title} {Cooling of
			neutron stars in soft {X}-ray transients with realistic crust composition},\
	}\href {https://doi.org/10.1016/j.jheap.2024.11.017} {\bibfield  {journal}
		{\bibinfo  {journal} {\jhea}\ }\textbf {\bibinfo {volume} {45}},\ \bibinfo
		{pages} {116} (\bibinfo {year} {2025})},\ \Eprint
	{https://arxiv.org/abs/2411.14395} {arXiv:2411.14395 [astro-ph.HE]}
	\BibitemShut {NoStop}%
	\bibitem [{\citenamefont {{Degenaar}}\ and\ \citenamefont
		{{Wijnands}}(2011)}]{DW11_AccretionHeatedCrust_J17480_2446}%
	\BibitemOpen
	\bibfield  {author} {\bibinfo {author} {\bibfnamefont {N.}~\bibnamefont
			{{Degenaar}}}\ and\ \bibinfo {author} {\bibfnamefont {R.}~\bibnamefont
			{{Wijnands}}},\ }\bibfield  {title} {\bibinfo {title} {The accretion-heated
			crust of the transiently accreting 11-{Hz} {X}-ray pulsar in the globular
			cluster {Terzan} 5},\ }\href
	{https://doi.org/10.1111/j.1745-3933.2011.01054.x} {\bibfield  {journal}
		{\bibinfo  {journal} {\mnras}\ }\textbf {\bibinfo {volume} {414}},\ \bibinfo
		{pages} {L50} (\bibinfo {year} {2011})},\ \Eprint
	{https://arxiv.org/abs/1103.1640} {arXiv:1103.1640 [astro-ph.HE]}
	\BibitemShut {NoStop}%
	\bibitem [{\citenamefont {{Degenaar}}\ \emph {et~al.}(2011)\citenamefont
		{{Degenaar}}, \citenamefont {{Brown}},\ and\ \citenamefont
		{{Wijnands}}}]{DBW11_CrustCooling_J17480_2446}%
	\BibitemOpen
	\bibfield  {author} {\bibinfo {author} {\bibfnamefont {N.}~\bibnamefont
			{{Degenaar}}}, \bibinfo {author} {\bibfnamefont {E.~F.}\ \bibnamefont
			{{Brown}}},\ and\ \bibinfo {author} {\bibfnamefont {R.}~\bibnamefont
			{{Wijnands}}},\ }\bibfield  {title} {\bibinfo {title} {Evidence for crust
			cooling in the transiently accreting 11-{Hz} {X}-ray pulsar in the globular
			cluster {Terzan} 5},\ }\href
	{https://doi.org/10.1111/j.1745-3933.2011.01164.x} {\bibfield  {journal}
		{\bibinfo  {journal} {\mnras}\ }\textbf {\bibinfo {volume} {418}},\ \bibinfo
		{pages} {L152} (\bibinfo {year} {2011})},\ \Eprint
	{https://arxiv.org/abs/1107.5317} {arXiv:1107.5317 [astro-ph.HE]}
	\BibitemShut {NoStop}%
	\bibitem [{\citenamefont {{Lau}}\ \emph {et~al.}(2018)\citenamefont {{Lau}},
		\citenamefont {{Beard}}, \citenamefont {{Gupta}}, \citenamefont {{Schatz}},
		\citenamefont {{Afanasjev}}, \citenamefont {{Brown}}, \citenamefont
		{{Deibel}}, \citenamefont {{Gasques}}, \citenamefont {{Hitt}}, \citenamefont
		{{Hix}}, \citenamefont {{Keek}}, \citenamefont {{M{\"o}ller}}, \citenamefont
		{{Shternin}}, \citenamefont {{Steiner}}, \citenamefont {{Wiescher}},\ and\
		\citenamefont {{Xu}}}]{Lau_ea18}%
	\BibitemOpen
	\bibfield  {author} {\bibinfo {author} {\bibfnamefont {R.}~\bibnamefont
			{{Lau}}}, \bibinfo {author} {\bibfnamefont {M.}~\bibnamefont {{Beard}}},
		\bibinfo {author} {\bibfnamefont {S.~S.}\ \bibnamefont {{Gupta}}}, \bibinfo
		{author} {\bibfnamefont {H.}~\bibnamefont {{Schatz}}}, \bibinfo {author}
		{\bibfnamefont {A.~V.}\ \bibnamefont {{Afanasjev}}}, \bibinfo {author}
		{\bibfnamefont {E.~F.}\ \bibnamefont {{Brown}}}, \bibinfo {author}
		{\bibfnamefont {A.}~\bibnamefont {{Deibel}}}, \bibinfo {author}
		{\bibfnamefont {L.~R.}\ \bibnamefont {{Gasques}}}, \bibinfo {author}
		{\bibfnamefont {G.~W.}\ \bibnamefont {{Hitt}}}, \bibinfo {author}
		{\bibfnamefont {W.~R.}\ \bibnamefont {{Hix}}}, \bibinfo {author}
		{\bibfnamefont {L.}~\bibnamefont {{Keek}}}, \bibinfo {author} {\bibfnamefont
			{P.}~\bibnamefont {{M{\"o}ller}}}, \bibinfo {author} {\bibfnamefont {P.~S.}\
			\bibnamefont {{Shternin}}}, \bibinfo {author} {\bibfnamefont {A.~W.}\
			\bibnamefont {{Steiner}}}, \bibinfo {author} {\bibfnamefont {M.}~\bibnamefont
			{{Wiescher}}},\ and\ \bibinfo {author} {\bibfnamefont {Y.}~\bibnamefont
			{{Xu}}},\ }\bibfield  {title} {\bibinfo {title} {Nuclear reactions in the
			crusts of accreting neutron stars},\ }\href
	{https://doi.org/10.3847/1538-4357/aabfe0} {\bibfield  {journal} {\bibinfo
			{journal} {\apj}\ }\textbf {\bibinfo {volume} {859}},\ \bibinfo {eid} {62}
		(\bibinfo {year} {2018})},\ \Eprint {https://arxiv.org/abs/1803.03818}
	{arXiv:1803.03818 [astro-ph.HE]} \BibitemShut {NoStop}%
	\bibitem [{\citenamefont {{Shchechilin}}\ \emph {et~al.}(2021)\citenamefont
		{{Shchechilin}}, \citenamefont {{Gusakov}},\ and\ \citenamefont
		{{Chugunov}}}]{Shchechilin_ea21}%
	\BibitemOpen
	\bibfield  {author} {\bibinfo {author} {\bibfnamefont {N.~N.}\ \bibnamefont
			{{Shchechilin}}}, \bibinfo {author} {\bibfnamefont {M.~E.}\ \bibnamefont
			{{Gusakov}}},\ and\ \bibinfo {author} {\bibfnamefont {A.~I.}\ \bibnamefont
			{{Chugunov}}},\ }\bibfield  {title} {\bibinfo {title} {Deep crustal heating
			for realistic compositions of thermonuclear ashes},\ }\href
	{https://doi.org/10.1093/mnras/stab2415} {\bibfield  {journal} {\bibinfo
			{journal} {\mnras}\ }\textbf {\bibinfo {volume} {507}},\ \bibinfo {pages}
		{3860} (\bibinfo {year} {2021})},\ \Eprint {https://arxiv.org/abs/2105.01991}
	{arXiv:2105.01991 [nucl-th]} \BibitemShut {NoStop}%
	\bibitem [{\citenamefont {Jones}(1999)}]{Jones99_AmorphousCrust}%
	\BibitemOpen
	\bibfield  {author} {\bibinfo {author} {\bibfnamefont {P.~B.}\ \bibnamefont
			{Jones}},\ }\bibfield  {title} {\bibinfo {title} {Amorphous and heterogeneous
			phase of neutron star matter},\ }\href
	{https://doi.org/10.1103/PhysRevLett.83.3589} {\bibfield  {journal} {\bibinfo
			{journal} {\prl}\ }\textbf {\bibinfo {volume} {83}},\ \bibinfo {pages}
		{3589} (\bibinfo {year} {1999})}\BibitemShut {NoStop}%
	\bibitem [{\citenamefont {{Jones}}(2001)}]{Jones01}%
	\BibitemOpen
	\bibfield  {author} {\bibinfo {author} {\bibfnamefont {P.~B.}\ \bibnamefont
			{{Jones}}},\ }\bibfield  {title} {\bibinfo {title} {First-principles
			point-defect calculations for solid neutron star matter},\ }\href
	{https://doi.org/10.1046/j.1365-8711.2001.03990.x} {\bibfield  {journal}
		{\bibinfo  {journal} {\mnras}\ }\textbf {\bibinfo {volume} {321}},\ \bibinfo
		{pages} {167} (\bibinfo {year} {2001})}\BibitemShut {NoStop}%
	\bibitem [{\citenamefont {{Sun}}\ \emph {et~al.}(2016)\citenamefont {{Sun}},
		\citenamefont {{Concustell}},\ and\ \citenamefont
		{{Greer}}}]{Sun_ea16_NatRM_MetallicGlass}%
	\BibitemOpen
	\bibfield  {author} {\bibinfo {author} {\bibfnamefont {Y.}~\bibnamefont
			{{Sun}}}, \bibinfo {author} {\bibfnamefont {A.}~\bibnamefont
			{{Concustell}}},\ and\ \bibinfo {author} {\bibfnamefont {A.~L.}\ \bibnamefont
			{{Greer}}},\ }\bibfield  {title} {\bibinfo {title} {{Thermomechanical
				processing of metallic glasses: extending the range of the glassy state}},\
	}\href {https://doi.org/10.1038/natrevmats.2016.39} {\bibfield  {journal}
		{\bibinfo  {journal} {\nrm}\ }\textbf {\bibinfo {volume} {1}},\ \bibinfo
		{eid} {16039} (\bibinfo {year} {2016})}\BibitemShut {NoStop}%
	\bibitem [{\citenamefont {{Ogata}}(1992)}]{Ogata92}%
	\BibitemOpen
	\bibfield  {author} {\bibinfo {author} {\bibfnamefont {S.}~\bibnamefont
			{{Ogata}}},\ }\bibfield  {title} {\bibinfo {title} {{Monte Carlo} simulation
			study of crystallization in rapidly supercooled one-component plasmas},\
	}\href {https://doi.org/10.1103/PhysRevA.45.1122} {\bibfield  {journal}
		{\bibinfo  {journal} {\pra}\ }\textbf {\bibinfo {volume} {45}},\ \bibinfo
		{pages} {1122} (\bibinfo {year} {1992})}\BibitemShut {NoStop}%
	\bibitem [{\citenamefont {{Hammerberg}}\ \emph {et~al.}(1994)\citenamefont
		{{Hammerberg}}, \citenamefont {{Holian}},\ and\ \citenamefont
		{{Ravelo}}}]{Hammerberg_ea94}%
	\BibitemOpen
	\bibfield  {author} {\bibinfo {author} {\bibfnamefont {J.~E.}\ \bibnamefont
			{{Hammerberg}}}, \bibinfo {author} {\bibfnamefont {B.~L.}\ \bibnamefont
			{{Holian}}},\ and\ \bibinfo {author} {\bibfnamefont {R.}~\bibnamefont
			{{Ravelo}}},\ }\bibfield  {title} {\bibinfo {title} {{Nucleation of
				long-range order in quenched {Yukawa} plasmas}},\ }\href
	{https://doi.org/10.1103/PhysRevE.50.1372} {\bibfield  {journal} {\bibinfo
			{journal} {\pre}\ }\textbf {\bibinfo {volume} {50}},\ \bibinfo {pages} {1372}
		(\bibinfo {year} {1994})}\BibitemShut {NoStop}%
	\bibitem [{\citenamefont {{Richet}}\ and\ \citenamefont
		{{Gillet}}(1997)}]{RH97_Pressure_induced_amorphization}%
	\BibitemOpen
	\bibfield  {author} {\bibinfo {author} {\bibfnamefont {P.}~\bibnamefont
			{{Richet}}}\ and\ \bibinfo {author} {\bibfnamefont {P.}~\bibnamefont
			{{Gillet}}},\ }\bibfield  {title} {\bibinfo {title} {{Pressure-induced
				amorphization of minerals: a review}},\ }\href
	{https://doi.org/10.1127/ejm/9/5/0907} {\bibfield  {journal} {\bibinfo
			{journal} {European Journal of Mineralogy}\ }\textbf {\bibinfo {volume}
			{9}},\ \bibinfo {pages} {907} (\bibinfo {year} {1997})}\BibitemShut {NoStop}%
	\bibitem [{\citenamefont {{Greaves}}\ \emph {et~al.}(2003)\citenamefont
		{{Greaves}}, \citenamefont {{Meneau}}, \citenamefont {{Sapelkin}},
		\citenamefont {{Colyer}}, \citenamefont {{ap Gwynn}}, \citenamefont
		{{Wade}},\ and\ \citenamefont
		{{Sankar}}}]{Greaves_ea03_NatMat_amorphization}%
	\BibitemOpen
	\bibfield  {author} {\bibinfo {author} {\bibfnamefont {G.~N.}\ \bibnamefont
			{{Greaves}}}, \bibinfo {author} {\bibfnamefont {F.}~\bibnamefont {{Meneau}}},
		\bibinfo {author} {\bibfnamefont {A.}~\bibnamefont {{Sapelkin}}}, \bibinfo
		{author} {\bibfnamefont {L.~M.}\ \bibnamefont {{Colyer}}}, \bibinfo {author}
		{\bibfnamefont {I.}~\bibnamefont {{ap Gwynn}}}, \bibinfo {author}
		{\bibfnamefont {S.}~\bibnamefont {{Wade}}},\ and\ \bibinfo {author}
		{\bibfnamefont {G.}~\bibnamefont {{Sankar}}},\ }\bibfield  {title} {\bibinfo
		{title} {{The rheology of collapsing zeolites amorphized by temperature and
				pressure}},\ }\href {https://doi.org/10.1038/nmat963} {\bibfield  {journal}
		{\bibinfo  {journal} {\natmater}\ }\textbf {\bibinfo {volume} {2}},\ \bibinfo
		{pages} {622} (\bibinfo {year} {2003})}\BibitemShut {NoStop}%
	\bibitem [{\citenamefont {{Bu}}\ \emph {et~al.}(2024)\citenamefont {{Bu}},
		\citenamefont {{Wu}}, \citenamefont {{Lei}}, \citenamefont {{Yuan}},
		\citenamefont {{Liu}}, \citenamefont {{Wang}}, \citenamefont {{Liu}},
		\citenamefont {{Wu}}, \citenamefont {{Liu}}, \citenamefont {{Wang}},
		\citenamefont {{Ritchie}}, \citenamefont {{Lu}},\ and\ \citenamefont
		{{Yang}}}]{Bu_ea24_NatCom_Amorphization}%
	\BibitemOpen
	\bibfield  {author} {\bibinfo {author} {\bibfnamefont {Y.}~\bibnamefont
			{{Bu}}}, \bibinfo {author} {\bibfnamefont {Y.}~\bibnamefont {{Wu}}}, \bibinfo
		{author} {\bibfnamefont {Z.}~\bibnamefont {{Lei}}}, \bibinfo {author}
		{\bibfnamefont {X.}~\bibnamefont {{Yuan}}}, \bibinfo {author} {\bibfnamefont
			{L.}~\bibnamefont {{Liu}}}, \bibinfo {author} {\bibfnamefont
			{P.}~\bibnamefont {{Wang}}}, \bibinfo {author} {\bibfnamefont
			{X.}~\bibnamefont {{Liu}}}, \bibinfo {author} {\bibfnamefont
			{H.}~\bibnamefont {{Wu}}}, \bibinfo {author} {\bibfnamefont {J.}~\bibnamefont
			{{Liu}}}, \bibinfo {author} {\bibfnamefont {H.}~\bibnamefont {{Wang}}},
		\bibinfo {author} {\bibfnamefont {R.~O.}\ \bibnamefont {{Ritchie}}}, \bibinfo
		{author} {\bibfnamefont {Z.}~\bibnamefont {{Lu}}},\ and\ \bibinfo {author}
		{\bibfnamefont {W.}~\bibnamefont {{Yang}}},\ }\bibfield  {title} {\bibinfo
		{title} {{Elastic strain-induced amorphization in high-entropy alloys}},\
	}\href {https://doi.org/10.1038/s41467-024-48619-0} {\bibfield  {journal}
		{\bibinfo  {journal} {\natcomm}\ }\textbf {\bibinfo {volume} {15}},\ \bibinfo
		{eid} {4599} (\bibinfo {year} {2024})}\BibitemShut {NoStop}%
	\bibitem [{\citenamefont {{Horowitz}}\ and\ \citenamefont
		{{Kadau}}(2009)}]{HK09_breaking}%
	\BibitemOpen
	\bibfield  {author} {\bibinfo {author} {\bibfnamefont {C.~J.}\ \bibnamefont
			{{Horowitz}}}\ and\ \bibinfo {author} {\bibfnamefont {K.}~\bibnamefont
			{{Kadau}}},\ }\bibfield  {title} {\bibinfo {title} {{Breaking Strain of
				Neutron Star Crust and Gravitational Waves}},\ }\href
	{https://doi.org/10.1103/PhysRevLett.102.191102} {\bibfield  {journal}
		{\bibinfo  {journal} {\prl}\ }\textbf {\bibinfo {volume} {102}},\ \bibinfo
		{eid} {191102} (\bibinfo {year} {2009})},\ \Eprint
	{https://arxiv.org/abs/0904.1986} {arXiv:0904.1986 [astro-ph.SR]}
	\BibitemShut {NoStop}%
	\bibitem [{\citenamefont {{Baiko}}\ and\ \citenamefont
		{{Chugunov}}(2022)}]{BC22}%
	\BibitemOpen
	\bibfield  {author} {\bibinfo {author} {\bibfnamefont {D.~A.}\ \bibnamefont
			{{Baiko}}}\ and\ \bibinfo {author} {\bibfnamefont {A.~I.}\ \bibnamefont
			{{Chugunov}}},\ }\bibfield  {title} {\bibinfo {title} {Ab initio
			thermodynamics of one-component plasma for astrophysics of white dwarfs and
			neutron stars},\ }\href {https://doi.org/10.1093/mnras/stab3613} {\bibfield
		{journal} {\bibinfo  {journal} {\mnras}\ }\textbf {\bibinfo {volume} {510}},\
		\bibinfo {pages} {2628} (\bibinfo {year} {2022})},\ \Eprint
	{https://arxiv.org/abs/2112.04822} {arXiv:2112.04822 [astro-ph.HE]}
	\BibitemShut {NoStop}%
	\bibitem [{\citenamefont {Lucco~Castello}\ and\ \citenamefont
		{Tolias}(2021)}]{CT21_GlassOCP}%
	\BibitemOpen
	\bibfield  {author} {\bibinfo {author} {\bibfnamefont {F.}~\bibnamefont
			{Lucco~Castello}}\ and\ \bibinfo {author} {\bibfnamefont {P.}~\bibnamefont
			{Tolias}},\ }\bibfield  {title} {\bibinfo {title} {Theoretical estimate of
			the glass transition line of {Yukawa} one-component plasmas},\ }\bibfield
	{journal} {\bibinfo  {journal} {Molecules}\ }\textbf {\bibinfo {volume}
		{26}},\ \href {https://doi.org/10.3390/molecules26030669}
	{10.3390/molecules26030669} (\bibinfo {year} {2021})\BibitemShut {NoStop}%
	\bibitem [{\citenamefont {{Yazdi}}\ \emph {et~al.}(2014)\citenamefont
		{{Yazdi}}, \citenamefont {{Ivlev}}, \citenamefont {{Khrapak}}, \citenamefont
		{{{Thomas}}}, \citenamefont {{Morfill}}, \citenamefont {{L{\"o}wen}},
		\citenamefont {{Wysocki}},\ and\ \citenamefont
		{{Sperl}}}]{Yazdi_ea14_Glass_Yuk}%
	\BibitemOpen
	\bibfield  {author} {\bibinfo {author} {\bibfnamefont {A.}~\bibnamefont
			{{Yazdi}}}, \bibinfo {author} {\bibfnamefont {A.}~\bibnamefont {{Ivlev}}},
		\bibinfo {author} {\bibfnamefont {S.}~\bibnamefont {{Khrapak}}}, \bibinfo
		{author} {\bibfnamefont {H.}~\bibnamefont {{{Thomas}}}}, \bibinfo {author}
		{\bibfnamefont {G.~E.}\ \bibnamefont {{Morfill}}}, \bibinfo {author}
		{\bibfnamefont {H.}~\bibnamefont {{L{\"o}wen}}}, \bibinfo {author}
		{\bibfnamefont {A.}~\bibnamefont {{Wysocki}}},\ and\ \bibinfo {author}
		{\bibfnamefont {M.}~\bibnamefont {{Sperl}}},\ }\bibfield  {title} {\bibinfo
		{title} {{Glass-transition properties of {Yukawa} potentials: From charged
				point particles to hard spheres}},\ }\href
	{https://doi.org/10.1103/PhysRevE.89.063105} {\bibfield  {journal} {\bibinfo
			{journal} {\pre}\ }\textbf {\bibinfo {volume} {89}},\ \bibinfo {eid} {063105}
		(\bibinfo {year} {2014})}\BibitemShut {NoStop}%
	\bibitem [{\citenamefont {{Chamel}}(2020)}]{Chamel20}%
	\BibitemOpen
	\bibfield  {author} {\bibinfo {author} {\bibfnamefont {N.}~\bibnamefont
			{{Chamel}}},\ }\bibfield  {title} {\bibinfo {title} {Analytical determination
			of the structure of the outer crust of a cold nonaccreted neutron star},\
	}\href {https://doi.org/10.1103/PhysRevC.101.032801} {\bibfield  {journal}
		{\bibinfo  {journal} {\prc}\ }\textbf {\bibinfo {volume} {101}},\ \bibinfo
		{eid} {032801} (\bibinfo {year} {2020})},\ \Eprint
	{https://arxiv.org/abs/2003.00983} {arXiv:2003.00983 [astro-ph.HE]}
	\BibitemShut {NoStop}%
	\bibitem [{\citenamefont {{Cackett}}\ \emph {et~al.}(2006)\citenamefont
		{{Cackett}}, \citenamefont {{Wijnands}}, \citenamefont {{Linares}},
		\citenamefont {{Miller}}, \citenamefont {{Homan}},\ and\ \citenamefont
		{{Lewin}}}]{Cackett_ea06_CoolingKS_MXB}%
	\BibitemOpen
	\bibfield  {author} {\bibinfo {author} {\bibfnamefont {E.~M.}\ \bibnamefont
			{{Cackett}}}, \bibinfo {author} {\bibfnamefont {R.}~\bibnamefont
			{{Wijnands}}}, \bibinfo {author} {\bibfnamefont {M.}~\bibnamefont
			{{Linares}}}, \bibinfo {author} {\bibfnamefont {J.~M.}\ \bibnamefont
			{{Miller}}}, \bibinfo {author} {\bibfnamefont {J.}~\bibnamefont {{Homan}}},\
		and\ \bibinfo {author} {\bibfnamefont {W.~H.~G.}\ \bibnamefont {{Lewin}}},\
	}\bibfield  {title} {\bibinfo {title} {{Cooling of the quasi-persistent
				neutron star X-ray transients {KS} 1731-260 and {MXB} 1659-29}},\ }\href
	{https://doi.org/10.1111/j.1365-2966.2006.10895.x} {\bibfield  {journal}
		{\bibinfo  {journal} {\mnras}\ }\textbf {\bibinfo {volume} {372}},\ \bibinfo
		{pages} {479} (\bibinfo {year} {2006})},\ \Eprint
	{https://arxiv.org/abs/astro-ph/0605490} {arXiv:astro-ph/0605490 [astro-ph]}
	\BibitemShut {NoStop}%
	\bibitem [{\citenamefont {{Shternin}}\ \emph {et~al.}(2007)\citenamefont
		{{Shternin}}, \citenamefont {{Yakovlev}}, \citenamefont {{Haensel}},\ and\
		\citenamefont {{Potekhin}}}]{syhp07}%
	\BibitemOpen
	\bibfield  {author} {\bibinfo {author} {\bibfnamefont {P.~S.}\ \bibnamefont
			{{Shternin}}}, \bibinfo {author} {\bibfnamefont {D.~G.}\ \bibnamefont
			{{Yakovlev}}}, \bibinfo {author} {\bibfnamefont {P.}~\bibnamefont
			{{Haensel}}},\ and\ \bibinfo {author} {\bibfnamefont {A.~Y.}\ \bibnamefont
			{{Potekhin}}},\ }\bibfield  {title} {\bibinfo {title} {Neutron star cooling
			after deep crustal heating in the {X}-ray transient {KS} 1731-260},\ }\href
	{https://doi.org/10.1111/j.1745-3933.2007.00386.x} {\bibfield  {journal}
		{\bibinfo  {journal} {\mnras}\ }\textbf {\bibinfo {volume} {382}},\ \bibinfo
		{pages} {L43} (\bibinfo {year} {2007})}\BibitemShut {NoStop}%
	\bibitem [{\citenamefont {{Jain}}\ \emph {et~al.}(2025)\citenamefont {{Jain}},
		\citenamefont {{Brown}}, \citenamefont {{Schatz}}, \citenamefont
		{{Afanasjev}}, \citenamefont {{Beard}}, \citenamefont {{Gasques}},
		\citenamefont {{Grace}}, \citenamefont {{Heger}}, \citenamefont {{Hitt}},
		\citenamefont {{Hix}}, \citenamefont {{Lau}}, \citenamefont {{Ong}},
		\citenamefont {{Wiescher}},\ and\ \citenamefont {{Xu}}}]{Jain_ea25_KS1731}%
	\BibitemOpen
	\bibfield  {author} {\bibinfo {author} {\bibfnamefont {R.}~\bibnamefont
			{{Jain}}}, \bibinfo {author} {\bibfnamefont {E.~F.}\ \bibnamefont {{Brown}}},
		\bibinfo {author} {\bibfnamefont {H.}~\bibnamefont {{Schatz}}}, \bibinfo
		{author} {\bibfnamefont {A.~V.}\ \bibnamefont {{Afanasjev}}}, \bibinfo
		{author} {\bibfnamefont {M.}~\bibnamefont {{Beard}}}, \bibinfo {author}
		{\bibfnamefont {L.~R.}\ \bibnamefont {{Gasques}}}, \bibinfo {author}
		{\bibfnamefont {J.}~\bibnamefont {{Grace}}}, \bibinfo {author} {\bibfnamefont
			{A.}~\bibnamefont {{Heger}}}, \bibinfo {author} {\bibfnamefont {G.~W.}\
			\bibnamefont {{Hitt}}}, \bibinfo {author} {\bibfnamefont {W.~R.}\
			\bibnamefont {{Hix}}}, \bibinfo {author} {\bibfnamefont {R.}~\bibnamefont
			{{Lau}}}, \bibinfo {author} {\bibfnamefont {W.~J.}\ \bibnamefont {{Ong}}},
		\bibinfo {author} {\bibfnamefont {M.}~\bibnamefont {{Wiescher}}},\ and\
		\bibinfo {author} {\bibfnamefont {Y.}~\bibnamefont {{Xu}}},\ }\bibfield
	{title} {\bibinfo {title} {Crust composition and the shallow heat source in
			{KS} 1731-260},\ }\href {https://doi.org/10.48550/arXiv.2504.09036}
	{\bibfield  {journal} {\bibinfo  {journal} {arXiv e-prints}\ ,\ \bibinfo
			{eid} {arXiv:2504.09036}} (\bibinfo {year} {2025})},\ \Eprint
	{https://arxiv.org/abs/2504.09036} {arXiv:2504.09036 [astro-ph.HE]}
	\BibitemShut {NoStop}%
	\bibitem [{\citenamefont {{Potekhin}}\ \emph {et~al.}(2023)\citenamefont
		{{Potekhin}}, \citenamefont {{Gusakov}},\ and\ \citenamefont
		{{Chugunov}}}]{Potekhin_ea23_transients}%
	\BibitemOpen
	\bibfield  {author} {\bibinfo {author} {\bibfnamefont {A.~Y.}\ \bibnamefont
			{{Potekhin}}}, \bibinfo {author} {\bibfnamefont {M.~E.}\ \bibnamefont
			{{Gusakov}}},\ and\ \bibinfo {author} {\bibfnamefont {A.~I.}\ \bibnamefont
			{{Chugunov}}},\ }\bibfield  {title} {\bibinfo {title} {Thermal evolution of
			neutron stars in soft {X}-ray transients with thermodynamically consistent
			models of the accreted crust},\ }\href
	{https://doi.org/10.1093/mnras/stad1309} {\bibfield  {journal} {\bibinfo
			{journal} {\mnras}\ }\textbf {\bibinfo {volume} {522}},\ \bibinfo {pages}
		{4830} (\bibinfo {year} {2023})},\ \Eprint {https://arxiv.org/abs/2303.08716}
	{arXiv:2303.08716 [astro-ph.HE]} \BibitemShut {NoStop}%
	\bibitem [{\citenamefont {{Shchechilin}}\ \emph {et~al.}(2023)\citenamefont
		{{Shchechilin}}, \citenamefont {{Gusakov}},\ and\ \citenamefont
		{{Chugunov}}}]{Shchechilin_ea23_InnerCrustComposition}%
	\BibitemOpen
	\bibfield  {author} {\bibinfo {author} {\bibfnamefont {N.~N.}\ \bibnamefont
			{{Shchechilin}}}, \bibinfo {author} {\bibfnamefont {M.~E.}\ \bibnamefont
			{{Gusakov}}},\ and\ \bibinfo {author} {\bibfnamefont {A.~I.}\ \bibnamefont
			{{Chugunov}}},\ }\bibfield  {title} {\bibinfo {title} {{Accreting neutron
				stars: composition of the upper layers of the inner crust}},\ }\href
	{https://doi.org/10.1093/mnras/stad1731} {\bibfield  {journal} {\bibinfo
			{journal} {\mnras}\ }\textbf {\bibinfo {volume} {523}},\ \bibinfo {pages}
		{4643} (\bibinfo {year} {2023})},\ \Eprint {https://arxiv.org/abs/2303.10003}
	{arXiv:2303.10003 [astro-ph.HE]} \BibitemShut {NoStop}%
	\bibitem [{\citenamefont {{Gusakov}}\ and\ \citenamefont
		{{Chugunov}}(2020)}]{GC20_DiffEq}%
	\BibitemOpen
	\bibfield  {author} {\bibinfo {author} {\bibfnamefont {M.~E.}\ \bibnamefont
			{{Gusakov}}}\ and\ \bibinfo {author} {\bibfnamefont {A.~I.}\ \bibnamefont
			{{Chugunov}}},\ }\bibfield  {title} {\bibinfo {title} {Thermodynamically
			consistent equation of state for an accreted neutron star crust},\ }\href
	{https://doi.org/10.1103/PhysRevLett.124.191101} {\bibfield  {journal}
		{\bibinfo  {journal} {\prl}\ }\textbf {\bibinfo {volume} {124}},\ \bibinfo
		{eid} {191101} (\bibinfo {year} {2020})},\ \Eprint
	{https://arxiv.org/abs/2004.04195} {arXiv:2004.04195 [astro-ph.HE]}
	\BibitemShut {NoStop}%
	\bibitem [{\citenamefont {{Gusakov}}\ and\ \citenamefont
		{{Chugunov}}(2024)}]{GC24_nHD_Shells}%
	\BibitemOpen
	\bibfield  {author} {\bibinfo {author} {\bibfnamefont {M.~E.}\ \bibnamefont
			{{Gusakov}}}\ and\ \bibinfo {author} {\bibfnamefont {A.~I.}\ \bibnamefont
			{{Chugunov}}},\ }\bibfield  {title} {\bibinfo {title} {{Thermodynamically
				consistent accreted crust of neutron stars: The role of proton shell
				effects}},\ }\href {https://doi.org/10.1103/PhysRevD.109.123032} {\bibfield
		{journal} {\bibinfo  {journal} {\prd}\ }\textbf {\bibinfo {volume} {109}},\
		\bibinfo {eid} {123032} (\bibinfo {year} {2024})},\ \Eprint
	{https://arxiv.org/abs/2406.05680} {arXiv:2406.05680 [nucl-th]} \BibitemShut
	{NoStop}%
	\bibitem [{\citenamefont {{Deibel}}\ \emph {et~al.}(2017)\citenamefont
		{{Deibel}}, \citenamefont {{Cumming}}, \citenamefont {{Brown}},\ and\
		\citenamefont {{Reddy}}}]{Diebel_ea17_LateCool_Pasta}%
	\BibitemOpen
	\bibfield  {author} {\bibinfo {author} {\bibfnamefont {A.}~\bibnamefont
			{{Deibel}}}, \bibinfo {author} {\bibfnamefont {A.}~\bibnamefont {{Cumming}}},
		\bibinfo {author} {\bibfnamefont {E.~F.}\ \bibnamefont {{Brown}}},\ and\
		\bibinfo {author} {\bibfnamefont {S.}~\bibnamefont {{Reddy}}},\ }\bibfield
	{title} {\bibinfo {title} {Late-time cooling of neutron star transients and
			the physics of the inner crust},\ }\href
	{https://doi.org/10.3847/1538-4357/aa6a19} {\bibfield  {journal} {\bibinfo
			{journal} {\apj}\ }\textbf {\bibinfo {volume} {839}},\ \bibinfo {eid} {95}
		(\bibinfo {year} {2017})},\ \Eprint {https://arxiv.org/abs/1609.07155}
	{arXiv:1609.07155 [astro-ph.HE]} \BibitemShut {NoStop}%
	\bibitem [{\citenamefont {{Chaikin}}\ \emph {et~al.}(2018)\citenamefont
		{{Chaikin}}, \citenamefont {{Kaminker}},\ and\ \citenamefont
		{{Yakovlev}}}]{Chaikin_ea18}%
	\BibitemOpen
	\bibfield  {author} {\bibinfo {author} {\bibfnamefont {E.~A.}\ \bibnamefont
			{{Chaikin}}}, \bibinfo {author} {\bibfnamefont {A.~D.}\ \bibnamefont
			{{Kaminker}}},\ and\ \bibinfo {author} {\bibfnamefont {D.~G.}\ \bibnamefont
			{{Yakovlev}}},\ }\bibfield  {title} {\bibinfo {title} {Afterburst thermal
			relaxation in neutron star crusts},\ }\href
	{https://doi.org/10.1007/s10509-018-3393-z} {\bibfield  {journal} {\bibinfo
			{journal} {\apss}\ }\textbf {\bibinfo {volume} {363}},\ \bibinfo {eid} {209}
		(\bibinfo {year} {2018})},\ \Eprint {https://arxiv.org/abs/1807.06855}
	{arXiv:1807.06855 [astro-ph.HE]} \BibitemShut {NoStop}%
		\bibitem [{\citenamefont {{Colpi}}\ \emph {et~al.}(2001)\citenamefont
		{{Colpi}}, \citenamefont {{Geppert}}, \citenamefont {{Page}},\ and\
		\citenamefont {{Possenti}}}]{Colpi_ea01_T_of_SXRT}%
	\BibitemOpen
	\bibfield  {author} {\bibinfo {author} {\bibfnamefont {M.}~\bibnamefont
			{{Colpi}}}, \bibinfo {author} {\bibfnamefont {U.}~\bibnamefont {{Geppert}}},
		\bibinfo {author} {\bibfnamefont {D.}~\bibnamefont {{Page}}},\ and\ \bibinfo
		{author} {\bibfnamefont {A.}~\bibnamefont {{Possenti}}},\ }\bibfield  {title}
	{\bibinfo {title} {{Charting the Temperature of the Hot Neutron Star in a
				Soft X-Ray Transient}},\ }\href {https://doi.org/10.1086/319107} {\bibfield
		{journal} {\bibinfo  {journal} {\apjl}\ }\textbf {\bibinfo {volume} {548}},\
		\bibinfo {pages} {L175} (\bibinfo {year} {2001})},\ \Eprint
	{https://arxiv.org/abs/astro-ph/0010572} {arXiv:astro-ph/0010572 [astro-ph]}
	\BibitemShut {NoStop}%
	\bibitem [{\citenamefont {{Yakovlev}}\ \emph {et~al.}(2003)\citenamefont
		{{Yakovlev}}, \citenamefont {{Levenfish}},\ and\ \citenamefont
		{{Haensel}}}]{Yakovlev_etal03_transients}%
	\BibitemOpen
	\bibfield  {author} {\bibinfo {author} {\bibfnamefont {D.~G.}\ \bibnamefont
			{{Yakovlev}}}, \bibinfo {author} {\bibfnamefont {K.~P.}\ \bibnamefont
			{{Levenfish}}},\ and\ \bibinfo {author} {\bibfnamefont {P.}~\bibnamefont
			{{Haensel}}},\ }\bibfield  {title} {\bibinfo {title} {{Thermal state of
				transiently accreting neutron stars}},\ }\href
	{https://doi.org/10.1051/0004-6361:20030830} {\bibfield  {journal} {\bibinfo
			{journal} {\aap}\ }\textbf {\bibinfo {volume} {407}},\ \bibinfo {pages} {265}
		(\bibinfo {year} {2003})},\ \Eprint {https://arxiv.org/abs/astro-ph/0209027}
	{arXiv:astro-ph/0209027 [astro-ph]} \BibitemShut {NoStop}%
	\bibitem [{\citenamefont {{Pringle}}\ and\ \citenamefont
		{{Rees}}(1972)}]{PR72}%
	\BibitemOpen
	\bibfield  {author} {\bibinfo {author} {\bibfnamefont {J.~E.}\ \bibnamefont
			{{Pringle}}}\ and\ \bibinfo {author} {\bibfnamefont {M.~J.}\ \bibnamefont
			{{Rees}}},\ }\bibfield  {title} {\bibinfo {title} {Accretion disc models for
			compact {X}-ray sources},\ }\href@noop {} {\bibfield  {journal} {\bibinfo
			{journal} {\aap}\ }\textbf {\bibinfo {volume} {21}},\ \bibinfo {pages} {1}
		(\bibinfo {year} {1972})}\BibitemShut {NoStop}%
	\bibitem [{\citenamefont {{Lamb}}\ \emph {et~al.}(1973)\citenamefont {{Lamb}},
		\citenamefont {{Pethick}},\ and\ \citenamefont {{Pines}}}]{L+73}%
	\BibitemOpen
	\bibfield  {author} {\bibinfo {author} {\bibfnamefont {F.~K.}\ \bibnamefont
			{{Lamb}}}, \bibinfo {author} {\bibfnamefont {C.~J.}\ \bibnamefont
			{{Pethick}}},\ and\ \bibinfo {author} {\bibfnamefont {D.}~\bibnamefont
			{{Pines}}},\ }\bibfield  {title} {\bibinfo {title} {A model for compact
			{X}-ray sources: Accretion by rotating magnetic stars},\ }\href
	{https://doi.org/10.1086/152325} {\bibfield  {journal} {\bibinfo  {journal}
			{\apj}\ }\textbf {\bibinfo {volume} {184}},\ \bibinfo {pages} {271} (\bibinfo
		{year} {1973})}\BibitemShut {NoStop}%
	\bibitem [{\citenamefont {{Cavecchi}}\ \emph {et~al.}(2011)\citenamefont
		{{Cavecchi}}, \citenamefont {{Patruno}}, \citenamefont {{Haskell}},
		\citenamefont {{Watts}}, \citenamefont {{Levin}}, \citenamefont {{Linares}},
		\citenamefont {{Altamirano}}, \citenamefont {{Wijnands}},\ and\ \citenamefont
		{{van der Klis}}}]{Cavecchi_ea11_J17480-2446_burst_oscill}%
	\BibitemOpen
	\bibfield  {author} {\bibinfo {author} {\bibfnamefont {Y.}~\bibnamefont
			{{Cavecchi}}}, \bibinfo {author} {\bibfnamefont {A.}~\bibnamefont
			{{Patruno}}}, \bibinfo {author} {\bibfnamefont {B.}~\bibnamefont
			{{Haskell}}}, \bibinfo {author} {\bibfnamefont {A.~L.}\ \bibnamefont
			{{Watts}}}, \bibinfo {author} {\bibfnamefont {Y.}~\bibnamefont {{Levin}}},
		\bibinfo {author} {\bibfnamefont {M.}~\bibnamefont {{Linares}}}, \bibinfo
		{author} {\bibfnamefont {D.}~\bibnamefont {{Altamirano}}}, \bibinfo {author}
		{\bibfnamefont {R.}~\bibnamefont {{Wijnands}}},\ and\ \bibinfo {author}
		{\bibfnamefont {M.}~\bibnamefont {{van der Klis}}},\ }\bibfield  {title}
	{\bibinfo {title} {Implications of burst oscillations from the slowly
			rotating accreting pulsar {IGR} {J}17480-2446 in the globular cluster
			{Terzan} 5},\ }\href {https://doi.org/10.1088/2041-8205/740/1/L8} {\bibfield
		{journal} {\bibinfo  {journal} {\apjl}\ }\textbf {\bibinfo {volume} {740}},\
		\bibinfo {eid} {L8} (\bibinfo {year} {2011})},\ \Eprint
	{https://arxiv.org/abs/1102.1548} {arXiv:1102.1548 [astro-ph.HE]}
	\BibitemShut {NoStop}%
	\bibitem [{\citenamefont {{Crociati}}\ \emph {et~al.}(2024)\citenamefont
		{{Crociati}}, \citenamefont {{Cignoni}}, \citenamefont {{Dalessandro}},
		\citenamefont {{Pallanca}}, \citenamefont {{Massari}}, \citenamefont
		{{Ferraro}}, \citenamefont {{Lanzoni}}, \citenamefont {{Origlia}},\ and\
		\citenamefont {{Valenti}}}]{Crociati_ea24_StarFormHist_Ter5}%
	\BibitemOpen
	\bibfield  {author} {\bibinfo {author} {\bibfnamefont {C.}~\bibnamefont
			{{Crociati}}}, \bibinfo {author} {\bibfnamefont {M.}~\bibnamefont
			{{Cignoni}}}, \bibinfo {author} {\bibfnamefont {E.}~\bibnamefont
			{{Dalessandro}}}, \bibinfo {author} {\bibfnamefont {C.}~\bibnamefont
			{{Pallanca}}}, \bibinfo {author} {\bibfnamefont {D.}~\bibnamefont
			{{Massari}}}, \bibinfo {author} {\bibfnamefont {F.~R.}\ \bibnamefont
			{{Ferraro}}}, \bibinfo {author} {\bibfnamefont {B.}~\bibnamefont
			{{Lanzoni}}}, \bibinfo {author} {\bibfnamefont {L.}~\bibnamefont
			{{Origlia}}},\ and\ \bibinfo {author} {\bibfnamefont {E.}~\bibnamefont
			{{Valenti}}},\ }\bibfield  {title} {\bibinfo {title} {The star formation
			history of the first bulge fossil fragment candidate {Terzan} 5},\ }\href
	{https://doi.org/10.1051/0004-6361/202451174} {\bibfield  {journal} {\bibinfo
			{journal} {\aap}\ }\textbf {\bibinfo {volume} {691}},\ \bibinfo {eid} {A311}
		(\bibinfo {year} {2024})},\ \Eprint {https://arxiv.org/abs/2410.16971}
	{arXiv:2410.16971 [astro-ph.GA]} \BibitemShut {NoStop}%
	\bibitem [{\citenamefont {{Manchester}}\ \emph {et~al.}(2005)\citenamefont
		{{Manchester}}, \citenamefont {{Hobbs}}, \citenamefont {{Teoh}},\ and\
		\citenamefont {{Hobbs}}}]{atnf_paper}%
	\BibitemOpen
	\bibfield  {author} {\bibinfo {author} {\bibfnamefont {R.~N.}\ \bibnamefont
			{{Manchester}}}, \bibinfo {author} {\bibfnamefont {G.~B.}\ \bibnamefont
			{{Hobbs}}}, \bibinfo {author} {\bibfnamefont {A.}~\bibnamefont {{Teoh}}},\
		and\ \bibinfo {author} {\bibfnamefont {M.}~\bibnamefont {{Hobbs}}},\
	}\bibfield  {title} {\bibinfo {title} {The {Australia} telescope national
			facility pulsar catalogue},\ }\href {https://doi.org/10.1086/428488}
	{\bibfield  {journal} {\bibinfo  {journal} {\aj}\ }\textbf {\bibinfo {volume}
			{129}},\ \bibinfo {pages} {1993} (\bibinfo {year} {2005})},\ \Eprint
	{https://arxiv.org/abs/astro-ph/0412641} {astro-ph/0412641} \BibitemShut
	{NoStop}%
	\bibitem [{atn()}]{atnf_web263}%
	\BibitemOpen
	\href@noop {} {\bibinfo {title} {The {ATNF} pulsar catalogue, version
			2.6.3}},\ \bibinfo {howpublished}
	{https://www.atnf.csiro.au/research/pulsar/psrcat/index.php?version=2.6.3}\BibitemShut
	{NoStop}%
	\bibitem [{Note1()}]{Note1}%
	\BibitemOpen
	\bibinfo {note} {For consistency, the magnetic fields of pulsars in Fig.\
		\ref {Fig:InitParams} are calculated as $B=2.6\times 10^{19} (P\protect \dot
		{P})^{1/2}$ G, where $P$ and $\protect \dot {P}$ are the period in seconds
		and its derivative; these data were taken from the ATNF catalog \cite
		{atnf_paper} (version 2.6.3 \cite {atnf_web263}; note, the numerical factor
		for $B$ is 1.7 times lower than that in the vacuum formula, used by default
		in the ATNF catalog for historical reasons). For pulsars in GCs, the data
		were updated according to P.\ Freire's catalog \cite
		{Freire_GC_Pulsar_Cat_y25m07d25} (downloaded on July 25, 2025); it allowed us
		to extract $\protect \dot {P}$ values for certain pulsars. The observed (not
		intrinsic) values of $\protect \dot {P}$ were used in all cases.}\BibitemShut
	{Stop}%
	\bibitem [{\citenamefont {{Faucher-Gigu{\`e}re}}\ and\ \citenamefont
		{{Kaspi}}(2006)}]{FK06}%
	\BibitemOpen
	\bibfield  {author} {\bibinfo {author} {\bibfnamefont {C.-A.}\ \bibnamefont
			{{Faucher-Gigu{\`e}re}}}\ and\ \bibinfo {author} {\bibfnamefont {V.~M.}\
			\bibnamefont {{Kaspi}}},\ }\bibfield  {title} {\bibinfo {title} {Birth and
			evolution of isolated radio pulsars},\ }\href
	{https://doi.org/10.1086/501516} {\bibfield  {journal} {\bibinfo  {journal}
			{\apj}\ }\textbf {\bibinfo {volume} {643}},\ \bibinfo {pages} {332} (\bibinfo
		{year} {2006})},\ \Eprint {https://arxiv.org/abs/astro-ph/0512585}
	{arXiv:astro-ph/0512585 [astro-ph]} \BibitemShut {NoStop}%
	\bibitem [{\citenamefont {{Kaspi}}\ and\ \citenamefont
		{{Kramer}}(2016)}]{KK16}%
	\BibitemOpen
	\bibfield  {author} {\bibinfo {author} {\bibfnamefont {V.~M.}\ \bibnamefont
			{{Kaspi}}}\ and\ \bibinfo {author} {\bibfnamefont {M.}~\bibnamefont
			{{Kramer}}},\ }\bibfield  {title} {\bibinfo {title} {Radio pulsars: The
			neutron star population \& fundamental physics},\ }\href
	{https://doi.org/10.48550/arXiv.1602.07738} {\bibfield  {journal} {\bibinfo
			{journal} {arXiv e-prints}\ ,\ \bibinfo {eid} {arXiv:1602.07738}} (\bibinfo
		{year} {2016})},\ \Eprint {https://arxiv.org/abs/1602.07738}
	{arXiv:1602.07738 [astro-ph.HE]} \BibitemShut {NoStop}%
	\bibitem [{\citenamefont {Igoshev}\ \emph {et~al.}(2022)\citenamefont
		{Igoshev}, \citenamefont {Frantsuzova}, \citenamefont {Gourgouliatos},
		\citenamefont {Tsichli}, \citenamefont {Konstantinou},\ and\ \citenamefont
		{Popov}}]{Igoshev_ea22_InitP_B}%
	\BibitemOpen
	\bibfield  {author} {\bibinfo {author} {\bibfnamefont {A.~P.}\ \bibnamefont
			{Igoshev}}, \bibinfo {author} {\bibfnamefont {A.}~\bibnamefont
			{Frantsuzova}}, \bibinfo {author} {\bibfnamefont {K.~N.}\ \bibnamefont
			{Gourgouliatos}}, \bibinfo {author} {\bibfnamefont {S.}~\bibnamefont
			{Tsichli}}, \bibinfo {author} {\bibfnamefont {L.}~\bibnamefont
			{Konstantinou}},\ and\ \bibinfo {author} {\bibfnamefont {S.~B.}\ \bibnamefont
			{Popov}},\ }\bibfield  {title} {\bibinfo {title} {Initial periods and
			magnetic fields of neutron stars},\ }\href
	{https://doi.org/10.1093/mnras/stac1648} {\bibfield  {journal} {\bibinfo
			{journal} {\mnras}\ }\textbf {\bibinfo {volume} {514}},\ \bibinfo {pages}
		{4606} (\bibinfo {year} {2022})},\ \Eprint
	{https://arxiv.org/abs/https://academic.oup.com/mnras/article-pdf/514/3/4606/44383044/stac1648.pdf}
	{https://academic.oup.com/mnras/article-pdf/514/3/4606/44383044/stac1648.pdf}
	\BibitemShut {NoStop}%
	\bibitem [{\citenamefont {{Pfahl}}\ \emph
		{et~al.}(2002{\natexlab{a}})\citenamefont {{Pfahl}}, \citenamefont
		{{Rappaport}},\ and\ \citenamefont {{Podsiadlowski}}}]{P+02}%
	\BibitemOpen
	\bibfield  {author} {\bibinfo {author} {\bibfnamefont {E.}~\bibnamefont
			{{Pfahl}}}, \bibinfo {author} {\bibfnamefont {S.}~\bibnamefont
			{{Rappaport}}},\ and\ \bibinfo {author} {\bibfnamefont {P.}~\bibnamefont
			{{Podsiadlowski}}},\ }\bibfield  {title} {\bibinfo {title} {A comprehensive
			study of neutron star retention in globular clusters},\ }\href
	{https://doi.org/10.1086/340494} {\bibfield  {journal} {\bibinfo  {journal}
			{\apj}\ }\textbf {\bibinfo {volume} {573}},\ \bibinfo {pages} {283} (\bibinfo
		{year} {2002}{\natexlab{a}})},\ \Eprint
	{https://arxiv.org/abs/astro-ph/0106141} {arXiv:astro-ph/0106141 [astro-ph]}
	\BibitemShut {NoStop}%
	\bibitem [{\citenamefont {{Kuranov}}\ and\ \citenamefont
		{{Postnov}}(2006)}]{KP06}%
	\BibitemOpen
	\bibfield  {author} {\bibinfo {author} {\bibfnamefont {A.~G.}\ \bibnamefont
			{{Kuranov}}}\ and\ \bibinfo {author} {\bibfnamefont {K.~A.}\ \bibnamefont
			{{Postnov}}},\ }\bibfield  {title} {\bibinfo {title} {Neutron stars in
			globular clusters: Formation and observational manifestations},\ }\href
	{https://doi.org/10.1134/S106377370606003X} {\bibfield  {journal} {\bibinfo
			{journal} {\astronlett}\ }\textbf {\bibinfo {volume} {32}},\ \bibinfo {pages}
		{393} (\bibinfo {year} {2006})},\ \Eprint
	{https://arxiv.org/abs/astro-ph/0605115} {arXiv:astro-ph/0605115 [astro-ph]}
	\BibitemShut {NoStop}%
	\bibitem [{\citenamefont {{Ivanova}}\ \emph {et~al.}(2008)\citenamefont
		{{Ivanova}}, \citenamefont {{Heinke}}, \citenamefont {{Rasio}}, \citenamefont
		{{Belczynski}},\ and\ \citenamefont {{Fregeau}}}]{Ivanova_etal08}%
	\BibitemOpen
	\bibfield  {author} {\bibinfo {author} {\bibfnamefont {N.}~\bibnamefont
			{{Ivanova}}}, \bibinfo {author} {\bibfnamefont {C.~O.}\ \bibnamefont
			{{Heinke}}}, \bibinfo {author} {\bibfnamefont {F.~A.}\ \bibnamefont
			{{Rasio}}}, \bibinfo {author} {\bibfnamefont {K.}~\bibnamefont
			{{Belczynski}}},\ and\ \bibinfo {author} {\bibfnamefont {J.~M.}\ \bibnamefont
			{{Fregeau}}},\ }\bibfield  {title} {\bibinfo {title} {Formation and evolution
			of compact binaries in globular clusters - {II}. binaries with neutron
			stars},\ }\href {https://doi.org/10.1111/j.1365-2966.2008.13064.x} {\bibfield
		{journal} {\bibinfo  {journal} {\mnras}\ }\textbf {\bibinfo {volume}
			{386}},\ \bibinfo {pages} {553} (\bibinfo {year} {2008})},\ \Eprint
	{https://arxiv.org/abs/0706.4096} {arXiv:0706.4096} \BibitemShut {NoStop}%
	\bibitem [{\citenamefont {{Podsiadlowski}}\ \emph {et~al.}(2005)\citenamefont
		{{Podsiadlowski}}, \citenamefont {{Dewi}}, \citenamefont {{Lesaffre}},
		\citenamefont {{Miller}}, \citenamefont {{Newton}},\ and\ \citenamefont
		{{Stone}}}]{P+05}%
	\BibitemOpen
	\bibfield  {author} {\bibinfo {author} {\bibfnamefont {P.}~\bibnamefont
			{{Podsiadlowski}}}, \bibinfo {author} {\bibfnamefont {J.~D.~M.}\ \bibnamefont
			{{Dewi}}}, \bibinfo {author} {\bibfnamefont {P.}~\bibnamefont {{Lesaffre}}},
		\bibinfo {author} {\bibfnamefont {J.~C.}\ \bibnamefont {{Miller}}}, \bibinfo
		{author} {\bibfnamefont {W.~G.}\ \bibnamefont {{Newton}}},\ and\ \bibinfo
		{author} {\bibfnamefont {J.~R.}\ \bibnamefont {{Stone}}},\ }\bibfield
	{title} {\bibinfo {title} {The double pulsar {J}0737-3039: testing the
			neutron star equation of state},\ }\href
	{https://doi.org/10.1111/j.1365-2966.2005.09253.x} {\bibfield  {journal}
		{\bibinfo  {journal} {\mnras}\ }\textbf {\bibinfo {volume} {361}},\ \bibinfo
		{pages} {1243} (\bibinfo {year} {2005})},\ \Eprint
	{https://arxiv.org/abs/astro-ph/0506566} {arXiv:astro-ph/0506566 [astro-ph]}
	\BibitemShut {NoStop}%
	\bibitem [{\citenamefont {{Boyles}}\ \emph {et~al.}(2011)\citenamefont
		{{Boyles}}, \citenamefont {{Lorimer}}, \citenamefont {{Turk}}, \citenamefont
		{{Mnatsakanov}}, \citenamefont {{Lynch}}, \citenamefont {{Ransom}},
		\citenamefont {{Freire}},\ and\ \citenamefont {{Belczynski}}}]{B+11}%
	\BibitemOpen
	\bibfield  {author} {\bibinfo {author} {\bibfnamefont {J.}~\bibnamefont
			{{Boyles}}}, \bibinfo {author} {\bibfnamefont {D.~R.}\ \bibnamefont
			{{Lorimer}}}, \bibinfo {author} {\bibfnamefont {P.~J.}\ \bibnamefont
			{{Turk}}}, \bibinfo {author} {\bibfnamefont {R.}~\bibnamefont
			{{Mnatsakanov}}}, \bibinfo {author} {\bibfnamefont {R.~S.}\ \bibnamefont
			{{Lynch}}}, \bibinfo {author} {\bibfnamefont {S.~M.}\ \bibnamefont
			{{Ransom}}}, \bibinfo {author} {\bibfnamefont {P.~C.}\ \bibnamefont
			{{Freire}}},\ and\ \bibinfo {author} {\bibfnamefont {K.}~\bibnamefont
			{{Belczynski}}},\ }\bibfield  {title} {\bibinfo {title} {Young radio pulsars
			in galactic globular clusters},\ }\href
	{https://doi.org/10.1088/0004-637X/742/1/51} {\bibfield  {journal} {\bibinfo
			{journal} {\apj}\ }\textbf {\bibinfo {volume} {742}},\ \bibinfo {eid} {51}
		(\bibinfo {year} {2011})},\ \Eprint {https://arxiv.org/abs/1108.4402}
	{arXiv:1108.4402 [astro-ph.SR]} \BibitemShut {NoStop}%
	\bibitem [{\citenamefont {{Tauris}}\ \emph {et~al.}(2013)\citenamefont
		{{Tauris}}, \citenamefont {{Sanyal}}, \citenamefont {{Yoon}},\ and\
		\citenamefont {{Langer}}}]{Tauris_ea13_AIC}%
	\BibitemOpen
	\bibfield  {author} {\bibinfo {author} {\bibfnamefont {T.~M.}\ \bibnamefont
			{{Tauris}}}, \bibinfo {author} {\bibfnamefont {D.}~\bibnamefont {{Sanyal}}},
		\bibinfo {author} {\bibfnamefont {S.~C.}\ \bibnamefont {{Yoon}}},\ and\
		\bibinfo {author} {\bibfnamefont {N.}~\bibnamefont {{Langer}}},\ }\bibfield
	{title} {\bibinfo {title} {Evolution towards and beyond accretion-induced
			collapse of massive white dwarfs and formation of millisecond pulsars},\
	}\href {https://doi.org/10.1051/0004-6361/201321662} {\bibfield  {journal}
		{\bibinfo  {journal} {\aap}\ }\textbf {\bibinfo {volume} {558}},\ \bibinfo
		{eid} {A39} (\bibinfo {year} {2013})},\ \Eprint
	{https://arxiv.org/abs/1308.4887} {arXiv:1308.4887 [astro-ph.SR]}
	\BibitemShut {NoStop}%
	\bibitem [{\citenamefont {{Dessart}}\ \emph {et~al.}(2007)\citenamefont
		{{Dessart}}, \citenamefont {{Burrows}}, \citenamefont {{Livne}},\ and\
		\citenamefont {{Ott}}}]{D+07}%
	\BibitemOpen
	\bibfield  {author} {\bibinfo {author} {\bibfnamefont {L.}~\bibnamefont
			{{Dessart}}}, \bibinfo {author} {\bibfnamefont {A.}~\bibnamefont
			{{Burrows}}}, \bibinfo {author} {\bibfnamefont {E.}~\bibnamefont {{Livne}}},\
		and\ \bibinfo {author} {\bibfnamefont {C.~D.}\ \bibnamefont {{Ott}}},\
	}\bibfield  {title} {\bibinfo {title} {Magnetically driven explosions of
			rapidly rotating white dwarfs following accretion-induced collapse},\ }\href
	{https://doi.org/10.1086/521701} {\bibfield  {journal} {\bibinfo  {journal}
			{\apj}\ }\textbf {\bibinfo {volume} {669}},\ \bibinfo {pages} {585} (\bibinfo
		{year} {2007})},\ \Eprint {https://arxiv.org/abs/0705.3678} {arXiv:0705.3678
		[astro-ph]} \BibitemShut {NoStop}%
	\bibitem [{\citenamefont {Cheong}\ \emph {et~al.}(2025)\citenamefont {Cheong},
		\citenamefont {Pitik}, \citenamefont {Longo~Micchi},\ and\ \citenamefont
		{Radice}}]{C+25}%
	\BibitemOpen
	\bibfield  {author} {\bibinfo {author} {\bibfnamefont {P.~C.-K.}\
			\bibnamefont {Cheong}}, \bibinfo {author} {\bibfnamefont {T.}~\bibnamefont
			{Pitik}}, \bibinfo {author} {\bibfnamefont {L.~F.}\ \bibnamefont
			{Longo~Micchi}},\ and\ \bibinfo {author} {\bibfnamefont {D.}~\bibnamefont
			{Radice}},\ }\bibfield  {title} {\bibinfo {title} {Gamma-ray bursts and
			kilonovae from the accretion-induced collapse of white dwarfs},\ }\href
	{https://doi.org/10.3847/2041-8213/ada1cc} {\bibfield  {journal} {\bibinfo
			{journal} {\apjl}\ }\textbf {\bibinfo {volume} {978}},\ \bibinfo {pages}
		{L38} (\bibinfo {year} {2025})}\BibitemShut {NoStop}%
	\bibitem [{\citenamefont {King}\ \emph {et~al.}(2001)\citenamefont {King},
		\citenamefont {Pringle},\ and\ \citenamefont {Wickramasinghe}}]{K+01}%
	\BibitemOpen
	\bibfield  {author} {\bibinfo {author} {\bibfnamefont {A.~R.}\ \bibnamefont
			{King}}, \bibinfo {author} {\bibfnamefont {J.~E.}\ \bibnamefont {Pringle}},\
		and\ \bibinfo {author} {\bibfnamefont {D.~T.}\ \bibnamefont
			{Wickramasinghe}},\ }\bibfield  {title} {\bibinfo {title} {Type {Ia}
			supernovae and remnant neutron stars},\ }\href
	{https://doi.org/10.1046/j.1365-8711.2001.04184.x} {\bibfield  {journal}
		{\bibinfo  {journal} {\mnras}\ }\textbf {\bibinfo {volume} {320}},\ \bibinfo
		{pages} {L45} (\bibinfo {year} {2001})},\ \Eprint
	{https://arxiv.org/abs/https://academic.oup.com/mnras/article-pdf/320/3/L45/3800744/320-3-L45.pdf}
	{https://academic.oup.com/mnras/article-pdf/320/3/L45/3800744/320-3-L45.pdf}
	\BibitemShut {NoStop}%
	\bibitem [{\citenamefont {Levan}\ \emph {et~al.}(2006)\citenamefont {Levan},
		\citenamefont {Wynn}, \citenamefont {Chapman}, \citenamefont {Davies},
		\citenamefont {King}, \citenamefont {Priddey},\ and\ \citenamefont
		{Tanvir}}]{L+06}%
	\BibitemOpen
	\bibfield  {author} {\bibinfo {author} {\bibfnamefont {A.~J.}\ \bibnamefont
			{Levan}}, \bibinfo {author} {\bibfnamefont {G.~A.}\ \bibnamefont {Wynn}},
		\bibinfo {author} {\bibfnamefont {R.}~\bibnamefont {Chapman}}, \bibinfo
		{author} {\bibfnamefont {M.~B.}\ \bibnamefont {Davies}}, \bibinfo {author}
		{\bibfnamefont {A.~R.}\ \bibnamefont {King}}, \bibinfo {author}
		{\bibfnamefont {R.~S.}\ \bibnamefont {Priddey}},\ and\ \bibinfo {author}
		{\bibfnamefont {N.~R.}\ \bibnamefont {Tanvir}},\ }\bibfield  {title}
	{\bibinfo {title} {Short gamma-ray bursts in old populations: magnetars from
			white dwarf-white dwarf mergers},\ }\href
	{https://doi.org/10.1111/j.1745-3933.2006.00144.x} {\bibfield  {journal}
		{\bibinfo  {journal} {\mnrasl}\ }\textbf {\bibinfo {volume} {368}},\ \bibinfo
		{pages} {L1} (\bibinfo {year} {2006})},\ \Eprint
	{https://arxiv.org/abs/https://academic.oup.com/mnrasl/article-pdf/368/1/L1/54689589/mnrasl\_368\_1\_l1.pdf}
	{https://academic.oup.com/mnrasl/article-pdf/368/1/L1/54689589/mnrasl\_368\_1\_l1.pdf}
	\BibitemShut {NoStop}%
	\bibitem [{\citenamefont {{Ruiter}}\ \emph {et~al.}(2019)\citenamefont
		{{Ruiter}}, \citenamefont {{Ferrario}}, \citenamefont {{Belczynski}},
		\citenamefont {{Seitenzahl}}, \citenamefont {{Crocker}},\ and\ \citenamefont
		{{Karakas}}}]{Ruiter_ea19_AIC}%
	\BibitemOpen
	\bibfield  {author} {\bibinfo {author} {\bibfnamefont {A.~J.}\ \bibnamefont
			{{Ruiter}}}, \bibinfo {author} {\bibfnamefont {L.}~\bibnamefont
			{{Ferrario}}}, \bibinfo {author} {\bibfnamefont {K.}~\bibnamefont
			{{Belczynski}}}, \bibinfo {author} {\bibfnamefont {I.~R.}\ \bibnamefont
			{{Seitenzahl}}}, \bibinfo {author} {\bibfnamefont {R.~M.}\ \bibnamefont
			{{Crocker}}},\ and\ \bibinfo {author} {\bibfnamefont {A.~I.}\ \bibnamefont
			{{Karakas}}},\ }\bibfield  {title} {\bibinfo {title} {On the formation of
			neutron stars via accretion-induced collapse in binaries},\ }\href
	{https://doi.org/10.1093/mnras/stz001} {\bibfield  {journal} {\bibinfo
			{journal} {\mnras}\ }\textbf {\bibinfo {volume} {484}},\ \bibinfo {pages}
		{698} (\bibinfo {year} {2019})},\ \Eprint {https://arxiv.org/abs/1802.02437}
	{arXiv:1802.02437 [astro-ph.SR]} \BibitemShut {NoStop}%
	\bibitem [{\citenamefont {{Nomoto}}(1984)}]{N84}%
	\BibitemOpen
	\bibfield  {author} {\bibinfo {author} {\bibfnamefont {K.}~\bibnamefont
			{{Nomoto}}},\ }\bibfield  {title} {\bibinfo {title} {Evolution of 8-10 solar
			mass stars toward electron capture supernovae. {I} - formation of
			electron-degenerate {O+Ne+Mg} cores},\ }\href
	{https://doi.org/10.1086/161749} {\bibfield  {journal} {\bibinfo  {journal}
			{\apj}\ }\textbf {\bibinfo {volume} {277}},\ \bibinfo {pages} {791} (\bibinfo
		{year} {1984})}\BibitemShut {NoStop}%
	\bibitem [{\citenamefont {{Nomoto}}(1987)}]{N87}%
	\BibitemOpen
	\bibfield  {author} {\bibinfo {author} {\bibfnamefont {K.}~\bibnamefont
			{{Nomoto}}},\ }\bibfield  {title} {\bibinfo {title} {Evolution of 8--10
			m$_\odot$ stars toward electron capture supernovae. {II}. collapse of an
			{O+Ne+Mg} core},\ }\href {https://doi.org/10.1086/165716} {\bibfield
		{journal} {\bibinfo  {journal} {\apj}\ }\textbf {\bibinfo {volume} {322}},\
		\bibinfo {pages} {206} (\bibinfo {year} {1987})}\BibitemShut {NoStop}%
	\bibitem [{\citenamefont {Podsiadlowski}\ \emph {et~al.}(2004)\citenamefont
		{Podsiadlowski}, \citenamefont {Langer}, \citenamefont {Poelarends},
		\citenamefont {Rappaport}, \citenamefont {Heger},\ and\ \citenamefont
		{Pfahl}}]{P+04}%
	\BibitemOpen
	\bibfield  {author} {\bibinfo {author} {\bibfnamefont {P.}~\bibnamefont
			{Podsiadlowski}}, \bibinfo {author} {\bibfnamefont {N.}~\bibnamefont
			{Langer}}, \bibinfo {author} {\bibfnamefont {A.~J.~T.}\ \bibnamefont
			{Poelarends}}, \bibinfo {author} {\bibfnamefont {S.}~\bibnamefont
			{Rappaport}}, \bibinfo {author} {\bibfnamefont {A.}~\bibnamefont {Heger}},\
		and\ \bibinfo {author} {\bibfnamefont {E.}~\bibnamefont {Pfahl}},\ }\bibfield
	{title} {\bibinfo {title} {The effects of binary evolution on the dynamics
			of core collapse and neutron star kicks},\ }\href
	{https://doi.org/10.1086/421713} {\bibfield  {journal} {\bibinfo  {journal}
			{\apj}\ }\textbf {\bibinfo {volume} {612}},\ \bibinfo {pages} {1044}
		(\bibinfo {year} {2004})}\BibitemShut {NoStop}%
	\bibitem [{\citenamefont {{Kitaura}}\ \emph {et~al.}(2006)\citenamefont
		{{Kitaura}}, \citenamefont {{Janka}},\ and\ \citenamefont
		{{{Hill}ebrandt}}}]{K+06}%
	\BibitemOpen
	\bibfield  {author} {\bibinfo {author} {\bibfnamefont {F.~S.}\ \bibnamefont
			{{Kitaura}}}, \bibinfo {author} {\bibfnamefont {H.~T.}\ \bibnamefont
			{{Janka}}},\ and\ \bibinfo {author} {\bibfnamefont {W.}~\bibnamefont
			{{{Hill}ebrandt}}},\ }\bibfield  {title} {\bibinfo {title} {Explosions of
			{O-Ne-Mg} cores, the {Crab} supernova, and subluminous type {II-P}
			supernovae},\ }\href {https://doi.org/10.1051/0004-6361:20054703} {\bibfield
		{journal} {\bibinfo  {journal} {\aap}\ }\textbf {\bibinfo {volume} {450}},\
		\bibinfo {pages} {345} (\bibinfo {year} {2006})},\ \Eprint
	{https://arxiv.org/abs/astro-ph/0512065} {arXiv:astro-ph/0512065 [astro-ph]}
	\BibitemShut {NoStop}%
	\bibitem [{\citenamefont {Poelarends}\ \emph {et~al.}(2008)\citenamefont
		{Poelarends}, \citenamefont {Herwig}, \citenamefont {Langer},\ and\
		\citenamefont {Heger}}]{P+08}%
	\BibitemOpen
	\bibfield  {author} {\bibinfo {author} {\bibfnamefont {A.~J.~T.}\
			\bibnamefont {Poelarends}}, \bibinfo {author} {\bibfnamefont
			{F.}~\bibnamefont {Herwig}}, \bibinfo {author} {\bibfnamefont
			{N.}~\bibnamefont {Langer}},\ and\ \bibinfo {author} {\bibfnamefont
			{A.}~\bibnamefont {Heger}},\ }\bibfield  {title} {\bibinfo {title} {The
			supernova channel of super-{AGB} stars},\ }\href
	{https://doi.org/10.1086/520872} {\bibfield  {journal} {\bibinfo  {journal}
			{\apj}\ }\textbf {\bibinfo {volume} {675}},\ \bibinfo {pages} {614} (\bibinfo
		{year} {2008})}\BibitemShut {NoStop}%
	\bibitem [{\citenamefont {{Janka}}\ \emph {et~al.}(2008)\citenamefont
		{{Janka}}, \citenamefont {{M{\"u}ller}}, \citenamefont {{Kitaura}},\ and\
		\citenamefont {{Buras}}}]{J+08}%
	\BibitemOpen
	\bibfield  {author} {\bibinfo {author} {\bibfnamefont {H.~T.}\ \bibnamefont
			{{Janka}}}, \bibinfo {author} {\bibfnamefont {B.}~\bibnamefont
			{{M{\"u}ller}}}, \bibinfo {author} {\bibfnamefont {F.~S.}\ \bibnamefont
			{{Kitaura}}},\ and\ \bibinfo {author} {\bibfnamefont {R.}~\bibnamefont
			{{Buras}}},\ }\bibfield  {title} {\bibinfo {title} {Dynamics of shock
			propagation and nucleosynthesis conditions in {O-Ne-Mg} core supernovae},\
	}\href {https://doi.org/10.1051/0004-6361:20079334} {\bibfield  {journal}
		{\bibinfo  {journal} {\aap}\ }\textbf {\bibinfo {volume} {485}},\ \bibinfo
		{pages} {199} (\bibinfo {year} {2008})},\ \Eprint
	{https://arxiv.org/abs/0712.4237} {arXiv:0712.4237 [astro-ph]} \BibitemShut
	{NoStop}%
	\bibitem [{\citenamefont {Jones}\ \emph {et~al.}(2013)\citenamefont {Jones},
		\citenamefont {Hirschi}, \citenamefont {Nomoto}, \citenamefont {Fischer},
		\citenamefont {Timmes}, \citenamefont {Herwig}, \citenamefont {Paxton},
		\citenamefont {Toki}, \citenamefont {Suzuki}, \citenamefont
		{Martinez-Pinedo}, \citenamefont {Lam},\ and\ \citenamefont
		{Bertolli}}]{J+13}%
	\BibitemOpen
	\bibfield  {author} {\bibinfo {author} {\bibfnamefont {S.}~\bibnamefont
			{Jones}}, \bibinfo {author} {\bibfnamefont {R.}~\bibnamefont {Hirschi}},
		\bibinfo {author} {\bibfnamefont {K.}~\bibnamefont {Nomoto}}, \bibinfo
		{author} {\bibfnamefont {T.}~\bibnamefont {Fischer}}, \bibinfo {author}
		{\bibfnamefont {F.~X.}\ \bibnamefont {Timmes}}, \bibinfo {author}
		{\bibfnamefont {F.}~\bibnamefont {Herwig}}, \bibinfo {author} {\bibfnamefont
			{B.}~\bibnamefont {Paxton}}, \bibinfo {author} {\bibfnamefont
			{H.}~\bibnamefont {Toki}}, \bibinfo {author} {\bibfnamefont {T.}~\bibnamefont
			{Suzuki}}, \bibinfo {author} {\bibfnamefont {G.}~\bibnamefont
			{Martinez-Pinedo}}, \bibinfo {author} {\bibfnamefont {Y.~H.}\ \bibnamefont
			{Lam}},\ and\ \bibinfo {author} {\bibfnamefont {M.~G.}\ \bibnamefont
			{Bertolli}},\ }\bibfield  {title} {\bibinfo {title} {Advanced burning stages
			and fate of $8-10 {M}_\odot$ stars},\ }\href
	{https://doi.org/10.1088/0004-637X/772/2/150} {\bibfield  {journal} {\bibinfo
			{journal} {\apj}\ }\textbf {\bibinfo {volume} {772}},\ \bibinfo {pages}
		{150} (\bibinfo {year} {2013})}\BibitemShut {NoStop}%
	\bibitem [{\citenamefont {{Moriya}}\ \emph {et~al.}(2014)\citenamefont
		{{Moriya}}, \citenamefont {{Tominaga}}, \citenamefont {{Langer}},
		\citenamefont {{Nomoto}}, \citenamefont {{Blinnikov}},\ and\ \citenamefont
		{{Sorokina}}}]{M+14}%
	\BibitemOpen
	\bibfield  {author} {\bibinfo {author} {\bibfnamefont {T.~J.}\ \bibnamefont
			{{Moriya}}}, \bibinfo {author} {\bibfnamefont {N.}~\bibnamefont
			{{Tominaga}}}, \bibinfo {author} {\bibfnamefont {N.}~\bibnamefont
			{{Langer}}}, \bibinfo {author} {\bibfnamefont {K.}~\bibnamefont {{Nomoto}}},
		\bibinfo {author} {\bibfnamefont {S.~I.}\ \bibnamefont {{Blinnikov}}},\ and\
		\bibinfo {author} {\bibfnamefont {E.~I.}\ \bibnamefont {{Sorokina}}},\
	}\bibfield  {title} {\bibinfo {title} {Electron-capture supernovae exploding
			within their progenitor wind},\ }\href
	{https://doi.org/10.1051/0004-6361/201424264} {\bibfield  {journal} {\bibinfo
			{journal} {\aap}\ }\textbf {\bibinfo {volume} {569}},\ \bibinfo {eid} {A57}
		(\bibinfo {year} {2014})},\ \Eprint {https://arxiv.org/abs/1407.4563}
	{arXiv:1407.4563 [astro-ph.HE]} \BibitemShut {NoStop}%
	\bibitem [{\citenamefont {Doherty}\ \emph {et~al.}(2017)\citenamefont
		{Doherty}, \citenamefont {Gil-Pons}, \citenamefont {Siess},\ and\
		\citenamefont {Lattanzio}}]{D+17}%
	\BibitemOpen
	\bibfield  {author} {\bibinfo {author} {\bibfnamefont {C.~L.}\ \bibnamefont
			{Doherty}}, \bibinfo {author} {\bibfnamefont {P.}~\bibnamefont {Gil-Pons}},
		\bibinfo {author} {\bibfnamefont {L.}~\bibnamefont {Siess}},\ and\ \bibinfo
		{author} {\bibfnamefont {J.~C.}\ \bibnamefont {Lattanzio}},\ }\bibfield
	{title} {\bibinfo {title} {Super-{AGB} stars and their role as electron
			capture supernova progenitors},\ }\href
	{https://doi.org/10.1017/pasa.2017.52} {\bibfield  {journal} {\bibinfo
			{journal} {\pasa}\ }\textbf {\bibinfo {volume} {34}},\ \bibinfo {pages}
		{e056} (\bibinfo {year} {2017})}\BibitemShut {NoStop}%
	\bibitem [{\citenamefont {{Gessner}}\ and\ \citenamefont
		{{Janka}}(2018)}]{GessnerJanka18_HD_ECSN}%
	\BibitemOpen
	\bibfield  {author} {\bibinfo {author} {\bibfnamefont {A.}~\bibnamefont
			{{Gessner}}}\ and\ \bibinfo {author} {\bibfnamefont {H.-T.}\ \bibnamefont
			{{Janka}}},\ }\bibfield  {title} {\bibinfo {title} {Hydrodynamical
			neutron-star kicks in electron-capture supernovae and implications for the
			{Crab} supernova},\ }\href {https://doi.org/10.3847/1538-4357/aadbae}
	{\bibfield  {journal} {\bibinfo  {journal} {\apj}\ }\textbf {\bibinfo
			{volume} {865}},\ \bibinfo {eid} {61} (\bibinfo {year} {2018})},\ \Eprint
	{https://arxiv.org/abs/1802.05274} {arXiv:1802.05274 [astro-ph.HE]}
	\BibitemShut {NoStop}%
	\bibitem [{\citenamefont {Zha}\ \emph {et~al.}(2022)\citenamefont {Zha},
		\citenamefont {O'Connor}, \citenamefont {Couch}, \citenamefont {Leung},\ and\
		\citenamefont {Nomoto}}]{Z+21}%
	\BibitemOpen
	\bibfield  {author} {\bibinfo {author} {\bibfnamefont {S.}~\bibnamefont
			{Zha}}, \bibinfo {author} {\bibfnamefont {E.~P.}\ \bibnamefont {O'Connor}},
		\bibinfo {author} {\bibfnamefont {S.~M.}\ \bibnamefont {Couch}}, \bibinfo
		{author} {\bibfnamefont {S.-C.}\ \bibnamefont {Leung}},\ and\ \bibinfo
		{author} {\bibfnamefont {K.}~\bibnamefont {Nomoto}},\ }\bibfield  {title}
	{\bibinfo {title} {Hydrodynamic simulations of electron-capture supernovae:
			progenitor and dimension dependence},\ }\href
	{https://doi.org/10.1093/mnras/stac1035} {\bibfield  {journal} {\bibinfo
			{journal} {\mnras}\ }\textbf {\bibinfo {volume} {513}},\ \bibinfo {pages}
		{1317} (\bibinfo {year} {2022})},\ \Eprint
	{https://arxiv.org/abs/https://academic.oup.com/mnras/article-pdf/513/1/1317/43514034/stac1035.pdf}
	{https://academic.oup.com/mnras/article-pdf/513/1/1317/43514034/stac1035.pdf}
	\BibitemShut {NoStop}%
	\bibitem [{\citenamefont {Zhang}\ \emph {et~al.}(2020)\citenamefont {Zhang},
		\citenamefont {Wang}, \citenamefont {Jozsef}, \citenamefont {Zhai},
		\citenamefont {Zhang}, \citenamefont {Filippenko}, \citenamefont {Brink},
		\citenamefont {Zheng}, \citenamefont {Wyrzykowski}, \citenamefont
		{Mikolajczyk}, \citenamefont {Huang}, \citenamefont {Rui}, \citenamefont
		{Mo}, \citenamefont {Sai}, \citenamefont {Zhang}, \citenamefont {Wang},
		\citenamefont {DerKacy}, \citenamefont {Baron}, \citenamefont {Sarneczky},
		\citenamefont {Bodi}, \citenamefont {Csornyei}, \citenamefont {Hanyecz},
		\citenamefont {Ignacz}, \citenamefont {Kalup}, \citenamefont {Kriskovics},
		\citenamefont {Konyves-Toth}, \citenamefont {Ordasi}, \citenamefont {Pal},
		\citenamefont {Sodor}, \citenamefont {Szakats}, \citenamefont {Vida},\ and\
		\citenamefont {Zsidi}}]{Z+20}%
	\BibitemOpen
	\bibfield  {author} {\bibinfo {author} {\bibfnamefont {J.}~\bibnamefont
			{Zhang}}, \bibinfo {author} {\bibfnamefont {X.}~\bibnamefont {Wang}},
		\bibinfo {author} {\bibfnamefont {V.}~\bibnamefont {Jozsef}}, \bibinfo
		{author} {\bibfnamefont {Q.}~\bibnamefont {Zhai}}, \bibinfo {author}
		{\bibfnamefont {T.}~\bibnamefont {Zhang}}, \bibinfo {author} {\bibfnamefont
			{A.~V.}\ \bibnamefont {Filippenko}}, \bibinfo {author} {\bibfnamefont
			{T.~G.}\ \bibnamefont {Brink}}, \bibinfo {author} {\bibfnamefont
			{W.}~\bibnamefont {Zheng}}, \bibinfo {author} {\bibfnamefont
			{L.}~\bibnamefont {Wyrzykowski}}, \bibinfo {author} {\bibfnamefont
			{P.}~\bibnamefont {Mikolajczyk}}, \bibinfo {author} {\bibfnamefont
			{F.}~\bibnamefont {Huang}}, \bibinfo {author} {\bibfnamefont
			{L.}~\bibnamefont {Rui}}, \bibinfo {author} {\bibfnamefont {J.}~\bibnamefont
			{Mo}}, \bibinfo {author} {\bibfnamefont {H.}~\bibnamefont {Sai}}, \bibinfo
		{author} {\bibfnamefont {X.}~\bibnamefont {Zhang}}, \bibinfo {author}
		{\bibfnamefont {H.}~\bibnamefont {Wang}}, \bibinfo {author} {\bibfnamefont
			{J.~M.}\ \bibnamefont {DerKacy}}, \bibinfo {author} {\bibfnamefont
			{E.}~\bibnamefont {Baron}}, \bibinfo {author} {\bibfnamefont
			{K.}~\bibnamefont {Sarneczky}}, \bibinfo {author} {\bibfnamefont
			{A.}~\bibnamefont {Bodi}}, \bibinfo {author} {\bibfnamefont {G.}~\bibnamefont
			{Csornyei}}, \bibinfo {author} {\bibfnamefont {O.}~\bibnamefont {Hanyecz}},
		\bibinfo {author} {\bibfnamefont {B.}~\bibnamefont {Ignacz}}, \bibinfo
		{author} {\bibfnamefont {C.}~\bibnamefont {Kalup}}, \bibinfo {author}
		{\bibfnamefont {L.}~\bibnamefont {Kriskovics}}, \bibinfo {author}
		{\bibfnamefont {R.}~\bibnamefont {Konyves-Toth}}, \bibinfo {author}
		{\bibfnamefont {A.}~\bibnamefont {Ordasi}}, \bibinfo {author} {\bibfnamefont
			{A.}~\bibnamefont {Pal}}, \bibinfo {author} {\bibfnamefont {A.}~\bibnamefont
			{Sodor}}, \bibinfo {author} {\bibfnamefont {R.}~\bibnamefont {Szakats}},
		\bibinfo {author} {\bibfnamefont {K.}~\bibnamefont {Vida}},\ and\ \bibinfo
		{author} {\bibfnamefont {G.}~\bibnamefont {Zsidi}},\ }\bibfield  {title}
	{\bibinfo {title} {{SN} 2018zd: an unusual stellar explosion as part of the
			diverse type {II} supernova landscape},\ }\href
	{https://doi.org/10.1093/mnras/staa2273} {\bibfield  {journal} {\bibinfo
			{journal} {\mnras}\ }\textbf {\bibinfo {volume} {498}},\ \bibinfo {pages}
		{84} (\bibinfo {year} {2020})},\ \Eprint
	{https://arxiv.org/abs/https://academic.oup.com/mnras/article-pdf/498/1/84/35319733/staa2273.pdf}
	{https://academic.oup.com/mnras/article-pdf/498/1/84/35319733/staa2273.pdf}
	\BibitemShut {NoStop}%
	\bibitem [{\citenamefont {{Hiramatsu}}\ \emph {et~al.}(2021)\citenamefont
		{{Hiramatsu}}, \citenamefont {{Howell}}, \citenamefont {{Van Dyk}},
		\citenamefont {{Goldberg}}, \citenamefont {{Maeda}}, \citenamefont
		{{Moriya}}, \citenamefont {{Tominaga}}, \citenamefont {{Nomoto}},
		\citenamefont {{Hosseinzadeh}}, \citenamefont {{Arcavi}}, \citenamefont
		{{McCully}}, \citenamefont {{Burke}}, \citenamefont {{Bostroem}},
		\citenamefont {{Valenti}}, \citenamefont {{Dong}}, \citenamefont {{Brown}},
		\citenamefont {{Andrews}}, \citenamefont {{Bilinski}}, \citenamefont
		{{Williams}}, \citenamefont {{Smith}}, \citenamefont {{Smith}}, \citenamefont
		{{Sand}}, \citenamefont {{Anand}}, \citenamefont {{Xu}}, \citenamefont
		{{Filippenko}}, \citenamefont {{Bersten}}, \citenamefont {{Folatelli}},
		\citenamefont {{Kelly}}, \citenamefont {{Noguchi}},\ and\ \citenamefont
		{{Itagaki}}}]{H+21}%
	\BibitemOpen
	\bibfield  {author} {\bibinfo {author} {\bibfnamefont {D.}~\bibnamefont
			{{Hiramatsu}}}, \bibinfo {author} {\bibfnamefont {D.~A.}\ \bibnamefont
			{{Howell}}}, \bibinfo {author} {\bibfnamefont {S.~D.}\ \bibnamefont {{Van
					Dyk}}}, \bibinfo {author} {\bibfnamefont {J.~A.}\ \bibnamefont {{Goldberg}}},
		\bibinfo {author} {\bibfnamefont {K.}~\bibnamefont {{Maeda}}}, \bibinfo
		{author} {\bibfnamefont {T.~J.}\ \bibnamefont {{Moriya}}}, \bibinfo {author}
		{\bibfnamefont {N.}~\bibnamefont {{Tominaga}}}, \bibinfo {author}
		{\bibfnamefont {K.}~\bibnamefont {{Nomoto}}}, \bibinfo {author}
		{\bibfnamefont {G.}~\bibnamefont {{Hosseinzadeh}}}, \bibinfo {author}
		{\bibfnamefont {I.}~\bibnamefont {{Arcavi}}}, \bibinfo {author}
		{\bibfnamefont {C.}~\bibnamefont {{McCully}}}, \bibinfo {author}
		{\bibfnamefont {J.}~\bibnamefont {{Burke}}}, \bibinfo {author} {\bibfnamefont
			{K.~A.}\ \bibnamefont {{Bostroem}}}, \bibinfo {author} {\bibfnamefont
			{S.}~\bibnamefont {{Valenti}}}, \bibinfo {author} {\bibfnamefont
			{Y.}~\bibnamefont {{Dong}}}, \bibinfo {author} {\bibfnamefont {P.~J.}\
			\bibnamefont {{Brown}}}, \bibinfo {author} {\bibfnamefont {J.~E.}\
			\bibnamefont {{Andrews}}}, \bibinfo {author} {\bibfnamefont {C.}~\bibnamefont
			{{Bilinski}}}, \bibinfo {author} {\bibfnamefont {G.~G.}\ \bibnamefont
			{{Williams}}}, \bibinfo {author} {\bibfnamefont {P.~S.}\ \bibnamefont
			{{Smith}}}, \bibinfo {author} {\bibfnamefont {N.}~\bibnamefont {{Smith}}},
		\bibinfo {author} {\bibfnamefont {D.~J.}\ \bibnamefont {{Sand}}}, \bibinfo
		{author} {\bibfnamefont {G.~S.}\ \bibnamefont {{Anand}}}, \bibinfo {author}
		{\bibfnamefont {C.}~\bibnamefont {{Xu}}}, \bibinfo {author} {\bibfnamefont
			{A.~V.}\ \bibnamefont {{Filippenko}}}, \bibinfo {author} {\bibfnamefont
			{M.~C.}\ \bibnamefont {{Bersten}}}, \bibinfo {author} {\bibfnamefont
			{G.}~\bibnamefont {{Folatelli}}}, \bibinfo {author} {\bibfnamefont {P.~L.}\
			\bibnamefont {{Kelly}}}, \bibinfo {author} {\bibfnamefont {T.}~\bibnamefont
			{{Noguchi}}},\ and\ \bibinfo {author} {\bibfnamefont {K.}~\bibnamefont
			{{Itagaki}}},\ }\bibfield  {title} {\bibinfo {title} {The electron-capture
			origin of supernova 2018zd},\ }\href
	{https://doi.org/10.1038/s41550-021-01384-2} {\bibfield  {journal} {\bibinfo
			{journal} {\natastron}\ }\textbf {\bibinfo {volume} {5}},\ \bibinfo {pages}
		{903} (\bibinfo {year} {2021})},\ \Eprint {https://arxiv.org/abs/2011.02176}
	{arXiv:2011.02176 [astro-ph.HE]} \BibitemShut {NoStop}%
	\bibitem [{\citenamefont {Sato}\ \emph {et~al.}(2024)\citenamefont {Sato},
		\citenamefont {Tominaga}, \citenamefont {Blinnikov}, \citenamefont
		{Potashov}, \citenamefont {Moriya},\ and\ \citenamefont {Hiramatsu}}]{S+24}%
	\BibitemOpen
	\bibfield  {author} {\bibinfo {author} {\bibfnamefont {M.}~\bibnamefont
			{Sato}}, \bibinfo {author} {\bibfnamefont {N.}~\bibnamefont {Tominaga}},
		\bibinfo {author} {\bibfnamefont {S.~I.}\ \bibnamefont {Blinnikov}}, \bibinfo
		{author} {\bibfnamefont {M.~S.}\ \bibnamefont {Potashov}}, \bibinfo {author}
		{\bibfnamefont {T.~J.}\ \bibnamefont {Moriya}},\ and\ \bibinfo {author}
		{\bibfnamefont {D.}~\bibnamefont {Hiramatsu}},\ }\bibfield  {title} {\bibinfo
		{title} {A robust light-curve diagnostic for electron-capture supernovae and
			low-mass fe-core-collapse supernovae},\ }\href
	{https://doi.org/10.3847/1538-4357/ad50cb} {\bibfield  {journal} {\bibinfo
			{journal} {\apj}\ }\textbf {\bibinfo {volume} {970}},\ \bibinfo {pages} {163}
		(\bibinfo {year} {2024})}\BibitemShut {NoStop}%
	\bibitem [{\citenamefont {{Canal}}\ and\ \citenamefont
		{{Schatzman}}(1976)}]{CS76}%
	\BibitemOpen
	\bibfield  {author} {\bibinfo {author} {\bibfnamefont {R.}~\bibnamefont
			{{Canal}}}\ and\ \bibinfo {author} {\bibfnamefont {E.}~\bibnamefont
			{{Schatzman}}},\ }\bibfield  {title} {\bibinfo {title} {{Non explosive
				collapse of white dwarfs.}},\ }\href@noop {} {\bibfield  {journal} {\bibinfo
			{journal} {\aap}\ }\textbf {\bibinfo {volume} {46}},\ \bibinfo {pages} {229}
		(\bibinfo {year} {1976})}\BibitemShut {NoStop}%
	\bibitem [{\citenamefont {{Saio}}\ and\ \citenamefont {{Nomoto}}(1985)}]{SN85}%
	\BibitemOpen
	\bibfield  {author} {\bibinfo {author} {\bibfnamefont {H.}~\bibnamefont
			{{Saio}}}\ and\ \bibinfo {author} {\bibfnamefont {K.}~\bibnamefont
			{{Nomoto}}},\ }\bibfield  {title} {\bibinfo {title} {{Evolution of a merging
				pair of C + O white dwarfs to form a single neutron star}},\ }\href@noop {}
	{\bibfield  {journal} {\bibinfo  {journal} {\aap}\ }\textbf {\bibinfo
			{volume} {150}},\ \bibinfo {pages} {L21} (\bibinfo {year}
		{1985})}\BibitemShut {NoStop}%
	\bibitem [{\citenamefont {{Nomoto}}\ and\ \citenamefont
		{{Kondo}}(1991)}]{NK91}%
	\BibitemOpen
	\bibfield  {author} {\bibinfo {author} {\bibfnamefont {K.}~\bibnamefont
			{{Nomoto}}}\ and\ \bibinfo {author} {\bibfnamefont {Y.}~\bibnamefont
			{{Kondo}}},\ }\bibfield  {title} {\bibinfo {title} {{Conditions for
				Accretion-induced Collapse of White Dwarfs}},\ }\href
	{https://doi.org/10.1086/185922} {\bibfield  {journal} {\bibinfo  {journal}
			{\apjl}\ }\textbf {\bibinfo {volume} {367}},\ \bibinfo {pages} {L19}
		(\bibinfo {year} {1991})}\BibitemShut {NoStop}%
	\bibitem [{\citenamefont {Hobbs}\ \emph {et~al.}(2005)\citenamefont {Hobbs},
		\citenamefont {Lorimer}, \citenamefont {Lyne},\ and\ \citenamefont
		{Kramer}}]{H+05}%
	\BibitemOpen
	\bibfield  {author} {\bibinfo {author} {\bibfnamefont {G.}~\bibnamefont
			{Hobbs}}, \bibinfo {author} {\bibfnamefont {D.~R.}\ \bibnamefont {Lorimer}},
		\bibinfo {author} {\bibfnamefont {A.~G.}\ \bibnamefont {Lyne}},\ and\
		\bibinfo {author} {\bibfnamefont {M.}~\bibnamefont {Kramer}},\ }\bibfield
	{title} {\bibinfo {title} {A statistical study of 233 pulsar proper
			motions},\ }\href {https://doi.org/10.1111/j.1365-2966.2005.09087.x}
	{\bibfield  {journal} {\bibinfo  {journal} {\mnras}\ }\textbf {\bibinfo
			{volume} {360}},\ \bibinfo {pages} {974} (\bibinfo {year} {2005})},\ \Eprint
	{https://arxiv.org/abs/https://academic.oup.com/mnras/article-pdf/360/3/974/3209564/360-3-974.pdf}
	{https://academic.oup.com/mnras/article-pdf/360/3/974/3209564/360-3-974.pdf}
	\BibitemShut {NoStop}%
	\bibitem [{\citenamefont {{{Hill}ebrandt}}(1982)}]{H82}%
	\BibitemOpen
	\bibfield  {author} {\bibinfo {author} {\bibfnamefont {W.}~\bibnamefont
			{{{Hill}ebrandt}}},\ }\bibfield  {title} {\bibinfo {title} {An exploding $10
			{M}_\odot$ star: a model for the {Crab} supernova},\ }\href@noop {}
	{\bibfield  {journal} {\bibinfo  {journal} {\aap}\ }\textbf {\bibinfo
			{volume} {110}},\ \bibinfo {pages} {L3} (\bibinfo {year} {1982})}\BibitemShut
	{NoStop}%
	\bibitem [{\citenamefont {{Nomoto}}\ \emph {et~al.}(1982)\citenamefont
		{{Nomoto}}, \citenamefont {{Sparks}}, \citenamefont {{Fesen}}, \citenamefont
		{{Gull}}, \citenamefont {{Miyaji}},\ and\ \citenamefont {{Sugimoto}}}]{N+82}%
	\BibitemOpen
	\bibfield  {author} {\bibinfo {author} {\bibfnamefont {K.}~\bibnamefont
			{{Nomoto}}}, \bibinfo {author} {\bibfnamefont {W.~M.}\ \bibnamefont
			{{Sparks}}}, \bibinfo {author} {\bibfnamefont {R.~A.}\ \bibnamefont
			{{Fesen}}}, \bibinfo {author} {\bibfnamefont {T.~R.}\ \bibnamefont {{Gull}}},
		\bibinfo {author} {\bibfnamefont {S.}~\bibnamefont {{Miyaji}}},\ and\
		\bibinfo {author} {\bibfnamefont {D.}~\bibnamefont {{Sugimoto}}},\ }\bibfield
	{title} {\bibinfo {title} {The {Crab} nebula's progenitor},\ }\href
	{https://doi.org/10.1038/299803a0} {\bibfield  {journal} {\bibinfo  {journal}
			{\nat}\ }\textbf {\bibinfo {volume} {299}},\ \bibinfo {pages} {803} (\bibinfo
		{year} {1982})}\BibitemShut {NoStop}%
	\bibitem [{\citenamefont {Smith}(2013)}]{S13}%
	\BibitemOpen
	\bibfield  {author} {\bibinfo {author} {\bibfnamefont {N.}~\bibnamefont
			{Smith}},\ }\bibfield  {title} {\bibinfo {title} {The {Crab} nebula and the
			class of type {IIn-P} supernovae caused by sub-energetic electron-capture
			explosions},\ }\href {https://doi.org/10.1093/mnras/stt1004} {\bibfield
		{journal} {\bibinfo  {journal} {\mnras}\ }\textbf {\bibinfo {volume} {434}},\
		\bibinfo {pages} {102} (\bibinfo {year} {2013})},\ \Eprint
	{https://arxiv.org/abs/https://academic.oup.com/mnras/article-pdf/434/1/102/18498497/stt1004.pdf}
	{https://academic.oup.com/mnras/article-pdf/434/1/102/18498497/stt1004.pdf}
	\BibitemShut {NoStop}%
	\bibitem [{\citenamefont {{van den Heuvel}}(2004)}]{vdH04}%
	\BibitemOpen
	\bibfield  {author} {\bibinfo {author} {\bibfnamefont {E.~P.~J.}\
			\bibnamefont {{van den Heuvel}}},\ }\bibfield  {title} {\bibinfo {title}
		{{X}-ray binaries and their descendants: Binary radio pulsars; evidence for
			three classes of neutron stars?},\ }in\ \href
	{https://doi.org/10.48550/arXiv.astro-ph/0407451} {\emph {\bibinfo
			{booktitle} {5th INTEGRAL Workshop on the INTEGRAL Universe}}},\ \bibinfo
	{series} {ESA Special Publication}, Vol.\ \bibinfo {volume} {552},\ \bibinfo
	{editor} {edited by\ \bibinfo {editor} {\bibfnamefont {V.}~\bibnamefont
			{{Schoenfelder}}}, \bibinfo {editor} {\bibfnamefont {G.}~\bibnamefont
			{{Lichti}}},\ and\ \bibinfo {editor} {\bibfnamefont {C.}~\bibnamefont
			{{Winkler}}}}\ (\bibinfo {year} {2004})\ p.\ \bibinfo {pages} {185},\ \Eprint
	{https://arxiv.org/abs/astro-ph/0407451} {arXiv:astro-ph/0407451 [astro-ph]}
	\BibitemShut {NoStop}%
	\bibitem [{\citenamefont {{van den Heuvel}}(2007)}]{vdH07}%
	\BibitemOpen
	\bibfield  {author} {\bibinfo {author} {\bibfnamefont {E.~P.~J.}\
			\bibnamefont {{van den Heuvel}}},\ }\bibfield  {title} {\bibinfo {title}
		{Double neutron stars: Evidence for two different neutron-star formation
			mechanisms},\ }in\ \href {https://doi.org/10.1063/1.2774916} {\emph {\bibinfo
			{booktitle} {The Multicolored Landscape of Compact Objects and Their
				Explosive Origins}}},\ \bibinfo {series} {American Institute of Physics
		Conference Series}, Vol.\ \bibinfo {volume} {924},\ \bibinfo {editor} {edited
		by\ \bibinfo {editor} {\bibfnamefont {T.}~\bibnamefont {{di Salvo}}},
		\bibinfo {editor} {\bibfnamefont {G.~L.}\ \bibnamefont {{Israel}}}, \bibinfo
		{editor} {\bibfnamefont {L.}~\bibnamefont {{Piersant}}}, \bibinfo {editor}
		{\bibfnamefont {L.}~\bibnamefont {{Burderi}}}, \bibinfo {editor}
		{\bibfnamefont {G.}~\bibnamefont {{Matt}}}, \bibinfo {editor} {\bibfnamefont
			{A.}~\bibnamefont {{Tornambe}}},\ and\ \bibinfo {editor} {\bibfnamefont
			{M.~T.}\ \bibnamefont {{Menna}}}}\ (\bibinfo {year} {2007})\ pp.\ \bibinfo
	{pages} {598--606},\ \Eprint {https://arxiv.org/abs/0704.1215}
	{arXiv:0704.1215 [astro-ph]} \BibitemShut {NoStop}%
	\bibitem [{\citenamefont {Schwab}\ \emph {et~al.}(2010)\citenamefont {Schwab},
		\citenamefont {Podsiadlowski},\ and\ \citenamefont {Rappaport}}]{S+10}%
	\BibitemOpen
	\bibfield  {author} {\bibinfo {author} {\bibfnamefont {J.}~\bibnamefont
			{Schwab}}, \bibinfo {author} {\bibfnamefont {P.}~\bibnamefont
			{Podsiadlowski}},\ and\ \bibinfo {author} {\bibfnamefont {S.}~\bibnamefont
			{Rappaport}},\ }\bibfield  {title} {\bibinfo {title} {Further evidence for
			the bimodal distribution of neutron-star masses},\ }\href
	{https://doi.org/10.1088/0004-637X/719/1/722} {\bibfield  {journal} {\bibinfo
			{journal} {\apj}\ }\textbf {\bibinfo {volume} {719}},\ \bibinfo {pages}
		{722} (\bibinfo {year} {2010})}\BibitemShut {NoStop}%
	\bibitem [{\citenamefont {{Ferdman}}\ \emph {et~al.}(2013)\citenamefont
		{{Ferdman}}, \citenamefont {{Stairs}}, \citenamefont {{Kramer}},
		\citenamefont {{Breton}}, \citenamefont {{McLaughlin}}, \citenamefont
		{{Freire}}, \citenamefont {{Possenti}}, \citenamefont {{Stappers}},
		\citenamefont {{Kaspi}}, \citenamefont {{Manchester}},\ and\ \citenamefont
		{{Lyne}}}]{F+13}%
	\BibitemOpen
	\bibfield  {author} {\bibinfo {author} {\bibfnamefont {R.~D.}\ \bibnamefont
			{{Ferdman}}}, \bibinfo {author} {\bibfnamefont {I.~H.}\ \bibnamefont
			{{Stairs}}}, \bibinfo {author} {\bibfnamefont {M.}~\bibnamefont {{Kramer}}},
		\bibinfo {author} {\bibfnamefont {R.~P.}\ \bibnamefont {{Breton}}}, \bibinfo
		{author} {\bibfnamefont {M.~A.}\ \bibnamefont {{McLaughlin}}}, \bibinfo
		{author} {\bibfnamefont {P.~C.~C.}\ \bibnamefont {{Freire}}}, \bibinfo
		{author} {\bibfnamefont {A.}~\bibnamefont {{Possenti}}}, \bibinfo {author}
		{\bibfnamefont {B.~W.}\ \bibnamefont {{Stappers}}}, \bibinfo {author}
		{\bibfnamefont {V.~M.}\ \bibnamefont {{Kaspi}}}, \bibinfo {author}
		{\bibfnamefont {R.~N.}\ \bibnamefont {{Manchester}}},\ and\ \bibinfo {author}
		{\bibfnamefont {A.~G.}\ \bibnamefont {{Lyne}}},\ }\bibfield  {title}
	{\bibinfo {title} {The double pulsar: Evidence for neutron star formation
			without an iron core-collapse supernova},\ }\href
	{https://doi.org/10.1088/0004-637X/767/1/85} {\bibfield  {journal} {\bibinfo
			{journal} {\apj}\ }\textbf {\bibinfo {volume} {767}},\ \bibinfo {eid} {85}
		(\bibinfo {year} {2013})},\ \Eprint {https://arxiv.org/abs/1302.2914}
	{arXiv:1302.2914 [astro-ph.SR]} \BibitemShut {NoStop}%
	\bibitem [{\citenamefont {{Pfahl}}\ \emph
		{et~al.}(2002{\natexlab{b}})\citenamefont {{Pfahl}}, \citenamefont
		{{Rappaport}}, \citenamefont {{Podsiadlowski}},\ and\ \citenamefont
		{{Spruit}}}]{Pf+02}%
	\BibitemOpen
	\bibfield  {author} {\bibinfo {author} {\bibfnamefont {E.}~\bibnamefont
			{{Pfahl}}}, \bibinfo {author} {\bibfnamefont {S.}~\bibnamefont
			{{Rappaport}}}, \bibinfo {author} {\bibfnamefont {P.}~\bibnamefont
			{{Podsiadlowski}}},\ and\ \bibinfo {author} {\bibfnamefont {H.}~\bibnamefont
			{{Spruit}}},\ }\bibfield  {title} {\bibinfo {title} {A new class of high-mass
			{X}-ray binaries: Implications for core collapse and neutron star recoil},\
	}\href {https://doi.org/10.1086/340794} {\bibfield  {journal} {\bibinfo
			{journal} {\apj}\ }\textbf {\bibinfo {volume} {574}},\ \bibinfo {pages} {364}
		(\bibinfo {year} {2002}{\natexlab{b}})},\ \Eprint
	{https://arxiv.org/abs/astro-ph/0109521} {arXiv:astro-ph/0109521 [astro-ph]}
	\BibitemShut {NoStop}%
	\bibitem [{\citenamefont {{Knigge}}\ \emph {et~al.}(2011)\citenamefont
		{{Knigge}}, \citenamefont {{Coe}},\ and\ \citenamefont
		{{Podsiadlowski}}}]{K+11}%
	\BibitemOpen
	\bibfield  {author} {\bibinfo {author} {\bibfnamefont {C.}~\bibnamefont
			{{Knigge}}}, \bibinfo {author} {\bibfnamefont {M.~J.}\ \bibnamefont
			{{Coe}}},\ and\ \bibinfo {author} {\bibfnamefont {P.}~\bibnamefont
			{{Podsiadlowski}}},\ }\bibfield  {title} {\bibinfo {title} {Two populations
			of {X}-ray pulsars produced by two types of supernova},\ }\href
	{https://doi.org/10.1038/nature10529} {\bibfield  {journal} {\bibinfo
			{journal} {\nat}\ }\textbf {\bibinfo {volume} {479}},\ \bibinfo {pages} {372}
		(\bibinfo {year} {2011})},\ \Eprint {https://arxiv.org/abs/1111.2051}
	{arXiv:1111.2051 [astro-ph.SR]} \BibitemShut {NoStop}%
	\bibitem [{\citenamefont {Wu}\ \emph {et~al.}(2024)\citenamefont {Wu},
		\citenamefont {Pan}, \citenamefont {Qian}, \citenamefont {Ransom},
		\citenamefont {Eatough}, \citenamefont {Wang}, \citenamefont {Freire},
		\citenamefont {Liu}, \citenamefont {Yan}, \citenamefont {Luo}, \citenamefont
		{Zhang}, \citenamefont {Li}, \citenamefont {Yin}, \citenamefont {Li},
		\citenamefont {Li}, \citenamefont {Dai}, \citenamefont {Li}, \citenamefont
		{Zhang}, \citenamefont {Liu},\ and\ \citenamefont {Pan}}]{Wu+24_3PSR_M15}%
	\BibitemOpen
	\bibfield  {author} {\bibinfo {author} {\bibfnamefont {Y.}~\bibnamefont
			{Wu}}, \bibinfo {author} {\bibfnamefont {Z.}~\bibnamefont {Pan}}, \bibinfo
		{author} {\bibfnamefont {L.}~\bibnamefont {Qian}}, \bibinfo {author}
		{\bibfnamefont {S.~M.}\ \bibnamefont {Ransom}}, \bibinfo {author}
		{\bibfnamefont {R.~P.}\ \bibnamefont {Eatough}}, \bibinfo {author}
		{\bibfnamefont {B.}~\bibnamefont {Wang}}, \bibinfo {author} {\bibfnamefont
			{P.~C.~C.}\ \bibnamefont {Freire}}, \bibinfo {author} {\bibfnamefont
			{K.}~\bibnamefont {Liu}}, \bibinfo {author} {\bibfnamefont {Z.}~\bibnamefont
			{Yan}}, \bibinfo {author} {\bibfnamefont {J.}~\bibnamefont {Luo}}, \bibinfo
		{author} {\bibfnamefont {L.}~\bibnamefont {Zhang}}, \bibinfo {author}
		{\bibfnamefont {M.}~\bibnamefont {Li}}, \bibinfo {author} {\bibfnamefont
			{D.}~\bibnamefont {Yin}}, \bibinfo {author} {\bibfnamefont {B.}~\bibnamefont
			{Li}}, \bibinfo {author} {\bibfnamefont {Y.}~\bibnamefont {Li}}, \bibinfo
		{author} {\bibfnamefont {Y.}~\bibnamefont {Dai}}, \bibinfo {author}
		{\bibfnamefont {Y.}~\bibnamefont {Li}}, \bibinfo {author} {\bibfnamefont
			{X.}~\bibnamefont {Zhang}}, \bibinfo {author} {\bibfnamefont
			{T.}~\bibnamefont {Liu}},\ and\ \bibinfo {author} {\bibfnamefont
			{Y.}~\bibnamefont {Pan}},\ }\bibfield  {title} {\bibinfo {title} {The
			discovery of three pulsars in the globular cluster {M15} with {FAST}},\
	}\href {https://doi.org/10.3847/2041-8213/ad7b9e} {\bibfield  {journal}
		{\bibinfo  {journal} {\apjl}\ }\textbf {\bibinfo {volume} {974}},\ \bibinfo
		{pages} {L23} (\bibinfo {year} {2024})}\BibitemShut {NoStop}%
	\bibitem [{\citenamefont {Kremer}\ \emph {et~al.}(2024)\citenamefont {Kremer},
		\citenamefont {Ye}, \citenamefont {Heinke}, \citenamefont {Piro},
		\citenamefont {Ransom},\ and\ \citenamefont
		{Rasio}}]{Kremer_ea24_NoPartialRecycling}%
	\BibitemOpen
	\bibfield  {author} {\bibinfo {author} {\bibfnamefont {K.}~\bibnamefont
			{Kremer}}, \bibinfo {author} {\bibfnamefont {C.~S.}\ \bibnamefont {Ye}},
		\bibinfo {author} {\bibfnamefont {C.~O.}\ \bibnamefont {Heinke}}, \bibinfo
		{author} {\bibfnamefont {A.~L.}\ \bibnamefont {Piro}}, \bibinfo {author}
		{\bibfnamefont {S.~M.}\ \bibnamefont {Ransom}},\ and\ \bibinfo {author}
		{\bibfnamefont {F.~A.}\ \bibnamefont {Rasio}},\ }\bibfield  {title} {\bibinfo
		{title} {Can slow pulsars in {Milky Way} globular clusters form via partial
			recycling?},\ }\href {https://doi.org/10.3847/2041-8213/ad9a4e} {\bibfield
		{journal} {\bibinfo  {journal} {\apjl}\ }\textbf {\bibinfo {volume} {977}},\
		\bibinfo {pages} {L42} (\bibinfo {year} {2024})}\BibitemShut {NoStop}%
	\bibitem [{\citenamefont {{Kremer}}\ \emph {et~al.}(2023)\citenamefont
		{{Kremer}}, \citenamefont {{Fuller}}, \citenamefont {{Piro}},\ and\
		\citenamefont {{Ransom}}}]{Kremer_ea23_MIC_for_young_Pulsars_in_GCs}%
	\BibitemOpen
	\bibfield  {author} {\bibinfo {author} {\bibfnamefont {K.}~\bibnamefont
			{{Kremer}}}, \bibinfo {author} {\bibfnamefont {J.}~\bibnamefont {{Fuller}}},
		\bibinfo {author} {\bibfnamefont {A.~L.}\ \bibnamefont {{Piro}}},\ and\
		\bibinfo {author} {\bibfnamefont {S.~M.}\ \bibnamefont {{Ransom}}},\
	}\bibfield  {title} {\bibinfo {title} {Connecting the young pulsars in {Milky
				Way} globular clusters with white dwarf mergers and the {M81} fast radio
			burst},\ }\href {https://doi.org/10.1093/mnrasl/slad088} {\bibfield
		{journal} {\bibinfo  {journal} {\mnras}\ }\textbf {\bibinfo {volume} {525}},\
		\bibinfo {pages} {L22} (\bibinfo {year} {2023})},\ \Eprint
	{https://arxiv.org/abs/2305.11933} {arXiv:2305.11933 [astro-ph.HE]}
	\BibitemShut {NoStop}%
	\bibitem [{\citenamefont {{Dan}}\ \emph {et~al.}(2014)\citenamefont {{Dan}},
		\citenamefont {{Rosswog}}, \citenamefont {{Br{\"u}ggen}},\ and\ \citenamefont
		{{Podsiadlowski}}}]{D+14}%
	\BibitemOpen
	\bibfield  {author} {\bibinfo {author} {\bibfnamefont {M.}~\bibnamefont
			{{Dan}}}, \bibinfo {author} {\bibfnamefont {S.}~\bibnamefont {{Rosswog}}},
		\bibinfo {author} {\bibfnamefont {M.}~\bibnamefont {{Br{\"u}ggen}}},\ and\
		\bibinfo {author} {\bibfnamefont {P.}~\bibnamefont {{Podsiadlowski}}},\
	}\bibfield  {title} {\bibinfo {title} {{The structure and fate of white dwarf
				merger remnants}},\ }\href {https://doi.org/10.1093/mnras/stt1766} {\bibfield
		{journal} {\bibinfo  {journal} {\mnras}\ }\textbf {\bibinfo {volume}
			{438}},\ \bibinfo {pages} {14} (\bibinfo {year} {2014})},\ \Eprint
	{https://arxiv.org/abs/1308.1667} {arXiv:1308.1667 [astro-ph.HE]}
	\BibitemShut {NoStop}%
	\bibitem [{\citenamefont {{Kashyap}}\ \emph {et~al.}(2018)\citenamefont
		{{Kashyap}}, \citenamefont {{Haque}}, \citenamefont {{Lor{\'e}n-Aguilar}},
		\citenamefont {{Garc{\'\i}a-Berro}},\ and\ \citenamefont {{Fisher}}}]{K+18}%
	\BibitemOpen
	\bibfield  {author} {\bibinfo {author} {\bibfnamefont {R.}~\bibnamefont
			{{Kashyap}}}, \bibinfo {author} {\bibfnamefont {T.}~\bibnamefont {{Haque}}},
		\bibinfo {author} {\bibfnamefont {P.}~\bibnamefont {{Lor{\'e}n-Aguilar}}},
		\bibinfo {author} {\bibfnamefont {E.}~\bibnamefont {{Garc{\'\i}a-Berro}}},\
		and\ \bibinfo {author} {\bibfnamefont {R.}~\bibnamefont {{Fisher}}},\
	}\bibfield  {title} {\bibinfo {title} {{Double-degenerate Carbon-Oxygen and
				Oxygen-Neon White Dwarf Mergers: A New Mechanism for Faint and Rapid Type Ia
				Supernovae}},\ }\href {https://doi.org/10.3847/1538-4357/aaedb7} {\bibfield
		{journal} {\bibinfo  {journal} {\apj}\ }\textbf {\bibinfo {volume} {869}},\
		\bibinfo {eid} {140} (\bibinfo {year} {2018})},\ \Eprint
	{https://arxiv.org/abs/1811.00013} {arXiv:1811.00013 [astro-ph.SR]}
	\BibitemShut {NoStop}%
	\bibitem [{\citenamefont {{Dessart}}\ \emph {et~al.}(2006)\citenamefont
		{{Dessart}}, \citenamefont {{Burrows}}, \citenamefont {{Ott}}, \citenamefont
		{{Livne}}, \citenamefont {{Yoon}},\ and\ \citenamefont {{Langer}}}]{D+06}%
	\BibitemOpen
	\bibfield  {author} {\bibinfo {author} {\bibfnamefont {L.}~\bibnamefont
			{{Dessart}}}, \bibinfo {author} {\bibfnamefont {A.}~\bibnamefont
			{{Burrows}}}, \bibinfo {author} {\bibfnamefont {C.~D.}\ \bibnamefont
			{{Ott}}}, \bibinfo {author} {\bibfnamefont {E.}~\bibnamefont {{Livne}}},
		\bibinfo {author} {\bibfnamefont {S.~C.}\ \bibnamefont {{Yoon}}},\ and\
		\bibinfo {author} {\bibfnamefont {N.}~\bibnamefont {{Langer}}},\ }\bibfield
	{title} {\bibinfo {title} {Multidimensional simulations of the
			accretion-induced collapse of white dwarfs to neutron stars},\ }\href
	{https://doi.org/10.1086/503626} {\bibfield  {journal} {\bibinfo  {journal}
			{\apj}\ }\textbf {\bibinfo {volume} {644}},\ \bibinfo {pages} {1063}
		(\bibinfo {year} {2006})},\ \Eprint {https://arxiv.org/abs/astro-ph/0601603}
	{arXiv:astro-ph/0601603 [astro-ph]} \BibitemShut {NoStop}%
	\bibitem [{\citenamefont {{Kuroda}}\ \emph {et~al.}(2025)\citenamefont
		{{Kuroda}}, \citenamefont {{Kawaguchi}},\ and\ \citenamefont
		{{Shibata}}}]{Kuroda_ea25_rotWD_collapse}%
	\BibitemOpen
	\bibfield  {author} {\bibinfo {author} {\bibfnamefont {T.}~\bibnamefont
			{{Kuroda}}}, \bibinfo {author} {\bibfnamefont {K.}~\bibnamefont
			{{Kawaguchi}}},\ and\ \bibinfo {author} {\bibfnamefont {M.}~\bibnamefont
			{{Shibata}}},\ }\bibfield  {title} {\bibinfo {title} {{Collapse of rotating
				white dwarfs and multimessenger signals}},\ }\href
	{https://doi.org/10.1093/mnras/staf1065} {\bibfield  {journal} {\bibinfo
			{journal} {\mnras}\ }\textbf {\bibinfo {volume} {541}},\ \bibinfo {pages}
		{1649} (\bibinfo {year} {2025})},\ \Eprint {https://arxiv.org/abs/2503.17082}
	{arXiv:2503.17082 [astro-ph.HE]} \BibitemShut {NoStop}%
	\bibitem [{\citenamefont {{Ogata}}\ and\ \citenamefont
		{{Ichimaru}}(1990)}]{oi90}%
	\BibitemOpen
	\bibfield  {author} {\bibinfo {author} {\bibfnamefont {S.}~\bibnamefont
			{{Ogata}}}\ and\ \bibinfo {author} {\bibfnamefont {S.}~\bibnamefont
			{{Ichimaru}}},\ }\bibfield  {title} {\bibinfo {title} {First-principles
			calculations of shear moduli for {Monte Carlo}-simulated {Coulomb} solids},\
	}\href {https://doi.org/10.1103/PhysRevA.42.4867} {\bibfield  {journal}
		{\bibinfo  {journal} {\pra}\ }\textbf {\bibinfo {volume} {42}},\ \bibinfo
		{pages} {4867} (\bibinfo {year} {1990})}\BibitemShut {NoStop}%
	\bibitem [{\citenamefont {{Strohmayer}}\ \emph {et~al.}(1991)\citenamefont
		{{Strohmayer}}, \citenamefont {{Ogata}}, \citenamefont {{Iyetomi}},
		\citenamefont {{Ichimaru}},\ and\ \citenamefont {{van
				Horn}}}]{Strohmayer_etal91}%
	\BibitemOpen
	\bibfield  {author} {\bibinfo {author} {\bibfnamefont {T.}~\bibnamefont
			{{Strohmayer}}}, \bibinfo {author} {\bibfnamefont {S.}~\bibnamefont
			{{Ogata}}}, \bibinfo {author} {\bibfnamefont {H.}~\bibnamefont {{Iyetomi}}},
		\bibinfo {author} {\bibfnamefont {S.}~\bibnamefont {{Ichimaru}}},\ and\
		\bibinfo {author} {\bibfnamefont {H.~M.}\ \bibnamefont {{van Horn}}},\
	}\bibfield  {title} {\bibinfo {title} {The shear modulus of the neutron star
			crust and nonradial oscillations of neutron stars},\ }\href
	{https://doi.org/10.1086/170231} {\bibfield  {journal} {\bibinfo  {journal}
			{\apj}\ }\textbf {\bibinfo {volume} {375}},\ \bibinfo {pages} {679} (\bibinfo
		{year} {1991})}\BibitemShut {NoStop}%
	\bibitem [{\citenamefont {{Baiko}}(2011)}]{Baiko11}%
	\BibitemOpen
	\bibfield  {author} {\bibinfo {author} {\bibfnamefont {D.~A.}\ \bibnamefont
			{{Baiko}}},\ }\bibfield  {title} {\bibinfo {title} {Shear modulus of neutron
			star crust},\ }\href {https://doi.org/10.1111/j.1365-2966.2011.18819.x}
	{\bibfield  {journal} {\bibinfo  {journal} {\mnras}\ }\textbf {\bibinfo
			{volume} {416}},\ \bibinfo {pages} {22} (\bibinfo {year} {2011})},\ \Eprint
	{https://arxiv.org/abs/1104.0173} {arXiv:1104.0173 [astro-ph.SR]}
	\BibitemShut {NoStop}%
	\bibitem [{\citenamefont {{Kobyakov}}\ and\ \citenamefont
		{{Pethick}}(2015)}]{kp15}%
	\BibitemOpen
	\bibfield  {author} {\bibinfo {author} {\bibfnamefont {D.}~\bibnamefont
			{{Kobyakov}}}\ and\ \bibinfo {author} {\bibfnamefont {C.~J.}\ \bibnamefont
			{{Pethick}}},\ }\bibfield  {title} {\bibinfo {title} {Elastic properties of
			polycrystalline dense matter},\ }\href
	{https://doi.org/10.1093/mnrasl/slv027} {\bibfield  {journal} {\bibinfo
			{journal} {\mnras}\ }\textbf {\bibinfo {volume} {449}},\ \bibinfo {pages}
		{L110} (\bibinfo {year} {2015})},\ \Eprint {https://arxiv.org/abs/1502.02461}
	{arXiv:1502.02461 [astro-ph.SR]} \BibitemShut {NoStop}%
	\bibitem [{\citenamefont {{Baiko}}\ and\ \citenamefont
		{{Kozhberov}}(2017)}]{BK17}%
	\BibitemOpen
	\bibfield  {author} {\bibinfo {author} {\bibfnamefont {D.~A.}\ \bibnamefont
			{{Baiko}}}\ and\ \bibinfo {author} {\bibfnamefont {A.~A.}\ \bibnamefont
			{{Kozhberov}}},\ }\bibfield  {title} {\bibinfo {title} {Anisotropic crystal
			structure of magnetized neutron star crust},\ }\href
	{https://doi.org/10.1093/mnras/stx1270} {\bibfield  {journal} {\bibinfo
			{journal} {\mnras}\ }\textbf {\bibinfo {volume} {470}},\ \bibinfo {pages}
		{517} (\bibinfo {year} {2017})},\ \Eprint {https://arxiv.org/abs/1704.05322}
	{arXiv:1704.05322 [astro-ph.HE]} \BibitemShut {NoStop}%
	\bibitem [{\citenamefont {{Chugunov}}(2021)}]{C21_elastCoins}%
	\BibitemOpen
	\bibfield  {author} {\bibinfo {author} {\bibfnamefont {A.~I.}\ \bibnamefont
			{{Chugunov}}},\ }\bibfield  {title} {\bibinfo {title} {Neutron star crust in
			{Voigt} approximation: general symmetry of the stress-strain tensor and an
			universal estimate for the effective shear modulus},\ }\href
	{https://doi.org/10.1093/mnrasl/slaa173} {\bibfield  {journal} {\bibinfo
			{journal} {\mnras}\ }\textbf {\bibinfo {volume} {500}},\ \bibinfo {pages}
		{L17} (\bibinfo {year} {2021})},\ \Eprint {https://arxiv.org/abs/2010.08398}
	{arXiv:2010.08398 [astro-ph.HE]} \BibitemShut {NoStop}%
	\bibitem [{\citenamefont {{Baiko}}\ and\ \citenamefont
		{{Chugunov}}(2018)}]{BC18}%
	\BibitemOpen
	\bibfield  {author} {\bibinfo {author} {\bibfnamefont {D.~A.}\ \bibnamefont
			{{Baiko}}}\ and\ \bibinfo {author} {\bibfnamefont {A.~I.}\ \bibnamefont
			{{Chugunov}}},\ }\bibfield  {title} {\bibinfo {title} {Breaking properties of
			neutron star crust},\ }\href {https://doi.org/10.1093/mnras/sty2259}
	{\bibfield  {journal} {\bibinfo  {journal} {\mnras}\ }\textbf {\bibinfo
			{volume} {480}},\ \bibinfo {pages} {5511} (\bibinfo {year} {2018})},\ \Eprint
	{https://arxiv.org/abs/1808.06415} {arXiv:1808.06415 [astro-ph.HE]}
	\BibitemShut {NoStop}%
	\bibitem [{\citenamefont {{Cumming}}\ \emph {et~al.}(2017)\citenamefont
		{{Cumming}}, \citenamefont {{Brown}}, \citenamefont {{Fattoyev}},
		\citenamefont {{Horowitz}}, \citenamefont {{Page}},\ and\ \citenamefont
		{{Reddy}}}]{Cumming_ea17}%
	\BibitemOpen
	\bibfield  {author} {\bibinfo {author} {\bibfnamefont {A.}~\bibnamefont
			{{Cumming}}}, \bibinfo {author} {\bibfnamefont {E.~F.}\ \bibnamefont
			{{Brown}}}, \bibinfo {author} {\bibfnamefont {F.~J.}\ \bibnamefont
			{{Fattoyev}}}, \bibinfo {author} {\bibfnamefont {C.~J.}\ \bibnamefont
			{{Horowitz}}}, \bibinfo {author} {\bibfnamefont {D.}~\bibnamefont {{Page}}},\
		and\ \bibinfo {author} {\bibfnamefont {S.}~\bibnamefont {{Reddy}}},\
	}\bibfield  {title} {\bibinfo {title} {Lower limit on the heat capacity of
			the neutron star core},\ }\href {https://doi.org/10.1103/PhysRevC.95.025806}
	{\bibfield  {journal} {\bibinfo  {journal} {\prc}\ }\textbf {\bibinfo
			{volume} {95}},\ \bibinfo {eid} {025806} (\bibinfo {year} {2017})},\ \Eprint
	{https://arxiv.org/abs/1608.07532} {arXiv:1608.07532 [astro-ph.HE]}
	\BibitemShut {NoStop}%
	\bibitem [{\citenamefont {{Haensel}}\ and\ \citenamefont
		{{Zdunik}}(1990)}]{HZ90}%
	\BibitemOpen
	\bibfield  {author} {\bibinfo {author} {\bibfnamefont {P.}~\bibnamefont
			{{Haensel}}}\ and\ \bibinfo {author} {\bibfnamefont {J.~L.}\ \bibnamefont
			{{Zdunik}}},\ }\bibfield  {title} {\bibinfo {title} {Non-equilibrium
			processes in the crust of an accreting neutron star},\ }\href@noop {}
	{\bibfield  {journal} {\bibinfo  {journal} {\aap}\ }\textbf {\bibinfo
			{volume} {227}},\ \bibinfo {pages} {431} (\bibinfo {year}
		{1990})}\BibitemShut {NoStop}%
	\bibitem [{\citenamefont {{Haensel}}\ and\ \citenamefont
		{{Zdunik}}(2003)}]{HZ03}%
	\BibitemOpen
	\bibfield  {author} {\bibinfo {author} {\bibfnamefont {P.}~\bibnamefont
			{{Haensel}}}\ and\ \bibinfo {author} {\bibfnamefont {J.~L.}\ \bibnamefont
			{{Zdunik}}},\ }\bibfield  {title} {\bibinfo {title} {Nuclear composition and
			heating in accreting neutron-star crusts},\ }\href
	{https://doi.org/10.1051/0004-6361:20030708} {\bibfield  {journal} {\bibinfo
			{journal} {\aap}\ }\textbf {\bibinfo {volume} {404}},\ \bibinfo {pages} {L33}
		(\bibinfo {year} {2003})},\ \Eprint {https://arxiv.org/abs/astro-ph/0305220}
	{astro-ph/0305220} \BibitemShut {NoStop}%
	\bibitem [{\citenamefont {{Haensel}}\ and\ \citenamefont
		{{Zdunik}}(2008)}]{HZ08}%
	\BibitemOpen
	\bibfield  {author} {\bibinfo {author} {\bibfnamefont {P.}~\bibnamefont
			{{Haensel}}}\ and\ \bibinfo {author} {\bibfnamefont {J.~L.}\ \bibnamefont
			{{Zdunik}}},\ }\bibfield  {title} {\bibinfo {title} {Models of crustal
			heating in accreting neutron stars},\ }\href
	{https://doi.org/10.1051/0004-6361:20078578} {\bibfield  {journal} {\bibinfo
			{journal} {\aap}\ }\textbf {\bibinfo {volume} {480}},\ \bibinfo {pages} {459}
		(\bibinfo {year} {2008})},\ \Eprint {https://arxiv.org/abs/0708.3996}
	{arXiv:0708.3996} \BibitemShut {NoStop}%
	\bibitem [{\citenamefont {{Fantina}}\ \emph {et~al.}(2018)\citenamefont
		{{Fantina}}, \citenamefont {{Zdunik}}, \citenamefont {{Chamel}},
		\citenamefont {{Pearson}}, \citenamefont {{Haensel}},\ and\ \citenamefont
		{{Goriely}}}]{Fantina_ea18}%
	\BibitemOpen
	\bibfield  {author} {\bibinfo {author} {\bibfnamefont {A.~F.}\ \bibnamefont
			{{Fantina}}}, \bibinfo {author} {\bibfnamefont {J.~L.}\ \bibnamefont
			{{Zdunik}}}, \bibinfo {author} {\bibfnamefont {N.}~\bibnamefont {{Chamel}}},
		\bibinfo {author} {\bibfnamefont {J.~M.}\ \bibnamefont {{Pearson}}}, \bibinfo
		{author} {\bibfnamefont {P.}~\bibnamefont {{Haensel}}},\ and\ \bibinfo
		{author} {\bibfnamefont {S.}~\bibnamefont {{Goriely}}},\ }\bibfield  {title}
	{\bibinfo {title} {Crustal heating in accreting neutron stars from the
			nuclear energy-density functional theory. {I}. {Proton} shell effects and
			neutron-matter constraint},\ }\href
	{https://doi.org/10.1051/0004-6361/201833605} {\bibfield  {journal} {\bibinfo
			{journal} {\aap}\ }\textbf {\bibinfo {volume} {620}},\ \bibinfo {eid} {A105}
		(\bibinfo {year} {2018})},\ \Eprint {https://arxiv.org/abs/1806.03861}
	{arXiv:1806.03861 [astro-ph.HE]} \BibitemShut {NoStop}%
	\bibitem [{\citenamefont {{Ootes}}\ \emph
		{et~al.}(2019{\natexlab{b}})\citenamefont {{Ootes}}, \citenamefont
		{{Wijnands}},\ and\ \citenamefont {{Page}}}]{Ootes_ea19_LongTerm}%
	\BibitemOpen
	\bibfield  {author} {\bibinfo {author} {\bibfnamefont {L.~S.}\ \bibnamefont
			{{Ootes}}}, \bibinfo {author} {\bibfnamefont {R.}~\bibnamefont
			{{Wijnands}}},\ and\ \bibinfo {author} {\bibfnamefont {D.}~\bibnamefont
			{{Page}}},\ }\bibfield  {title} {\bibinfo {title} {Long-term temperature
			evolution of neutron stars undergoing episodic accretion outbursts},\ }\href
	{https://doi.org/10.1051/0004-6361/201936035} {\bibfield  {journal} {\bibinfo
			{journal} {\aap}\ }\textbf {\bibinfo {volume} {630}},\ \bibinfo {eid} {A95}
		(\bibinfo {year} {2019}{\natexlab{b}})},\ \Eprint
	{https://arxiv.org/abs/1906.02554} {arXiv:1906.02554 [astro-ph.HE]}
	\BibitemShut {NoStop}%
	\bibitem [{\citenamefont {{Ghosh}}\ and\ \citenamefont {{Lamb}}(1979)}]{gl79b}%
	\BibitemOpen
	\bibfield  {author} {\bibinfo {author} {\bibfnamefont {P.}~\bibnamefont
			{{Ghosh}}}\ and\ \bibinfo {author} {\bibfnamefont {F.~K.}\ \bibnamefont
			{{Lamb}}},\ }\bibfield  {title} {\bibinfo {title} {Accretion by rotating
			magnetic neutron stars. {III} - accretion torques and period changes in
			pulsating {X}-ray sources},\ }\href {https://doi.org/10.1086/157498}
	{\bibfield  {journal} {\bibinfo  {journal} {\apj}\ }\textbf {\bibinfo
			{volume} {234}},\ \bibinfo {pages} {296} (\bibinfo {year}
		{1979})}\BibitemShut {NoStop}%
	\bibitem [{\citenamefont {{Srinivasan}}\ \emph {et~al.}(1990)\citenamefont
		{{Srinivasan}}, \citenamefont {{Bhattacharya}}, \citenamefont {{Muslimov}},\
		and\ \citenamefont {{Tsygan}}}]{Srinivasan_ea90_n5}%
	\BibitemOpen
	\bibfield  {author} {\bibinfo {author} {\bibfnamefont {G.}~\bibnamefont
			{{Srinivasan}}}, \bibinfo {author} {\bibfnamefont {D.}~\bibnamefont
			{{Bhattacharya}}}, \bibinfo {author} {\bibfnamefont {A.~G.}\ \bibnamefont
			{{Muslimov}}},\ and\ \bibinfo {author} {\bibfnamefont {A.~J.}\ \bibnamefont
			{{Tsygan}}},\ }\bibfield  {title} {\bibinfo {title} {A novel mechanism for
			the decay of neutron star magnetic fields},\ }\href@noop {} {\bibfield
		{journal} {\bibinfo  {journal} {\currsci}\ }\textbf {\bibinfo {volume}
			{59}},\ \bibinfo {pages} {31} (\bibinfo {year} {1990})}\BibitemShut {NoStop}%
	\bibitem [{\citenamefont {{Spitkovsky}}(2006)}]{Spitkovsky06}%
	\BibitemOpen
	\bibfield  {author} {\bibinfo {author} {\bibfnamefont {A.}~\bibnamefont
			{{Spitkovsky}}},\ }\bibfield  {title} {\bibinfo {title} {Time-dependent
			force-free pulsar magnetospheres: Axisymmetric and oblique rotators},\ }\href
	{https://doi.org/10.1086/507518} {\bibfield  {journal} {\bibinfo  {journal}
			{\apjl}\ }\textbf {\bibinfo {volume} {648}},\ \bibinfo {pages} {L51}
		(\bibinfo {year} {2006})},\ \Eprint {https://arxiv.org/abs/astro-ph/0603147}
	{astro-ph/0603147} \BibitemShut {NoStop}%
	\bibitem [{\citenamefont {{Freire}}()}]{Freire_GC_Pulsar_Cat_y25m07d25}%
	\BibitemOpen
	\bibfield  {author} {\bibinfo {author} {\bibfnamefont {P.}~\bibnamefont
			{{Freire}}},\ }\href@noop {} {\bibinfo {title} {Pulsars in globular clusters;
			downloaded on july 25, 2025}},\ \bibinfo {howpublished}
	{https://www3.mpifr-bonn.mpg.de/staff/pfreire/GCpsr.html}\BibitemShut
	{NoStop}%
\end{thebibliography}
%

\onecolumngrid
\section*{End Matter}
\twocolumngrid
{\it NS formation: observations and theory.}
The degenerate collapse may occur in three different 
configurations which are
an accretion-induced collapse (AIC) of an 
ONe
white dwarf (WD),
a merger-induced collapse
(MIC) of an 
ONe
WD upon a merger with another WD, 
or a collapse of the core of a super-AGB star (also known
as an electron-capture supernova, ECSN)
\cite{CS76,M+80,SN85,NK91,Ivanova_etal08,Wang_ea26_ECSN_review}.
NSs produced in ECSN or AIC are 
expected to have a low natal kick \cite{P+04,GessnerJanka18_HD_ECSN}.
By contrast,
normal rotation-powered pulsars, presumably originating in FeCC, have birth velocity distribution 
well described 
\cite{H+05} 
by a Maxwellian 
with a mean three-dimensional 
velocity of $\sim 400$ km/s.
In view of that, it has been long suspected 
\cite{BG90_NS_MSP_from_AIC,BvdH91,P+02,KP06,Ivanova_etal08} 
that majority of NSs in GCs originated via a degenerate collapse
as, for an NS to be retained in a GC after birth, it is required 
to receive a natal kick of no more than $\sim 50$ km/s. 
Besides that, there was a large body of work,  
suggesting that SN1054 which produced 
the poster-child rotation-powered pulsar Crab, was of the ECSN type 
based on composition and dynamics of its supernova remnant
\cite{H82,N+82,S13,M+14}.
However, the pulsar kick velocity of $\sim 160$ km/s was shown to be 
impossible to obtain by the preferred hydrodynamic 
mechanism in ECSN models \cite{GessnerJanka18_HD_ECSN}.

NSs produced in ECSN or AIC are expected to have a low mass: the gravitational 
mass of an NS formed by these mechanisms should be equal to the 
critical mass limit minus the binding energy,
which yields $\sim 1.25 \, M_\odot$ \cite{P+05}.
On account of this and the low natal kick argument,
ECSN appears to be required for the formation of double 
NS systems, where it has been proposed as the origin of the double pulsar 
member J0737-3039B and of a number of similar 
objects 
\cite{vdH04, P+04, P+05, vdH07, S+10, F+13}, 
as well as 
for certain classes of high-mass X-ray binaries \cite{Pf+02, K+11}. 

Unlike FeCC and ECSN,
AIC and MIC can produce NS with a substantial delay
(up to the Hubble time \cite{Ruiter_ea19_AIC})
with respect to a star formation episode.
In view of that, AIC was invoked to explain 
six seemingly young, moderately spinning pulsars, observed in GCs 
\cite{B+11,Wu+24_3PSR_M15,Kremer_ea24_NoPartialRecycling}.
All six have $B$-fields of $10^{11} - 2\times 10^{12}$ G
and if they were born during star formation episodes in GC (i.e. some Gyrs ago), 
they would have had enough time to spin down 
to much lower frequencies.
Their inferred young ages thus 
require delayed formation, implying AIC 
(or MIC \cite{Tauris_ea13_AIC,Kremer_ea23_MIC_for_young_Pulsars_in_GCs}) scenario.

Both, AIC and MIC can produce rapidly spinning NS by tapping just a fraction of the angular 
momentum reservoir available in a binary.
MIC likely results in a higher NS mass as only a minor amount of mass,
$\lesssim 0.1 \, M_\odot$,
is expected to be ejected in a merger \cite{D+14,K+18,Ruiter_ea19_AIC}.
Simulations of AIC of a super-Chandrasekhar mass 
WD supported by rapid rotation have been also performed \cite{D+06,Kuroda_ea25_rotWD_collapse}.

{\it Breaking parameter: observations and theory.}
Another hydrostatic equilibrium condition can be written at $\rho_2$, where 
the same accreted mass 
causes an additional stress which is
equilibrated by a critical compression of the crystal.
The respective change of the hydrostatic pressure 
can be written as a sum of two terms: the first one is associated with  
a change of electron density in an ion crystal critically contracted in 
the vertical direction. The second one is due to compression and pressure anisotropy 
of the ion lattice \cite{oi90,Strohmayer_etal91,Baiko11,kp15,BK17,C21_elastCoins}.
This leads to the equation:
\beq
P (\rho_1) = \delta_{\rm crit} \rho_2 \frac{{\rm d}P(\rho_2)}{{\rm d}\rho}
+\delta_{\rm crit} \, C n_2 \, \frac{Z_2^2 e^2}{a_2}~.
\label{Compression_rho2}
\eeq
In this case, 
$n_2=\rho_2/(A_2 m_u)$ is the ion number density at $\rho_2$, 
$a_2 = (4\pi n_2/3)^{-1/3}$
is the respective Wigner-Seitz radius, 
$Z_2$ and $A_2$ are the charge and mass numbers at $\rho_2$.
The constant $C\approx 0.24$ can be estimated using Voigt averaged 
elastic properties of the crust \cite{C21_elastCoins}.

Assuming
$P\propto n_\mathrm{e}^{4/3}$ \cite{hpy07},
we obtain
\beq
\left(\frac{Y_1 \rho_1}{Y_2 \rho_2}\right)^{4/3} = \frac{4}{3} \delta_{\rm crit}
+0.04 (Z_2/40)^{2/3}\delta_{\rm crit}~,
\label{hydro2_2}       
\eeq
and neglecting the last term, we arrive at
\beq
\delta_{\rm crit} \approx \frac{3}{4}  \left(\frac{Y_1\, \rho_1}{Y_2\,\rho_2}\right)^{4/3}~,
\label{delt_crit}
\eeq
where $Y_i=Y(\rho_i)$, $i=1,2$, $\delta_{\rm crit} \equiv 1-\xi_{\rm crit}$, 
and $\xi_{\rm crit}$ is the critical unidirectional
contraction factor of crystal matter \cite{BK17,BC18}.  

Substituting 
numerical values, 
we get $\delta_{\rm crit} \approx 0.047$. 
It has been established \cite{BC18} that the minimum $\delta_{\rm crit}$ over crystallographic
directions of contraction was 0.065. 
This result
did not take into account various effects which reduce $\delta_{\rm crit}$:
electron screening, ion vibrations, crystal imperfections.
In view of this, we consider 
$\delta_{\rm crit} \approx 0.047$ deduced from observations to be in 
an adequate agreement with the theory.
On top of that, there is likely a fair amount of uncertainty in the determination
of $\rho_1$ and $\rho_2$ from cooling simulations. For example, 
if $\rho_1$ is set to $3.8\times 10^{10}$ g/cc, this
would bring 
$\delta_{\rm crit}$ into a perfect agreement with the theoretical prediction \cite{BC18}.

{\it Thermal state of $\mathrm{NS_{J1748}}$.}
Quiescent state
of $\mathrm{NS_{J1748}}$ 
corresponds to internal temperature $T\sim 10^8$ K \cite{Ootes_ea19}.
Assuming a low-mass NS ($M \lesssim 1.4 \, M_\odot$)
and strongly superfluid nucleons, the thermal energy 
${\cal E}_{\rm th}$ of the NS can be estimated on the basis 
of its lepton heat capacity of
$\lesssim 2\times 10^{37}\left(T/10^8~\mathrm{K}\right)$ erg/K  \cite{Cumming_ea17}.
This yields ${\cal E}_{\rm th} \lesssim 10^{45}$ erg. 
Dividing ${\cal E}_{\rm th}$ by $\Delta M$ (Eq.\ \ref{dM}), we get an estimate 
of the required heating energy per accreted nucleon 
as $E_{\rm h,est} \lesssim 0.2$ MeV/nucleon.
It is a few times greater than the energy $E_{\rm h}$ predicted by 
theoretical deep crustal heating 
models \cite{HZ90,HZ03,HZ08,Fantina_ea18,Lau_ea18,Shchechilin_ea21} 
for the accreted layer 
constrained by the density $\rho<\rho_1$.

However, it is well known that short-term thermal 
evolution models require an additional shallow heating source $E_{\rm sh}$ to be 
consistent with observations.
For $\mathrm{NS_{J1748}}$,
the energy of this 
source reaches  
$E_{\rm sh}\sim (1.5-2)$ MeV/nucleon \cite{Ootes_ea19,Potekhin_ea25_transients}. 
In modeling \cite{Potekhin_ea25_transients}, 
the source was placed 
at $\rho=1.4\times 10^9$ g/cc, and
from this location, a significant fraction of shallow heating energy 
goes into NS interior \cite{Ootes_ea19_LongTerm,Potekhin_ea25_transients}.
Thus, we conclude that under the assumptions of a relatively low
NS mass,
strongly suppressed nucleon 
heat capacity, and 
partial core heating by the shallow heating mechanism, there is no contradiction 
between the observed thermal state
of $\mathrm{NS_{J1748}}$
and the $\Delta M$ value Eq.\ (\ref{dM}).

If $M$ increases, $\Delta M$ decreases (cf.\ Eq.\ \ref{dM}). Besides that,
nucleons might not be superfluid in the central part of a massive NS in some models 
of superfluid critical temperatures \cite{Potekhin_ea23_transients}.
All of these effects amplify $E_{\rm h,est}$ making it harder to reconcile
heating requirements with available $E_{\rm h}+f E_{\rm sh}$ ($f<1$ is a fraction of 
shallow heating, which goes into NS interior).
In principle, quantitative simulation of accretion history in 
J1748
with known $\Delta M$
should help further constrain NS parameters and crustal heating models. 
Although more work is needed, we take this as a tentative observational evidence 
of a relatively low mass of $\mathrm{NS_{J1748}}$ 
and of 
a relatively strong nucleon superfluidity in its core.

{\it Accretion torque and birth parameters of $\mathrm{NS_{J1748}}$.} 
A similar $\Delta \nu$ estimate
can be obtained using
the standard theory of the accretion torque \cite{PR72,L+73}.
It can be written as 
\beq
N = \dot{M}_{\rm b} (G M r_{\rm A})^{1/2}~,
\label{torq}
\eeq
where $r_{\rm A}$ is the Alfven radius
\beq
r_{\rm A} = \mu^{4/7} (G M \dot{M}_{\rm b}^2)^{-1/7}.
\label{rA}
\eeq

The equatorial magnetic field at the surface of this NS at the present 
time is estimated
as 
$B_{\rm obs} \sim 2 \times 10^8 - 2.4 \times 10^{10}$ G 
\cite{Papitto_ea11_IGR17480}
which corresponds to magnetic moment
$\mu_{\rm obs} \sim 2\times 10^{26} - 2.4\times 10^{28}$ G cm$^3$ 
(following a tradition, we assume NS radius equal to 10 km while 
converting magnetic field to magnetic moment and vice versa).
We integrate Eq.\ (\ref{torq}), neglecting any variation of 
$\dot{M}^{-1/7}_{\rm b}$ within and between accretion episodes. 
Using canonical NS mass $M=1.4 \, M_\odot$ and moment of inertia $I=10^{45}$ g cm$^2$,
we obtain the total spin-up $\Delta \nu\sim 0.06$ Hz for the upper limit of the magnetic moment. 
Although this is almost twice
the estimate based on the observed spin-up rate 
during the outburst, 
it is still almost $\sim 200$ times
lower than $\nu_{\rm obs}$. 
Note, that more elaborate accretion torque models predict lower 
magnetosphere radius and lower torque \cite{gl79b}.

Let us denote magnetic
field (moment) at the end of the spin-down epoch as $B_n$ ($\mu_n$),
where $n$ is the braking index,
and consider two representative 
spin-down models $n=3$ and $n=5$.
The $n=3$ model assumes no magnetic field decay during spin-down so that
magnetic field at birth 
$B_\ini=B_3$ ($\mu_\ini=\mu_3$). 
In the $n=5$ model \cite{Srinivasan_ea90_n5},
one assumes
that the magnetic field decays proportionally to the spin frequency
so that 
$\mu(t) = \mu_\ini~\nu(t)/\nu_\ini$. 

In both models, the source age is given by the formula:
\beq
t_n =
\frac{c^3 I}{4 K (n-1)\,\pi^2\,\mu_n^2 \nu_{\rm obs}^2}
\left(1-\frac{\nu_{\rm obs}^{n-1}}{\nu^{n-1}_\ini}\right)~,
\label{sd_age}
\eeq
where $c$ is the speed of light and the parameter $K$ describes 
the 
braking efficiency. 
Calculations of force-free 
magnetospheres \cite{Spitkovsky06} yield
$K\approx 1+\sin^2\alpha$, where $\alpha$ is the obliqueness.
For estimates, we take the lower limit, $K=1$, leading 
to an overestimation of $\mu_n$ (at a given age).

If $n=3$, the magnetic field is conserved, $B_\ini=B_3$.
Thus, if we take 
$t_3 \gtrsim 2$ Gyr
and assume a rapid rotation at birth ($\nu_\ini \gg \nu_{\rm obs}$),
we get the magnetic field at birth
$B_\ini\lesssim 7\times 10^9$ G.
For lower magnetic fields, $B_\ini\ll 7\times 10^9$ G,
$\nu_\ini$ approaches the present day value.
A higher initial magnetic field,
$B_\ini> 7\times 10^9$ G, would require a delay of the NS birth. 

In the $n=5$ model, 
assuming $\nu_\ini\gg \nu_{\rm obs}$ and 
$t_5 \gtrsim 2$ Gyr, one obtains $B_5\lesssim 5\times 10^9$ G, which 
implies a range of birth parameters
(region hatched by solid lines in Fig.\ \ref{Fig:InitParams}), for instance, $\nu_\ini=100$ Hz 
and $B_\ini\lesssim 4\times 10^{10}$ G. 
Birth of the NS during the main peak of star formation in Terzan 5, i.e.\ 
$12-13$ Gyr ago \cite{Crociati_ea24_StarFormHist_Ter5}, requires 
$B_5\lesssim 2\times 10^9$ G (in this case e.g.\ $\nu_\ini=100$ Hz  
implies $B_\ini\lesssim 2\times 10^{10}$ G).
A lower $B_5$ would correspond to a lower spin frequency at birth, whereas  
$B_5> 5\times 10^9$ G means NS age below 2 Gyr, i.e.\ a delay of NS 
birth with respect to the last star formation episode.

Note that both $B_3$ and $B_5$ are in full agreement with
$B_{\rm obs}$.

\end{document}